\documentclass[12pt]{article}   
\usepackage{epsfig}
\usepackage{amsfonts}
\usepackage{amsmath}
\usepackage{times}
\usepackage{array,hhline}
\title{Confining Effective Theories Based on \\ Instantons and Merons
\footnote{This work is supported in part by funds provided by the U.S.
Department of Energy (D.O.E.) under grant
DE-FG02-94ER40818.}}
\author{F. Lenz$^{{\rm a}}$,
J. W. Negele$^{{\rm b}}$ and M. Thies$^{{\rm a}}$ \\ \\
$^{{\rm a}}$ Institute for Theoretical Physics III \\
University of Erlangen-N\"urnberg \\
Staudstrasse 7, 91058 Erlangen, Germany\\ \\
$^{{\rm b}}$ Center for Theoretical Physics, \\
Laboratory for Nuclear Science, and Department of Physics \\
Massachusetts Institute of Technology \\
Cambridge, Massachusetts 02139, U.S.A. \\ \\
FAU-TP3-07/06, MIT-CTP-3422}
\begin{document}
\date{}
\maketitle
\vspace{-.1in}
\begin{abstract} \normalsize
An effective theory based on ensembles of  either  regular gauge instantons or merons
is shown to produce confinement  in SU(2) Yang-Mills theory. When the scale is set by the string tension,
the action density, topological susceptibility and low-lying glueball spectrum are  similar  to
those arising in lattice QCD. The physical mechanism producing confinement  is explained, and a number
of analytical insights into the effective theory are presented.
\end{abstract}
\newpage
\section{Introduction}
Understanding the mechanism of confinement in quantum chromodynamics (QCD) is a fundamental but elusive challenge. Lattice
calculations convincingly
demonstrate that QCD yields confinement, but do not yet elucidate its mechanism. Although the strong coupling expansion produces
an area law by disorder already in lowest order, this disorder alone is not the whole story as evidenced by the fact
that the strong coupling expansion also erroneously yields confinement in U(1)
gauge theory. 
Hence, in this work, we develop a confining effective theory based on degrees of freedom arising naturally from QCD. We explore
in detail how confinement arises, and the degree to which this effective theory approximates the results of SU(2) lattice gauge theory.

There has been a long and fruitful history of seeking to understand  nonperturbative QCD analytically  by expanding  the path integral
 for the partition function around stationary classical solutions and evaluating the fluctuations around these solutions. 
Central to this approach was the discovery of instantons \cite{Belavin:1975fg} that satisfy the classical Euclidean Yang-Mills equations
 and implement the tunneling required between different winding number sectors of the $\Theta$ vacuum. Since instantons can be
 transformed to the singular gauge in which the gauge field falls off like $\sim 1/x^3$, a dilute gas of these instantons is an approximate
 solution to the field equations. The contribution of this dilute instanton gas to the potential between two static color charges can be 
calculated analytically, and was shown not to produce confinement \cite{Callan:1978ye}.  Whereas an instanton that contributes to two
 opposite Wilson lines in a Wilson loop does contribute to a linear term to the potential, when the integral over color orientations of 
independent instantons that affect only one or the other Wilson line is performed, these uncorrelated contributions do not produce 
a linear term. Hence at  separation larger than the instanton size, the potential becomes a constant and fails to confine.  

When instantons failed to produce confinement, there was then hope that merons \cite{deAlfaro:1976qz}, solutions to the Yang-Mills
equations with topological charge 1/2, would do so.
Although a free meron has an action that diverges logarithmically with the volume, a pair of merons has a finite action that depends 
logarithmically on the separation. If the coupling is strong enough that the logarithmic potential is weaker than the entropy associated 
with the Euclidean separation ($\sim r^3 = {\rm e}^{3 \ln r}$), meron pairs could dissociate and form a meron gas. 
Since meron  fields only fall off as $\sim 1/x$, they are expected to be more effective in disordering than singular gauge instantons and
thus potentially be a more effective mechanism for confinement \cite{Callan:1978bm,Callan:1977gz,Callan:qs}.
Unfortunately the same $\sim 1/x$ behavior that makes them a candidate for confinement renders them intractable analytically, and as 
a result they have never been shown by analytic arguments to be responsible for confinement.

Motivated by the quark zero modes associated with instantons and the 't~Hooft interaction governing light quarks, highly successful 
instanton liquid models, which had to be implemented numerically,  were developed that provide clear insight into the role instantons
play in chiral symmetry breaking and reproduce many of the salient features of QCD with light quarks
\cite{Shuryak:1981ff,Schafer:1996wv,Diakonov:1983hh,Diakonov:1985eg}.
For our present purposes, it is best to think of an instanton liquid as an effective theory in which one parameterizes a set of gauge 
configurations in terms of the positions and color orientations of an ensemble of instantons to represent the salient degrees of freedom
 for gluons. Because the field equations are nonlinear, the superposition of a set of instanton solutions is not a classical solution to the
 field equations.  Hence,  rather than summing over a set of classical solutions, 
one is approximating the original functional integral for the partition function over gauge fields by performing the integral over these 
effective degrees of freedom, for example by using the Metropolis algorithm to sample configurations. Physically, this sampling will favor 
the appropriate configurations that are energetically and entropically optimal.  Because the gauge fields of singular gauge instantons fall
off as $\sim 1/x^3$, a random instanton liquid in which the positions and color orientations of instantons are chosen randomly does not differ
qualitatively from one generated by more sophisticated Metropolis sampling, so most of the qualitative phenomena occur in a random 
instanton liquid. As in the case of the dilute instanton gas, Wilson loops in a random instanton liquid also show no confinement at distances
larger than the instanton size.

In the present work, we consider a more general class of effective theories based on ensembles of regular gauge instantons or merons, 
whose fields fall off like $\sim 1/x$, so that the resulting superposition of gauge fields is not close to a classical solution. 
Throughout this work, we will refer to regular gauge instantons or merons as {\it pseudoparticles}. The term {\it instanton} will always 
mean a regular gauge instanton, and if we wish to refer to a singular gauge instanton, we will always do so explicitly. 
Although it is analytically intractable to perform the resulting integral of the partition function over the  pseudoparticle collective 
coordinates, it is straightforward to do so numerically using the Metropolis algorithm. Whereas most random choices of pseudoparticle
fields  create a large background field yielding an unphysically high action density, Metropolis sampling selects the particular
configurations in which the collective coordinates are correlated in such a way as to produce very low action density. 
In this way, one can achieve the  balance of energy and entropy in ensembles based on regular gauge pseudoparticles that is 
conceptually equivalent to what has been done in the past with singular gauge instantons.

This more general class of theories based on pseudoparticles combines both confinement and the successes of previous singular
gauge instanton liquid models.
As will be shown below, color correlations are required to superpose the long range pseudoparticle gauge fields to create the small 
background fields whose low action allows them to dominate the path integral. These also produce the correlations necessary to yield an 
area law for large Wilson loops and thus produce confinement. The short distance ``spikes" in the gluon field arising from the short 
distance behavior of small merons or instantons behave just like the spikes in conventional singular gauge instanton liquid models, 
and thus produce the physics associated with the 't~Hooft interaction and chiral symmetry breaking. 
Pseudoparticle ensembles were first used by the present authors to study meron ensembles \cite{LNT04}, where there is no singular 
gauge alternative. However, in the context we have just described, it is clear that one should also consider regular gauge instantons. 
Indeed, since the gauge for each instanton is defined relative to its position, the fields obtained by superposing $N$ singular gauge 
instantons is drastically different from the one obtained by transforming each of these instantons to the regular gauge
and superposing the resulting fields. Neither is a classical solution of the Yang-Mills equations, and neither is preferred on physical grounds. 
Hence, in exploring the full physics based on classical solutions, it is essential to study both alternatives. 
Although we will not develop it further in this work, Ref.~\cite{LNT04} discusses how one may expect center symmetry to be realized in
 pseudoparticle effective theories.
A closely related effective theory \cite{Wagner:2006qn} superposes pseudoparticle fields with dynamically determined amplitudes and 
generalizes to finite temperature, where a clear crossover in the Polyakov loop is observed from a nearly vanishing value at zero 
temperature to a magnitude close to unity at high temperature.

The primary goals of this work are to understand the mechanism by which this effective theory produces confinement and other 
observed properties of SU(2) gauge theory, and explore its success and limitations in quantitatively approximating the results of 
lattice QCD calculations of the action density, topological susceptibility, and glueball masses. Although we will discuss both 
instanton and meron pseudoparticles in detail and give extensive examples for both, the primary focus in comparing with the lattice 
QCD spectrum is on the confining instanton theory.  We will show that the confining instanton theory has essentially all the 
advantages of previous random and interacting singular gauge instanton models, with the added feature of incorporating confinement.  
To obtain insight into the effective pseudoparticle theory, we will strongly emphasize analytic arguments wherever possible, as 
indicated in the outline below.

The outline of this paper is as follows.  Section 2 describes the details of the pseudoparticle ensembles and some relevant analytical
properties. In particular, since one might be concerned that the long range fields of pseudoparticles are necessarily large, we 
present a simple example of how sets of merons can be chosen such that the leading order $1/x^4$ term in the action at large distances 
vanishes identically. Section 3 describes random and correlated pseudoparticle ensembles. For pedagogical purposes, we begin with an 
ensemble in which the pseudoparticle positions and color orientations are selected randomly without correlations. Although this 
ensemble has unphysical features, it enables us to make instructive analytical arguments. We then allow Metropolis dynamics to 
produce the appropriate color and spatial correlations. 
The resulting gauge fields are shown to have a proper thermodynamic limit and to be composed of small background fields plus 
``spikes" at the locations of pseudoparticles. 
By separately applying Metropolis dynamics to only color parameters and then both color and spatial parameters, we show that the
dominant correlations are in the color degrees of freedom.
Section 4 addresses Wilson loops and confinement. When scaled appropriately, Wilson loops are shown to lie on a universal scaling
curve, which is nearly independent of the pseudoparticle size and the coupling constant and which shows a clear area law indicating
confinement. The behavior of Wilson loops in both the small size and large size limits is explained analytically.  We also show that the 
distribution of Wilson loops follows the diffusion behavior of Ref.~\cite{blnt05} 
and that Wilson loops in higher representations exhibit Casimir scaling, both of which are in agreement with lattice QCD.
Section 5 presents calculations of physical observables.  Since the theory is confining, we can set the scale by the measured string 
tension and compare unambiguously with lattice calculations in which observables are also specified in terms of the measured string 
tension.  The two most robust observables are the gluon condensate and the topological susceptibility, and both instanton and meron 
ensembles are shown to yield values in units of the string tension in qualitative agreement with lattice QCD. We show that for many 
operators of interest, the effective theory produces Euclidean correlation functions with the proper physical behavior.  The correlation 
function of the topological charge density displays the proper negativity behavior, and operators constructed from field-strengths and
Wilson loops enable us to measure the masses of $0^+$, $1^-$, $1^+$, $2^+$, and $2^-$ glueball states. Aside from an overall scale factor
which we discuss, this glueball spectrum is in qualitative agreement with lattice QCD.
The conclusions are summarized in the final Section 6.
%####################################################################################
\section{Ensembles of Pseudoparticles}
\subsection{Pseudoparticle Properties and Definition of Pseudoparticle Ensembles}
In this work, we explore the idea that pseudoparticles, i.e., merons or regular gauge instantons, are the essential degrees of
freedom in SU(2) Yang-Mills gauge theory by writing the partition function as a
path integral of an effective action depending on the positions and color
orientations of an ensemble of  pseudoparticles,
\begin{equation}
   \label{pathintegral}
Z = \int {\rm d} z_i {\rm d} h_i {\rm e}^{- \frac{1}{g^2} S[A(z_i,h_i)]         }\, .
\end{equation}
We identify the effective action $S$ with the standard (Euclidean) action
\begin{equation}
   \label{act0}
S=\int_V {\rm d}^4 x s(x), \quad s(x)= \frac{1}{4} F_{\mu\nu}^a F_{\mu\nu}^a\, ,
\end{equation}
associated with the fields in the space-time region $V$.  Dual field-strength and topological charge density are defined by 
\begin{equation}
\tilde{F}_{\mu\nu}^a=\frac{1}{2}\,\epsilon_{\mu\nu\sigma\rho}F^{a\,\sigma\rho}\,,\quad {\tilde s}(x) =
\frac{1}{4}F_{\mu\nu}^a\tilde{F}_{\mu\nu}^a\, .
\label{tocd}
\end{equation}
The gauge field for a pseudoparticle  in Lorentz gauge with its center  at the
origin, after appropriate choice of the coordinate system in color space  and
after regularization of  the singularity,  is given by 
\begin{equation}
    \label{mer}
     a_{\mu}(x) = \xi\,\frac{\eta _{a\mu \nu } x_{\nu}}{x^{2} + \rho ^{2}}
\frac{\sigma^a}{2}\,,
\end{equation}
with $\xi=1$ denoting merons and $\xi=2$ instantons.
Color and space-time dependence are correlated via the 't~Hooft tensor
$$
\eta_{a \mu \nu} =\epsilon _{a\mu \nu} +\delta_{a \mu}\delta_{\nu 0}-
\delta _{a \nu}\delta_{\mu 0} .$$
For instantons and for merons of vanishing  size $\rho$, $a_{\mu}(x)$ is a solution of the  Euclidean
classical  field equations  \cite{Belavin:1975fg,deAlfaro:1976qz}. The gauge fields of anti-merons or anti-instantons differ in sign if
one of the space-time indices is 0. The field-strength of the pseudoparticles, its dual, the  action density, the topological charge
density and the topological charge are given by  
\begin{equation}
  \label{fssm}
  F_{\mu\nu}^a=\frac{\xi}{(x^2+\rho^2)^2}\Big[-\eta_{a\mu\nu}((2-\xi)x^2+2\rho^2)+(2-\xi)(\eta_{a\mu\rho}x^{\nu}-\eta_{a\nu\rho}x^{\mu})x^{\rho}\Big]\, ,
\end{equation}
\begin{equation}
  \label{dfssm}
 \tilde{F}_{\mu\nu}^a=\frac{\xi}{(x^2+\rho^2)^2}\Big[-2\rho^2 \eta_{a\mu\nu}-(2-\xi)(\eta_{a\mu\rho}x^{\nu}-\eta_{a\nu\rho}x^{\mu})x^{\rho}\Big] \, ,
\end{equation}
\begin{eqnarray}
  \label{smde1}
  s_0(x) &=& \frac{3\xi^2}{2(x^2+\rho^2)^4}\big[((2-\xi)x^2+2\rho^2)^2+4\rho^4\big]\\
  \tilde{s}_0(x) &=& \pm \frac{6\xi^2 \rho^2}{(x^2+\rho^2)^4}\big[2\rho^2+(2-\xi)x^2\big]\, ,
\label{smde2}
\end{eqnarray}
\begin{equation}
  \label{tocha}
  \nu=\frac{1}{8\pi^2}\int {\rm d}^4 x \tilde{s}_0(x) = \pm \frac{\xi}{2}\, . 
\end{equation}
Pseudoparticles and anti-pseudoparticles differ in the sign of the topological charge and its density.   
We note the difference in the  asymptotic behavior of the field-strength of instantons and merons. The meron field-strength decays 
asymptotically as $1/x^2$, giving rise to an infrared logarithmic singularity in the action. 
For vanishing size,  the action of a meron  is also logarithmically divergent  in the
ultraviolet, so for merons, the size parameter $\rho$ acts as an ultraviolet regulator.  

An important quantity in our study of confinement will be the Wilson loop 
\begin{equation}
  \label{wilo}
  W = \frac{1}{2} \,\mbox{tr}\left\{ P \exp {\rm i}g \oint_{\cal C}A_{\mu}(x){\rm d}x^\mu\right\}\,.
\end{equation}
The integral in (\ref{wilo})  is ordered along the closed path ${\cal C}$.
Here we 
compile the relevant result for Wilson loops in the field of single pseudoparticles. Pseudoparticle fields [Eq.~(\ref{mer})] in a ``spacelike'' 
plane with $x_{4}=0$ are of the form
$${\bf a} = \frac{\xi}{2}\frac{{\bf x}\times \mbox{\boldmath$\sigma$}}{{\bf x}^2+\rho^2}\,.$$
Analytical expressions can be derived if the center of the pseudoparticle fields is in the plane of the Wilson loop. In this case, 
for a loop in the ($x_1,x_2$) plane, 
$$a_{1}=\frac{\xi}{2}\frac{x_{2}}{x_{1}^2+x_{2}^2+\rho^2}\sigma_{3},\quad a_{2}=-\frac{\xi}{2}\frac{x_{1}}{x_{1}^2+x_{2}^2+\rho^2}\sigma_{3}$$
and the field is abelian pointing in the 3 direction in color space. For calculation of the Wilson loop we can use  Stokes' theorem with
$$( \mbox{rot} \,{\bf a}^{3})_{3} = \frac{\xi}{2}\frac{\rho^2}{(x_{1}^2+x_{2}^2+\rho^2)^2}$$
and obtain for a circular Wilson loop (radius $r$) with the center of the circle and of the pseudoparticle  coinciding
$$\oint a_{\mu}\, {\rm d}x^{\mu} = \int_{\Sigma} {\rm d}\mbox{\boldmath$\sigma$} \mbox{rot}\,{\bf a} =
\xi\pi\,\frac{r^2}{r^2+\rho^2}\sigma_{3}.$$ 
In the limit of vanishing size,
$$ \mbox{rot}\,{\bf a}_{3} = \xi\pi\,\delta^{(2)}({\bf x})\sigma_{3}, \quad \rho \rightarrow 0$$
and therefore the Wilson loop in a plane containing the singularity of the pseudoparticle field (position ${\bf x}$) becomes
\begin{equation}
\label{swil0}
W({\bf x})= (-1)^{\xi \theta ({\bf x})},\quad  \theta ({\bf x})=1\, (0) \; \mbox{if} \;{\bf x}\, \mbox{ is inside (outside) the loop}.\end{equation}
Instantons generate a trivial Wilson loop. This is true everywhere, not only if the center is located on the plane of the Wilson loop.
For $\rho=0$ the instanton field is a pure gauge,
\begin{equation}
  \label{puga1}
 A_{\mu}= i\Omega_I\partial_{\mu}\Omega_I^{\dagger} 
\end{equation}
with 
\begin{equation}
  \label{puga2}
\Omega_I = \frac{x^4+{\rm i}\mbox{\boldmath$\sigma$}{\bf x}}{x^2}, 
\end{equation}
and therefore loops not passing through the singularity generate 
$$W_{\rm Inst}\equiv 1 .$$
The Wilson loop of merons cannot be evaluated in closed form if the plane of the loop does not contain the singularity.\vskip .1cm
The  ensembles to be considered in this study contain field
configurations obtained by superposition of  pseudoparticles and anti-pseudoparticles of fixed
and equal number $N_P/2$,
\begin{equation}
    \label{supo}
    A_{\mu}(x)= \sum_{i=1}^{N_P} h(i) a_{\mu}(x-z(i)) h^{-1}(i).
\end{equation}
Such a configuration is specified by the position of the centers
$z(i)$ and their
color orientations
\begin{equation}
    \label{color}
    h(i) = h_0(i) + {\rm i} {\bf h}(i) \cdot \mbox{\boldmath$\sigma$} \ , \qquad
    h_0^2(i)  + {\bf h}^2(i) = 1 .
\end{equation}

In the ensembles to be discussed, the location of the merons is restricted to
a hypercube
$$   -1\le z_{\mu}(i)\le 1,\quad V=16 \, .$$
Our standard choice for the meron size and coupling constant is
 \begin{equation}
    \label{stco} 
\rho = 0.16\,, \quad g^2 =32\, . 
\end{equation}
We will see later that after rescaling,  our results have very little dependence on these values.
%###########################################################################
\subsection{Finite  Action Density of Pseudoparticles}
The infrared divergence in the action of a meron may appear as an obstacle in their use as building blocks of field configurations
with (infrared) finite action.  A similar concern arises in the superposition of (regular gauge) instantons. However, 
here we show that with an appropriate choice of the color orientation, the action of a cluster of merons of finite size can be made  finite.
Similar arguments apply for clusters of  regular gauge instantons. 

By dimensional arguments, the leading term of the action density 
of a collection of merons behaves asymptotically as
$$s\sim S_{\rm asy}\frac{1}{x^4}$$
and therefore leads to a total action that  is logarithmically divergent in the infrared. To calculate the coefficient $S_{\rm asy}$
we proceed by averaging over the orientation. Useful identities are:
$$\frac{1}{\Omega} \int {\rm d}\Omega \, \hat{x}_{\mu} \hat{x}_{\nu}\hat{x}_{\rho} \hat{x}_{\sigma} = \frac{1}{24}
\Big[\delta_{\mu\rho}\delta_{\nu\sigma}+\delta_{\mu\sigma}\delta_{\nu\rho}+\delta_{\mu\nu}\delta_{\rho\sigma}\Big]\,,$$
$$\frac{1}{\Omega} \int {\rm d}\Omega \, \hat{x}_{\mu} \hat{x}_{\nu} = \frac{1}{4}\delta_{\mu\nu}\,,\quad\frac{1}{\Omega} 
\int {\rm d}\Omega \, (\hat{x}_{\nu})^4 = \frac{1}{8}\,.$$ 
We introduce the notation 
\begin{eqnarray*}\eta(i)_{a\mu\nu} &=&\eta_{a\mu\nu} \quad \mbox{if $i$-th pseudoparticle is a meron}\,, \\ 
\quad  \eta(i)_{a\mu\nu}&=&\bar{\eta}_{a\mu\nu} \quad \mbox{if $i$-th pseudoparticle is an anti-meron}\,,\end{eqnarray*}
and find
$$\eta_{e\mu\rho}(i)\eta_{f\nu\rho}(j)\eta_{g\mu\nu}(k)= 4\delta_{ijk}\epsilon_{efg}\,,\quad\eta_{e\mu\rho}(i)
\eta_{e^{\prime}\mu\rho}(k)\eta_{f\nu\sigma}(j)\eta_{f^{\prime}\nu\sigma}(l)=  16 \delta_{ik}\delta_{jl}\delta_{e e^{\prime}}\delta_{f f^{\prime}}$$
$$\eta_{e\mu\rho}(i)\eta_{e^{\prime}\mu\sigma}(k)\eta_{f\nu\sigma}(j)\eta_{f^{\prime}\nu\rho}(l)=  
4\Big[ \delta_{ik}\delta_{jl}\delta_{e e^{\prime}}\delta_{f f^{\prime}}+  \delta_{il}\delta_{jk}\delta_{e f^{\prime}}\delta_{f e^{\prime}}
+ (1-2\delta_{ijkl})\delta_{ij}\delta_{kl}\delta_{e f}\delta_{ e^{\prime}f^{\prime}}\Big]\,,$$
$$\epsilon^{abc}\epsilon^{ab ^{\prime} c ^{\prime}}u^{be}(i)u^{b ^{\prime}e ^{\prime}}(k)
u^{bcf}(i)u^{c ^{\prime}f ^{\prime}}(l)=\big(u^{\dagger}(i)u(k)\big)_{ee ^{\prime}}\big(u^{\dagger}(j)u(l)\big)_{f f ^{\prime}}-
\big(u^{\dagger}(i)u(l)\big)_{e f ^{\prime}}\big(u^{\dagger}(j)u(k)\big)_{f e ^{\prime}}. $$
After a tedious calculation, we arrive at the following expression for the coefficient $S_{\rm asy}$:
\begin{equation}
  \label{sasy}
  S_{\rm asy}= 2 \mbox{tr}\,(\,V + \bar{V}\,)-6 \Big[\det U+\det \bar{U}\Big] -\frac{1}{4} \Big[ \mbox{tr}\, V^2 - ( \mbox{tr}\,V)^2 +\mbox{tr}\,\bar{V}^2 \,-
( \mbox{tr}\, \bar{V})^2\Big]-\frac{1}{3}\Big[\mbox{tr}\,V\bar{V}-\mbox{tr}\,V\mbox{tr}\,\bar{V}\Big]
\end{equation}
with
\begin{equation}
  \label{asyh}
  U= \sum _{i\in M}u(i),\; \bar{U}= \sum _{i\in \bar{M}}u(i),\quad V= U U^{\dagger},\quad \bar{V}= \bar{U} \bar{U}^{\dagger}.
\end{equation}
The color orientation of 4 merons can be chosen such as to make the leading asymptotic contribution to the action vanish. 
This corresponds to a neutral object. From the trivial identity
$$0=\left (\begin {array}{ccc} 1&0&0\\\noalign{\medskip}0&1
&0\\\noalign{\medskip}0&0&1\\
\end {array}\right )+\left (\begin {array}{ccc} 1&0&0\\\noalign{\medskip}0&-1
&0\\\noalign{\medskip}0&0&-1\\
\end {array}\right )+\left (\begin {array}{ccc} -1&0&0\\\noalign{\medskip}0&1
&0\\\noalign{\medskip}0&0&-1\\
\end {array}\right )+\left (\begin {array}{ccc} -1&0&0\\\noalign{\medskip}0&-1
&0\\\noalign{\medskip}0&0&1\\
\end {array}\right )$$
and by comparison with the representation of the color matrices $u$
\begin{equation}
  \label{grpasy}
 u^{bc} = \delta _{bc} \, (h^{2}_{0} - {\bf h}^{2} ) + 2 \, h_{0} \, h_{a} \, 
\epsilon ^{abc} + 2 h_{b} h_{c}\, ,
 \end{equation}
we read off that 4 merons with the color orientation of the $i$-th meron given by $h_i=1$
yield a vanishing $1/x^4$ term in the asymptotic expansion of $s$. Such configurations exhibit confinement of their building blocks,
analogous to the role of the gluons, since a complete dissociation of such a pseudoparticle cluster is accompanied by an infinite
increase in the action.

The pseudoparticle ensembles  have been generated by  Monte Carlo
sampling of the action in the path integral,  Eq.~(\ref{pathintegral}).
In  each step of a Metropolis update, the position and color orientation of a given
meron are tentatively changed, the induced changes in the action density are
evaluated at a set of mesh points distributed over the whole volume, and the
configuration is accepted or rejected based on the global change in action. The
long range nature of the meron fields makes the changes extend throughout the
whole system.
%#############################################################################
\section{Stochastic  and Correlated Pseudoparticle Ensembles}
Our goal is to calculate vacuum expectation values of observables via the path integral over the collective coordinates of our effective
theory with pseudoparticle degrees of freedom
\begin{equation}
   \label{pathintegral-obs}
\langle {\cal O} \rangle = \frac{1}{Z} \int {\rm d} z_i {\rm d} h_i {\rm e}^{- \frac{1}{g^2} S[A(z_i,h_i)]}
 {\cal O} [A(z_i,h_i)]  ,
\end{equation}
where $Z$ is the partition function defined in Eq.~(\ref{pathintegral}). As discussed in the introduction,  it is useful to proceed in  steps.
We begin with an ensemble in which the pseudoparticle positions and color orientations are selected according to a uniform distribution. 
This corresponds to the strong coupling limit, in which $g^2$ is infinite so that the action does not affect the weight of each configuration. 
It will be fruitful to think of the present effective theory in the context of corresponding lattice calculations, and this stochastic ensemble will 
be comparable to the strong coupling limit of lattice QCD.  As in lattice QCD,  various quantities can be computed analytically   
in the strong coupling limit. The comparison with the stochastic ensembles will be a cornerstone for understanding  the significantly more 
complex structure of the 
dynamically correlated ensembles. 
  
Dynamical correlations are included using the Metropolis algorithm to sample variables with a probability distribution given by the action. 
Thus, in the usual way, to calculate the path integral $\int {\rm d} x {\rm e}^{-S(x)} {\cal O}(x)$, we change $x$ with a microreversible change 
$\Delta x$, calculate the corresponding change in action, $\Delta S$, and accept the new value $x+\Delta x$ with probability $P$ = min 
$ (1, {\rm e}^{-\Delta S})$. 

The crucial  step beyond the stochastic ensemble is to include dynamical correlations in the pseudoparticle color orientations. Update 
of pseudoparticle positions is much less important, but will always be included unless we want to highlight the difference.  Thus both
 the positions $z(i)$  and the parameters $h(i)$ specifying the color orientations of all the pseudoparticles are equilibrated using Metropolis 
updates. We will show that these dynamical color correlations make a qualitative change in the physical observables. In order to allow 
for a variable ultraviolet scale, we will also consider ensembles of pseudoparticles with a uniformly distributed pseudoparticle size $\rho$. 
\subsection{Action Density}
\begin{figure}
\hspace{2cm}\epsfig{file=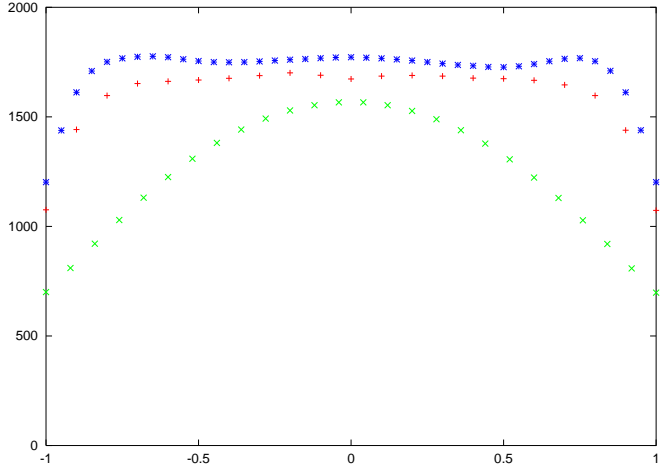, width=.4\linewidth,angle=-90}
\caption{ Action density  $g^2\langle s(x)  \rangle$ along the $x$-axis  for the stochastic (rescaled) (green $\times$) and  dynamically 
correlated  ensembles with stochastic  (blue $*$) and with dynamical positions (red $+$). Dynamical correlations decrease the value  at the 
center of the cube from  261 000 to 1700. }
\label{acde500}
\end{figure}

In the stochastic ensemble,  the action density can be evaluated analytically by assuming that the average over the color orientations
satisfies 
$$ u^{ac}(i)u^{bc}(j) \approx \delta_{ij}\delta_{ab}\, .$$
This leads to the following expression: 
\begin{eqnarray*}
4 s \approx \sum _{i=1}^{N_P}\Bigg[ \, F^{a} _{\mu \nu} (i) \; F^{a}_{\mu \nu} (i) + \sum _{j \not= i} \, \left[\left ( A^{b}_{\mu} (i) \; A^{b}_{\mu} (i) \right )
 \; \left ( 
A^{b}_{\nu} (j) \; A^{b}_{\nu} (j) \right ) 
 - \left ( A^{b}_{\mu} (i) \; A^{b}_{\nu} (i) \right ) \, 
\left ( A^{c}_{\nu} (j) \, A^{c}_{\mu} (j) \right )\right]\Bigg].
\end{eqnarray*}
Using  standard identities for the 't~Hooft symbol, we obtain, cf. Eq.~(\ref{mer})
$$
g^2\,A^{a}_{\mu} (i) \, A^{a}_{\nu} (i) = \frac{\xi^2}{[(x-z(i))^{2} +\rho^2]^2} \; \left (
\delta _{\mu \nu} \, (x-z(i))^{2} - (x-z_{\mu} (i)) \, (x-z_{\nu} (i)) \, \right )\,,
$$
which in the center of the cube ($x=0$) yields
\begin{equation}
  \label{atwo}
 g^2\, \sum _{i=1}^{N_P}A^{a}_{\mu} (i) \, A^{a}_{\nu} (i) \approx  \frac{3 \xi^2}{4}\delta_{\mu\nu}\sum _{i=1}^{N_P}\frac{z(i)^2} {[z(i)^{2} +\rho^2]^2} \, .
\end{equation} 
Furthermore, with the action density of a single meron or instanton, cf. Eq.~(\ref{smde1}),
\begin{equation}
  \label{Ftwo}
F_{\mu\nu}^a (i) F_{\mu\nu}^a (i) = \frac{6 \xi^2}{[(x-z(i))^{2} +\rho^2]^4}\left[ \Big((2-\xi)(x-z(i))^2 +2\rho^2\Big)^2 +4\rho^4\right] ,
\end{equation}
we derive for the action density  of the gauge field generated by $N_{P}$ pseudoparticles
\begin{eqnarray}
  \label{avacde}
 g^{2} s(x) &=&   \frac{3 \xi^2}{2}  \sum^{N_P}_{i=1} \;\Bigg[ \frac{1}{((z (i)-x)^{2} + \rho^{2})^{4}} \ 
\left[ \Big((2 -\xi)(z(i)-x)^2 +2\rho^2\Big)^2 +4\rho^4\right] \nonumber\\  
  &+& \frac{3\xi^2}{4} \sum_{j \neq i} \; \frac{(z(i)-x)^2}{((z(i)-x)^2 + \rho^{2})^{2}} \,
\frac{(z(j)-x)^2}{((z(j)-x)^2 + \rho ^{2} )^{2}} \,  \Bigg]  \,.
\end{eqnarray}
Performing the ensemble average we obtain for a system of $N_M$ merons
\begin{equation}
  \label{numac}
 g^{2} \langle s(x=0)\rangle  =  4.5 \; N_{M} + 1.03 \; N_M(N_M-1) . 
\end{equation}
Fig.~\ref{acde500}  shows the action density for the stochastic and for dynamically correlated ensembles.  We observe a qualitatively 
different behavior of the action density  as function of the position in the cube for stochastic and dynamically correlated ensembles. 
In agreement with the expression (\ref{avacde}), the action density of the stochastic ensemble peaks in the center of the cube and 
decreases by a factor of 2 when approaching the surface of the cube. The dynamically correlated ensembles, on the other hand, are 
approximately constant   for $|x_\mu|\le 0.8$ and thus seem to respect translational symmetry. This central region is used for the 
evaluation of observables. The values  of the action density of stochastic and dynamically correlated ensembles differ by two orders 
of magnitude. The  value of the action density of the stochastic ensemble in the center of the cube, 
\begin{displaymath}
g^{2}  \langle s(x=0)\rangle = 2.6 \cdot 10^{5}, 
\end{displaymath} 
is, within 1\%, correctly predicted by Eq.~(\ref{numac}).  \vskip.1cm
The one-pseudoparticle  contribution to the action density in (\ref{numac})  arising from the abelian part of the field-strength yields
an intensive contribution to the action density in the thermodynamic limit, up to a logarithmic correction.
In contrast, in the thermodynamic limit, assuming a 
finite density of pseudoparticles, the non-abelian contributions to the field-strength are dominant and give rise to an action density  
increasing linearly with the volume for sufficiently small $\rho$. Stochastic cancellations are not sufficient to achieve volume
independence. For a cube of side-length $2L$ with $L \rightarrow \infty$,
the action density of $N_{M}$ merons at the center of the cube is 
\begin{equation}
  \label{thdl}
 \langle s(x=0)\rangle \approx 1.29 \,\frac{N_{M}^2}{g^2 L^4}.  
\end{equation}
In our calculation, all interference terms have disappeared and the same result applies for an arbitrary mixture of a total of $N_{M}$ 
merons and anti-merons; furthermore, the same number of regular instantons  yields a 16 times larger action density. Clearly, the 
expression (\ref{thdl}) is not compatible with a proper thermodynamic limit.

\begin{figure}
\epsfig{file=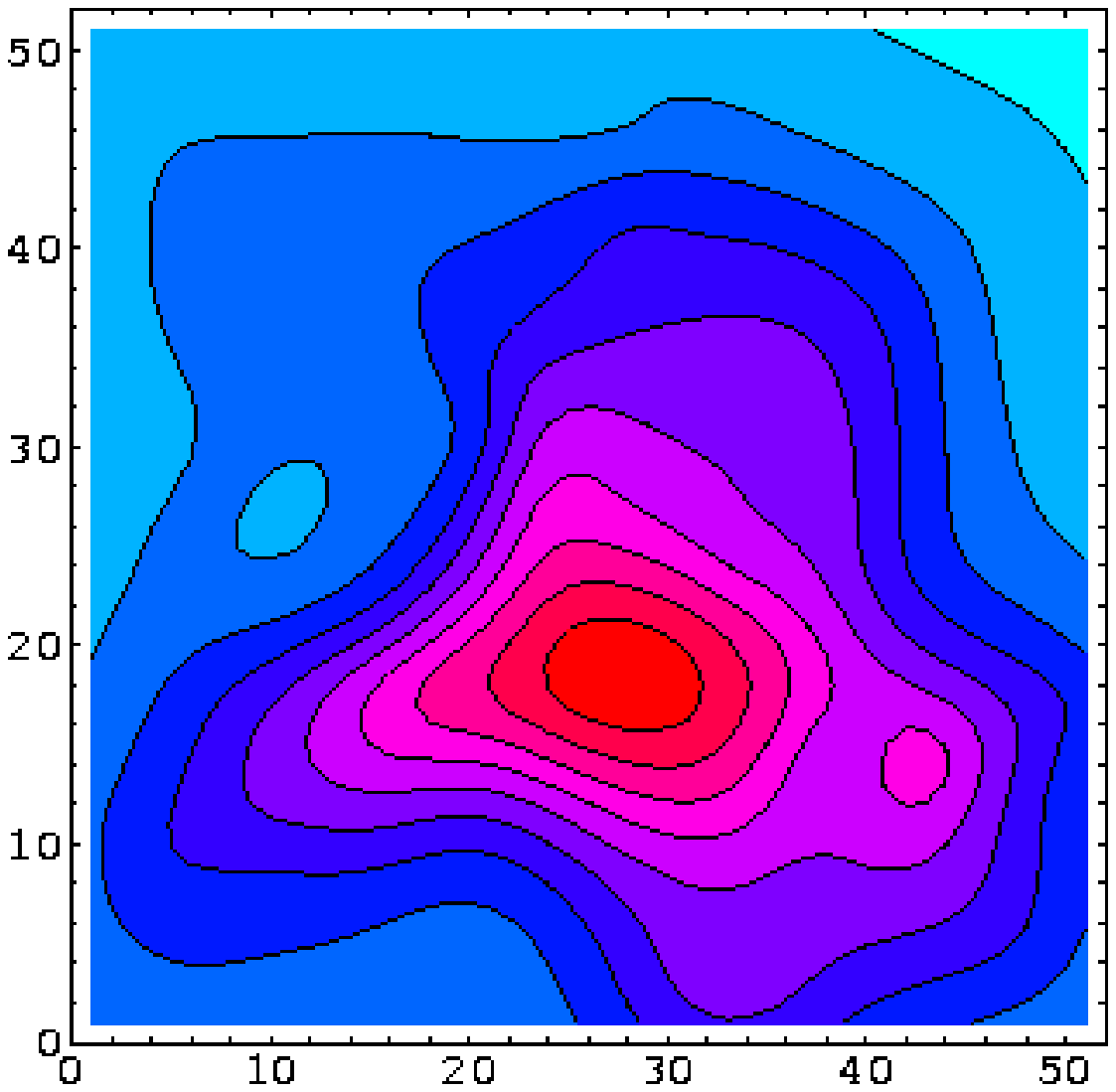, width=.5\linewidth}\epsfig{file=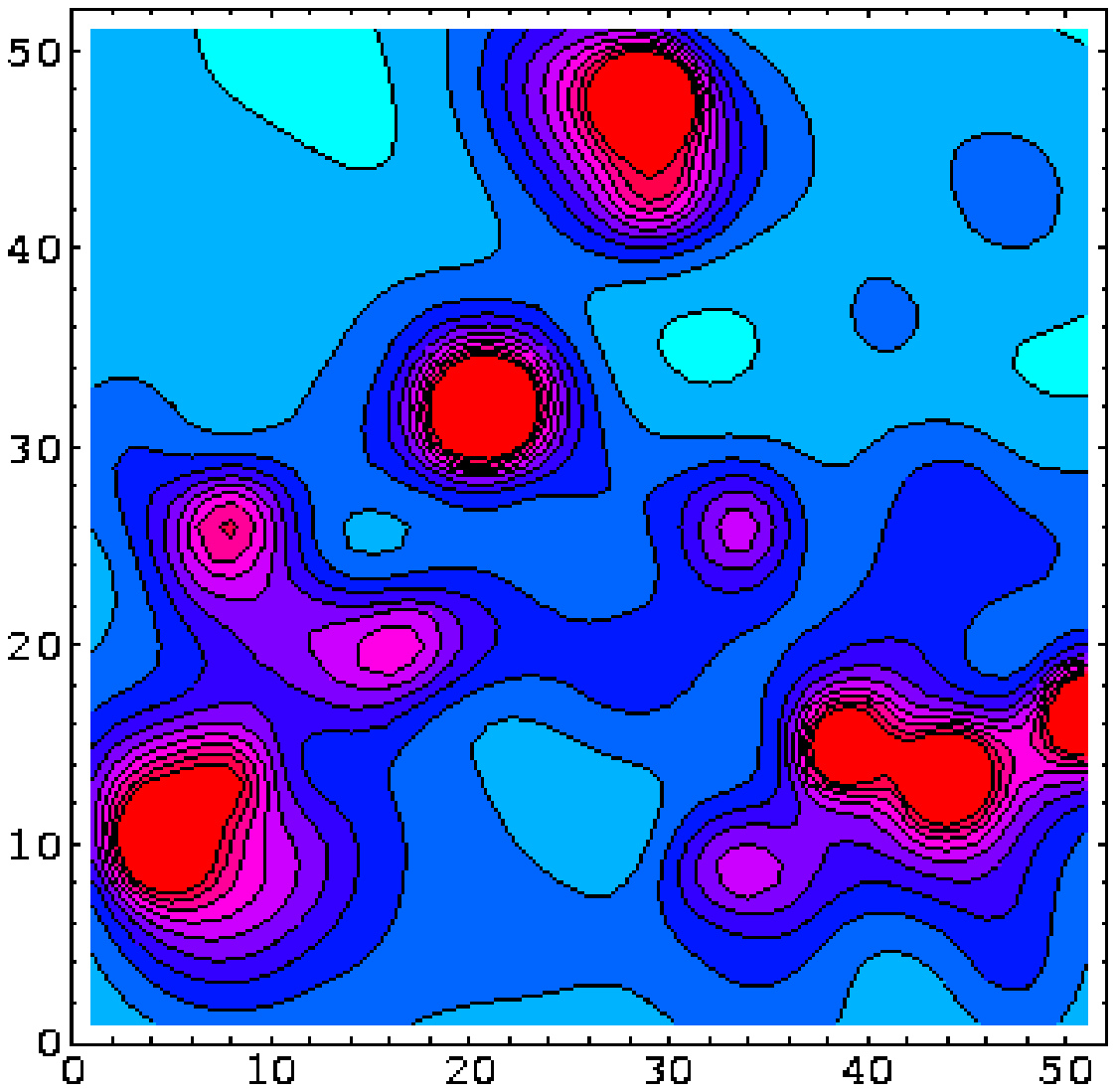, width=.53\linewidth}\\
%\captionof{figure}
\caption{Contour plot of the action density  in the ($x,y$) plane for a single  configuration of the stochastic (left) and of the 
dynamically correlated meron ensemble (right)
with $N_{M}=500$. Left: $ s[A] = 110 000,\; 15 000 < s(x) < 327 000.\quad$ Right: $ s[A] = 1530 ,\; 180 < s(x) < 11 300 $.}
\label{accool}
\end{figure}

\begin{figure}
\hskip -1cm\epsfig{file=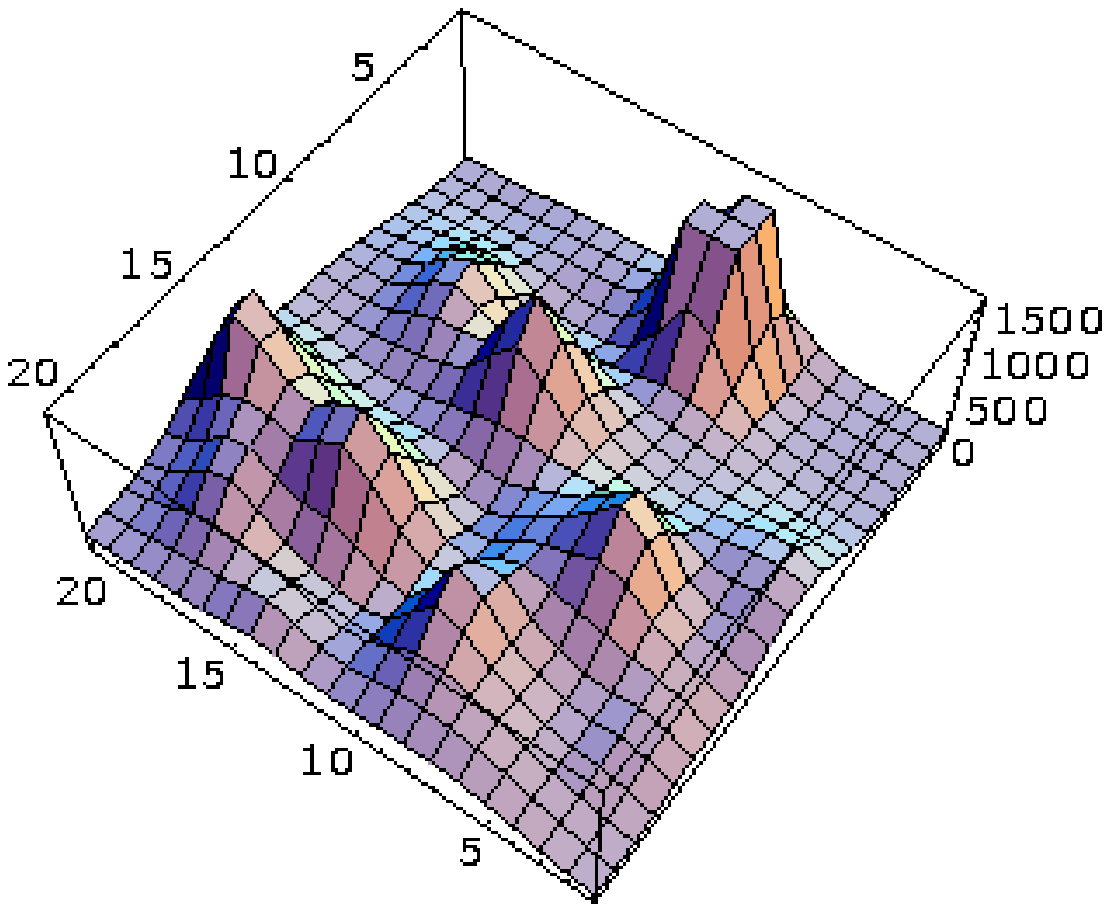, width=.6\linewidth}\hskip -1.5cm\epsfig{file=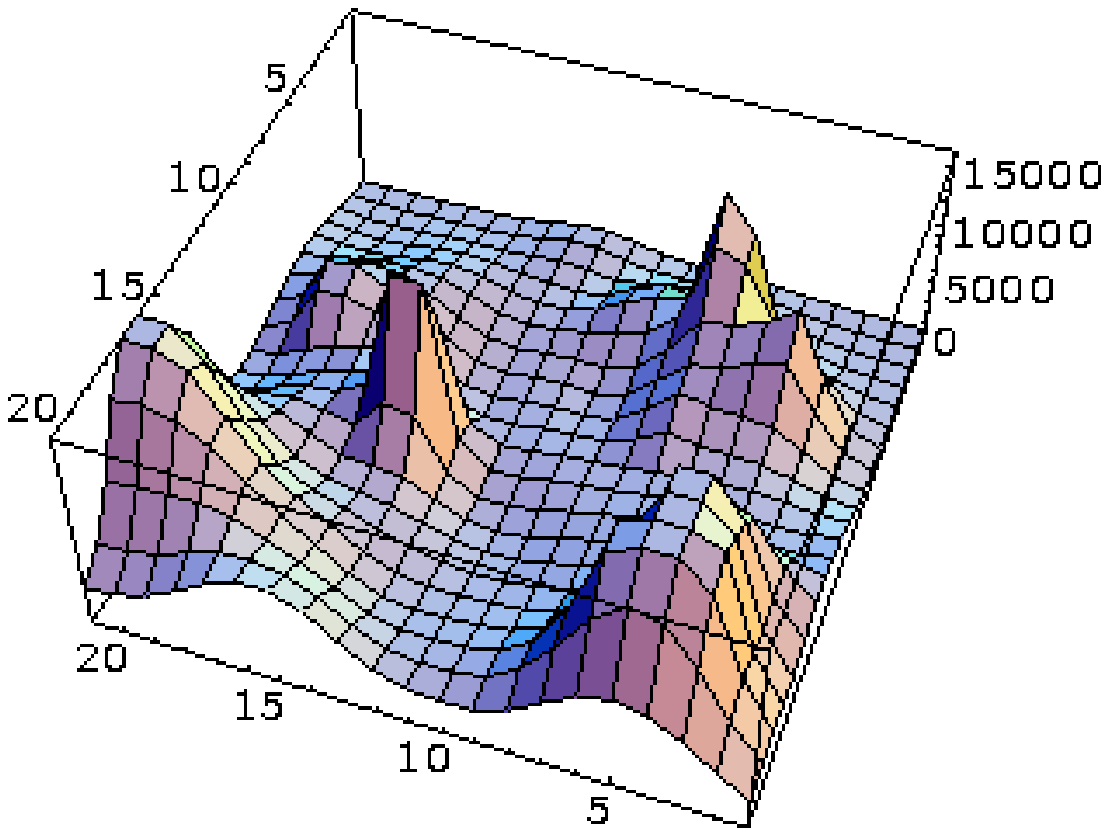, width=.6\linewidth}
%\captionof{figure}
\caption{The action density in the ($x,y$) plane for a single configuration of 200 merons (left) and 200 regular  instantons (right) 
respectively. Averaged over the plane the action density is 513 and 4700 respectively.}
\label{actionmer}
\end{figure}

For comparison, we refer to the action density in the standard dilute instanton liquid. In this model the building blocks are instantons 
in singular gauge with the gauge field decreasing asymptotically like
$$ A_{\mu}^a \sim \frac{1}{x^3}.$$
Thus for sufficiently low densities of instantons, $n_I \ll \rho^{-4}$, the action density is given by
\begin{equation}
  \label{acins}
g^2 \langle s \rangle  = 8\pi^2 n_{I}  
\end{equation}
and yields for  $N_{I}$ instantons in the volume $2^4$ 
\begin{equation}
  \label{acinsnu}
g^2 \langle s \rangle \approx 4.94 \,N_{I}.  
\end{equation}
The action associated with a field generated by stochastic superposition of 500 merons (regular instantons) is about 100 (16 000) times
 larger than the  action of 500 singular instantons.  As the expression (\ref{numac}) shows, the large value of the action density is not 
generated by  pseudoparticles in the neighborhood, rather it results from the superposition of  gauge fields with their $1/x$ asymptotic
behavior 
generated by distant pseudoparticles. The contour plot of the action density of a single configuration of the stochastic meron ensemble 
(left side of Fig.~\ref{accool}) confirms this interpretation of the results.  One clearly identifies a peak close to the center of the plane  and 
the decrease towards the surface. No remnant of the building blocks, the merons, can be identified. 
The dynamical correlations change completely the picture (right side of Fig.~\ref{accool}).  The weight determined by the action in 
Eq.~(\ref{pathintegral-obs})  favors  configurations where destructive interference between gauge fields of individual pseudoparticles 
prevent the formation of a large ``mean field''. In turn, the action density reflects clearly the structure of individual pseudoparticles.  
Screening of the gauge fields of the pseudoparticles is apparently also responsible for the restoration of translational symmetry.  
The comparison of the action density of merons and instantons in Fig.~\ref{actionmer} demonstrates the similarity in the structure of the 
configurations generated by superimposing and dynamically correlating meron and instanton fields, respectively. The larger amplitudes 
of the instanton fields give rise to correspondingly larger values of the action density. 
These results suggest  decomposing  the action density of the dynamically correlated ensembles into a single instanton ($s_I$) or 
meron ($s_M$) and a background ($s_B$) component. 
For instantons with their finite value of the action, the simplest Ansatz for the action density is [cf. Eqs.~(\ref{acins},\ref{acinsnu})]
\begin{equation}
\label{mfins}
g^2 \langle s \rangle = g^2(s_I+s_B) = 8\pi^2 n_{I} + g^2 s_B = 4.94 \,N_{I} + g^2 s_B .  
\end{equation}
For the dynamically correlated instanton ensembles of Table \ref{inst1} we find 
\begin{equation}
\label{bgin}
\frac{s_B}{\langle s \rangle} = 0.55  ...  0.65\,. 
\end{equation}
The logarithmic infrared singularity of the meron action requires a slight modification in the Ansatz for the separation of single meron 
and mean-field contributions \cite{LNT04}. The single meron contribution is obtained by integrating $s(x) $  
[Eq.~(\ref{smde1})] over a sphere of radius $r$,  which for small meron
size ($\rho\ll r$) becomes
\begin{equation}
  \label{sm}  
s_{M}=n\int ^{r} {\rm d}^4 x  s(x)\rightarrow 3\pi^2 n\left(\frac{5}{12}+\ln
\frac{r}{\rho}\right) .
\end{equation}
The  matching requirement on  $r$,
$$ s(r) = s_{B},$$
yields the following expression for the action density,
\begin{equation}
  \label{sm2}  
\langle  s\rangle = \Big[s_B+
n\frac{3\pi^2}{4}\Big(\frac{5}{3}-\ln \frac{2}{3}  s_B \rho^4
\Big)\Big].
\end{equation} 
For the dynamically correlated meron ensembles of Table \ref{medy12}  we find 
\begin{equation}
\label{bgme}
\frac{s_B}{\langle s \rangle} = 0.65  ...  0.75\,. 
\end{equation}
\subsection{Wilson Loops}
\begin{figure}
\hspace{1cm}\epsfig{file=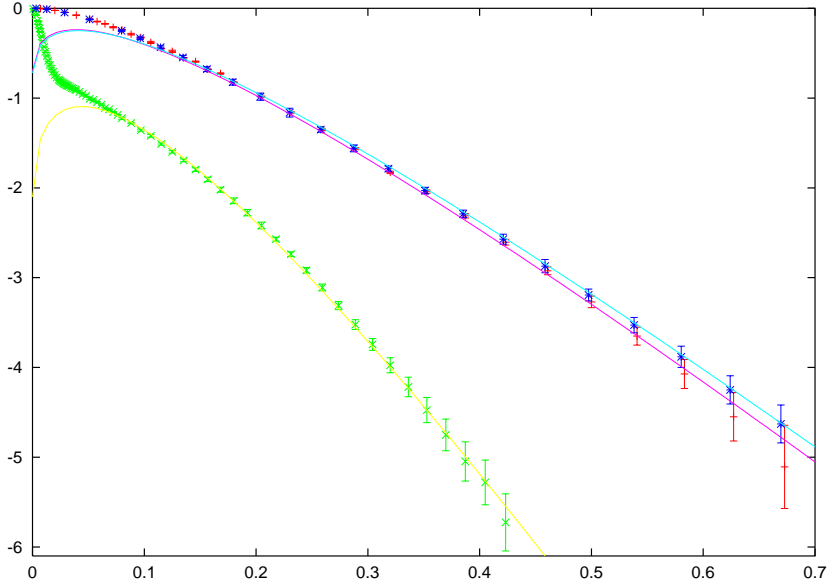, width=.5\linewidth,angle=-90}
%\captionof{figure}
\caption{Logarithm of a Wilson loop as a function of its area for three meron ensembles  with $N_M = 500$: A stochastic ensemble 
(green $\times$), an ensemble with dynamical color orientations (pink $-$) and an ensemble with dynamical color orientations and 
dynamical meron positions (blue $*$). The fit curves are described in the text.}
\label{sco500sd}
\end{figure}

The following discussion  focuses  on the comparison of Wilson loops (\ref{wilo}) in stochastic and dynamically correlated ensembles. A more thorough investigation of various issues concerning  Wilson loops will be
 presented in later sections. Our standard choice for the path  ${\cal C}$ in Eq. (\ref{wilo}) is a rectangular 
path located in the ($x_i,x_j$) planes with the center at the origin and with the ratio of the sides equal to 2. Thus, for a given configuration,
 we evaluate 12 different Wilson loops $W_{ij}$ and obtain our final results by taking the ensemble average and the average over the 12 
different orientations,
\begin{equation}
  \label{wilo1}
  \langle W \rangle  =\frac{1}{12}\sum_{i\neq j}\, \langle W_{ij}\rangle .
\end{equation}
The variance 
\begin{equation}
  \label{wilo2}
\Delta W ^2 = \frac{1}{12}\sum_{i\neq j}\Big( \langle W_{ij}\rangle -  \langle W \rangle\Big)^2  
\end{equation}
is used to estimate the statistical uncertainty in the determination of  $\langle W \rangle$.
Fig.~\ref{sco500sd}  shows the Wilson loops for the stochastic and correlated ensembles as function of the area. \vskip .1cm
For sufficiently large loop size, we parameterize the logarithm of the  Wilson loop as a sum of constant, perimeter (${\cal P}$)
and area (${\cal A}$) terms,
\begin{equation}
  \label{fit}
  \ln  \langle W \rangle = \omega + \tau \, {\cal P} -\sigma {\cal A} ,
\end{equation}
with
$$  {\cal A} = \frac{{\cal P}^2}{18} $$
for our standard choice of the path ${\cal C}$.  
The parameters of the fit  are  
\begin{equation}
  \label{stst500}
  \omega = -2.1,\; \tau = 2.3, \; \sigma = 22.9 
\end{equation}
for the stochastic ensemble, \begin{equation}
  \label{stdy500}
  \omega = -0.7,\; \tau = 1.1, \; \sigma = 11.8 
\end{equation}
for  the ensemble with dynamical color orientations but random  positions, 
and
\begin{equation}
  \label{dydy500}
  \omega = -0.7,\; \tau = 1.1, \; \sigma = 11.5
\end{equation}
for  the ensemble with dynamical color orientations and positions. As in the case of the action density, the change from the stochastic 
to the dynamically correlated ensemble 
is significant.  In comparison, only minor changes result when, in addition, the meron positions are also dynamical variables.
 For $ {\cal A} \ge 0.1$, the Wilson loops of the 3 ensembles are well reproduced by the parameterization (\ref{fit}).  In particular, the 
results strongly suggest that all 3 ensembles give rise to an area law for sufficiently large size of the loops with a value of the  string 
tension differing by a factor of 2 between stochastic and dynamically correlated ensembles. The large difference in the dimensionless 
ratios formed by action density and string tension,  
\begin{equation}
  \label{si2s}
\frac{g^2 s}{\sigma^2}\approx 500 \; (\mbox{stoch. ensemble})\,, \quad  \frac{g^2 s}{\sigma^2}\approx 11 \; (\mbox{dyn. corr. ensemble}),  
\end{equation}
reflects the essential difference in the dynamics of the ensembles. 
It is plausible that this strong decrease in the dimensionless ratio arises from suppression of fluctuations.
One expects that reducing the fluctuations has a larger effect on  the positive definite local action density
than on the non-local Wilson loop. As  will be seen  below, the large value of the action density expressed in units defined by the string
tension  rules out the stochastic ensemble as a viable  model of  the Yang-Mills dynamics.  
\newline

{\bf Small Wilson Loops in the Stochastic Ensemble}

In all the ensembles considered, the size dependence of small Wilson loops  can be determined analytically. Since the configurations 
are built up by a finite number of pseudoparticles,  fluctuations of the gauge fields cannot occur on arbitrarily small scales. In other 
words, for sufficiently small sizes, the gauge fields can be assumed to be spatially constant stochastic variables. For constant gauge 
fields, we use the identity
\begin{equation}
  \label{id1}
{\rm e}^{- {\rm i} \mbox{\boldmath$\scriptstyle{\sigma}$} {\bf C}_{2}} \; {\rm e}^{- {\rm i}\mbox{\boldmath$\scriptstyle{\sigma}$} {\bf C}_{1}} \; 
{\rm e}^{{\rm i}\mbox{\boldmath$\scriptstyle{\sigma}$} {\bf C}_{2}} \; {\rm e}^{{\rm i}\mbox{\boldmath$\scriptstyle{\sigma}$} {\bf C}_{1}}= 2 {\rm i}\,
\sin C_{1}\,\sin C_{2} \,  e ^{- {\rm i}\mbox{\boldmath$\scriptstyle{\sigma}$} {\bf C}_{2}} \big(\hat{{\bf C}}_{1} \times  \hat{{\bf C}}_{2}\big)\,
\mbox{\boldmath$\sigma$}\, {\rm e}^{{\rm i}\mbox{\boldmath$\scriptstyle{\sigma}$} {\bf C}_{2}}
\end{equation}
and obtain  with 
$$ {\bf C}_i = \frac{1}{2}g\ell_i{\bf A}_i $$
\begin{eqnarray}
 W_0=1 - 2 \sin^{2} \frac {1}{2} g \ell _{1} A_{1} \,\sin ^{2} \frac{1}{2} g \ell _{2}
A_{2} \, [1 - (\hat{{\bf A}}_{1} \hat{{\bf A}}_{2})^{2} ]  =  1 - \frac{4}{3} \sin ^{2}\frac{1}{2}  g \ell _{1} A_{1} \sin ^{2}\frac{1}{2}  g \ell _{2} A_{2} .
\label{smwlc}
\end{eqnarray}
In the last step we have averaged over the color orientations.
The space-time independent gauge fields ${\bf A}_i$ are generated by a superposition of a large number of the stochastic variables
 (positions and 
color orientations) specifying the individual pseudoparticles. We therefore expect  the space-time components  ${\bf A}_i$ to be 
Gaussian distributed, 
\begin{displaymath}
\rho (A_{i} ) {\rm d}^{3} A_i = \rho_{0} A^{2}_{i} {\rm d} A_{i} {\rm d} \Omega _{i}  \, {\rm e}^{- g^{2} A^{2}_{i} /2\sigma ^{2}_0}.
\end{displaymath}
We determine the width of the distribution by the expectation value of ${\bf A}^2$ [cf. Eq.~(\ref{atwo})]
\begin{equation}
\int \rho (A) g^{2} A^{2} {\rm d} A {\rm d} \Omega = 3\sigma ^{2}_0 = \frac{3 \xi ^2 N_{P}}{4} \, \langle 
\frac{x^{2}}{(x^{2} + \rho ^{2} )^{2}} \, \rangle
\label{atwo2}
\end{equation}
and  obtain 
\begin{equation}
  \label{swl1}
W_0 = 1 - \frac{1}{3}\left ( 1 - \left ( 1 -\ell_1^{2} \sigma^{2}_{0}
\right ) {\rm e}^{- \ell_1 ^{2} \sigma^{2}_{0} /2} \right )\left ( 1 - \left ( 1 -\ell_2 ^{2} \sigma^{2}_{0}
\right ) {\rm e}^{- \ell_2 ^{2} \sigma^{2}_{0} /2} \right ) \, ,
\end{equation}
with the values for the stochastic meron ensemble
\begin{equation}
\label{smwlre2}
\sigma_0 ^{2} = 120\quad \mbox {for} \quad N_{P}=N_{M} = 500 \, , \quad \rho =0.16\,. 
\end{equation}
The  typical size of a gauge field component (with fixed color and space labels) is thus of the order of 
$$ g A_i^a \sim \frac{1}{2} \sqrt{N_M} \frac{1}{L}\approx 10\, .$$
Depending on $N_P$, to next order we still may consider $x$-dependent fluctuations to be small as  compared to $C_i$ but to be of the
 order of 1 and therefore not suppressed. We write   
$$
 \tilde{{\bf C}} _{i} = {\bf C}_{i} + \mbox{\boldmath$\gamma$} _{i} \approx {\bf C}_{i} + \hat{\bf C}_i\gamma_i \, , \; | \mbox{\boldmath$\gamma$} _{i}
 | \ll | {\bf C}_{i} | .$$
To leading order, the directions of $ \tilde{{\bf C}} _{i}$ and ${\bf C}_{i}$ are identified but not their length.
We find 
\begin{eqnarray}
& &\frac{1}{2}\mbox{tr} \; {\rm e}^{- {\rm i} g\mbox{\boldmath$\scriptstyle{\sigma}$} \tilde{{\bf C}}_{2}} \; 
{\rm e}^{- {\rm i} g\mbox{\boldmath$\scriptstyle{\sigma}$} 
\tilde{{\bf C}}_{1}} \; {\rm e}^{{\rm i} g\mbox{\boldmath$\scriptstyle{\sigma}$} {\bf C}_{2}} \; 
{\rm e}^{{\rm i} g\mbox{\boldmath$\scriptstyle{\sigma}$} {\bf C}_{1}}
\label{WL1} \\
&=&  \cos \gamma_{1} \cos \gamma_{2} - \left ( \hat{{\bf C}}_{1} \cdot \hat{{\bf C}}_{2}
\right ) \sin \gamma _{1} \sin \gamma _{2} 
- 2 \sin C_{1} \sin \tilde{C}_{1} \cdot \sin C_{2} \sin \tilde{C}_{2} \; \left ( 1 - (\hat{{\bf C}}_{1} \hat{{\bf C}}_{2} )^{2} \; \right )\, .\nonumber\\
\nonumber
\end{eqnarray}
\begin{figure}
\begin{center}
\epsfig{file=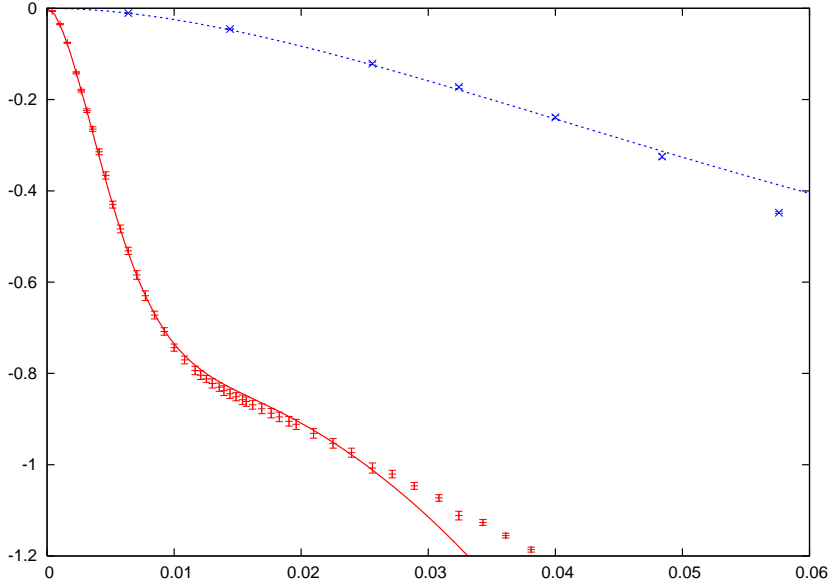, width=.5\linewidth,angle=-90}
%\captionof{figure}
\caption{Enlarged portion of Fig.~\ref{sco500sd} for small size Wilson loops. The curves show the result of the small
loop expansion 
[Eq.~(\ref{W1})] (red, lower curve)  for the stochastic ensemble and the result [Eqs.~(\ref{swl1},\ref{sig0})] (blue, upper curve)
for the ensemble with dynamical  meron positions and color orientations.} 
\label{smalwil1}
\end{center}
\end{figure}
By Taylor expansion of the gauge fields at the center of the cube, the fluctuations can be calculated. With 
$$A_{\mu}^a(x) = A_{\mu}^a(0) +\delta_{\mu}^a , \quad\gamma_{\mu}^a = \frac{1}{2}\ell_{\mu}\delta_{\mu}^a\, , $$
we obtain for the deviations along a  rectangular  loop in the $(1,2)$ plane 
$$\sum_{a=1}^3 \gamma_{\mu}^{a\,2} = \xi^2 (\ell_1\ell_2)^2
 \sum^{N_P}_{i=1} \frac{3\,z (i)^4 + 4\,z (i)^2\rho^2 +4\,\rho^4}{16\, \big(\,z (i)^{2} + \rho^{2}\,\big)^{4}}\quad \mu=1,2 \, .$$
Treating the deviations as independent stochastic variables with a Gaussian distribution 
$$\tilde{\rho}(\gamma)=\tilde{\rho}_{0}{\rm e}^{-\gamma_{\mu}^{a\, 2}/\sigma_{1}^2}$$
of  variance 
$$\sigma_1^2 = \frac{1}{(\ell_1\ell_2)^2}\frac{2}{3}\sum_{a} \gamma_{\mu}^{a\, 2} $$
we obtain    
$$\langle \cos \frac{\gamma_1}{2}\rangle=\langle \cos \frac{\gamma_2}{2}\rangle= {\rm e}^{-\sigma^2_1(\ell_1\ell_2)^2/4}.$$
In this approximation, the expectation value of the Wilson loop [Eq.~(\ref{swl1})] acquires the following correction [cf. Eq.~(\ref{WL1})],
 \begin{equation}
   \label{W1}
 W_1  =  W_0  -1 + {\rm e}^{-\sigma^2_1(\ell_1\ell_2)^2/2} \,,   
 \end{equation}
and we find for the stochastic meron ensemble
\begin{equation}
\label{smwlre}
\sigma_1 ^{2} = 118\quad \mbox {for} \quad N_{M} = 500 \, , \quad \rho =0.16\, .
\end{equation}
The agreement with the results of the numerical evaluation (cf.  Fig.~\ref{smalwil1})  confirms the role of the constant gauge fields for 
small size Wilson loops.
The above analysis of the small Wilson loops applies to some extent also to the dynamical ensembles where spatial variations of the
 gauge fields cannot happen either 
on arbitrarily small scales. Assuming independence of the fluctuation of the constant gauge fields, we still may use the expression 
(\ref{swl1})  and,  in the absence of a closed expression for the fluctuations of the gauge fields, determine the value of the width 
\begin{equation}
  \label{sig0}
  \sigma_0^2 = 10
\end{equation}
by a fit to our numerical results (cf. Fig.~\ref{smalwil1}). 
The reduction in $\sigma_0$,  i.e., in the size of the gauge field fluctuations, by one order of magnitude from the value (\ref{smwlre2}) 
of the stochastic ensemble  by the dynamics of the pseudoparticles is compatible with the reduction in the action density 
 (dominated by the $A^4$ term), cf. Fig.~\ref{acde500}. 
\subsection{Thermodynamic Limit}
Unlike the stochastic ensemble, the dynamically correlated ensembles exhibit with the restoration of  translational invariance a 
proper thermodynamic limit.  In the case of stochastic ensembles, when doubling the linear dimensions of the system $L$ with the 
density of pseudoparticles kept fixed, intensive quantities like the action density calculated in the stochastic ensemble do not
 remain invariant. The action density actually increases in this case by a factor 16 [cf. Eq.~(\ref{thdl})].
\begin{figure}[h]
\epsfig{file=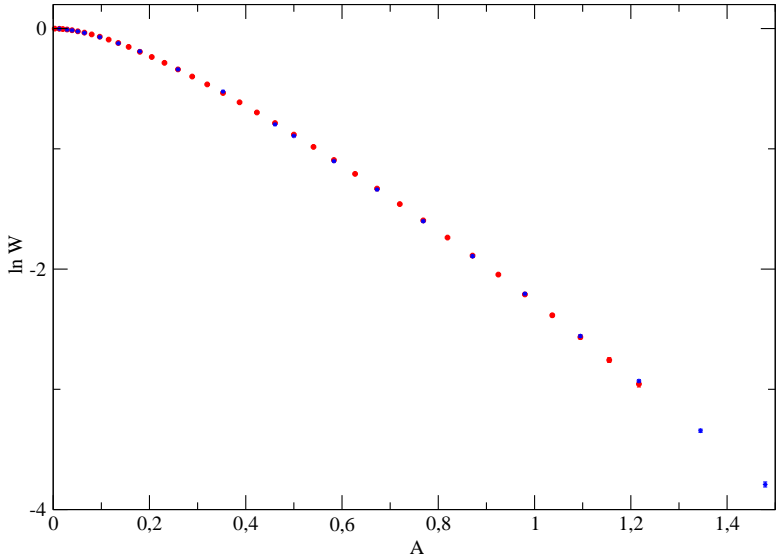, width=.6\linewidth,angle=-90}
%\captionof{figure}
\caption{Logarithm of Wilson loop as function of the area for meron ensembles  $N_M=50 \; (L=2)$  (circle) and $ N_M = 800\; 
(L=4)$ (star).}
\label{wiltdl}
\end{figure}
Results for the dynamically correlated ensembles indicate significant improvement. The realization of translational 
symmetry (cf. Fig.~\ref{acde500}), or the scaling properties to be discussed in the next section,  would be hard to understand without 
a proper behavior in the thermodynamic limit. Here we present the results of a direct numerical examination of the thermodynamic
 limit by doubling the side length $L$ and increasing the number of merons from 50 to 800. 
\begin{table}[h]
\begin{center} 
\begin{tabular}{|r|r|r|r|r|r|}  \hline
$N_{M}$&$\rho\,\,$&$\langle s\rangle\;$&  $\chi^{1/4}$&$C_s(0)$&$ C_{\tilde{s}}(0)$  \\ \hline \hline
50 &$0.16 $ &273 & 0.66  &$2.50\cdot 10^5 $&$1.86\cdot 10^5 $\\ \hline
800 ($L=4$) &0.16 & 284 & 0.66  &$2.53\cdot 10^5 $&$1.88\cdot 10^5 $\\ \hline
\hline
\end{tabular}
\end{center}
\caption{Test of thermodynamic limit for the vacuum expectation values of the action density $s$ defined in (\ref{act0}) and 
of the topological susceptibility $\chi$ (\ref{susc}), and of the values of the correlation function of the action density (\ref{C_s}) and  
the topological charge density (\ref{C_q}) at zero separation.}
\label{tdl1}
\end{table}
To achieve high accuracy in this important test, we have generated ensembles which in comparison with our ``standard'' ensembles 
 contain 4 times as many field configurations and where the  action density in the update process has been evaluated at 3 times
 more meshpoints.  The results in Fig.~\ref{wiltdl} and Table \ref{tdl1}  demonstrate the agreement in the Wilson loops  and
  in various  other observables  for these two ensembles. It is remarkable that the thermodynamic limit is established as a consequence
 of the dynamics of the pseudoparticles.  
%\end{center}
\section{Wilson Loops in Correlated Pseudoparticle Ensembles} 
This section contains  our main results concerning Wilson loops, an analysis of the structure of small Wilson loops and a discussion 
of the relation of our results to the general properties of Yang-Mills Wilson loops in  the small and large loop limit as known from
 lattice gauge studies.      
The results have been obtained on the basis of about 150 000 - 250 000 configurations in each of the ensembles discussed in 
this section. \vskip -.4cm
\subsection{Scaling Properties}
\begin{figure}
\begin{center}
\vskip-1cm
\epsfig{file=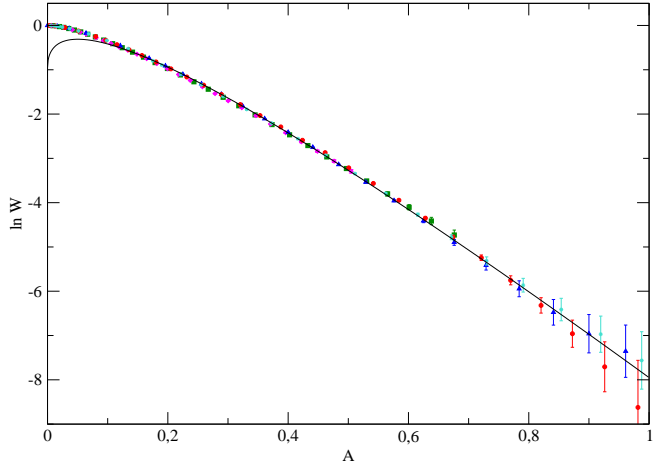, width=.5\linewidth,angle=-90}\vskip-.1cm
%\captionof{figure}
\caption{Logarithm of a Wilson loop as a function of its area for meron ensembles  $N_M = 800\ (L=4)$  (star), 100 (diamond), 
200 (square), 500 (circle), 1000 (triangle). The values of the scale parameter  $\lambda$ are given in Table \ref{medy12}.  Also shown 
is the  curve corresponding to the parameterization  (\ref{fitzins})   with the values of the parameters  given in (\ref{fitmer}).  }
\label{wilompar}
\end{center}
\end{figure}

\begin{figure}[h]
\begin{center}
\epsfig{file=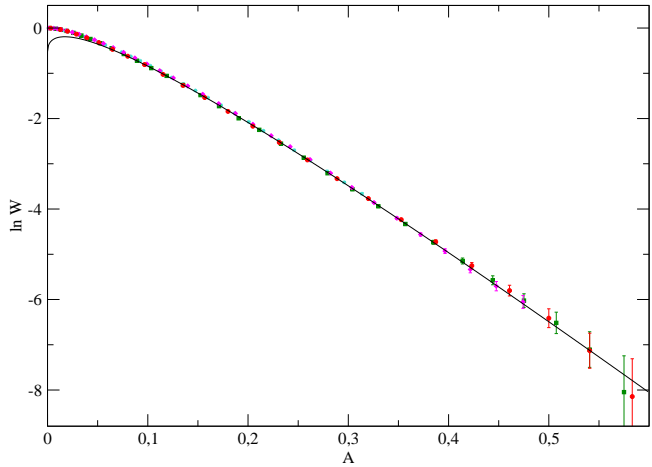, width=.5\linewidth,angle=-90}
%\captionof{figure}
\caption{Logarithm of a Wilson loop as a function of its area for instanton ensembles  $N_I = 50$ (star), 100 (diamond), 200 
(square), 500 (circle). The area has been rescaled  with $\lambda$ given in Table \ref{inst1}.  Also shown is the  curve corresponding
 to the parameterization  (\ref{fitzins})   with the values of the parameters  given in (\ref{fitins}).}
\label{wiloins}
\end{center}
\end{figure}

\begin{figure} 
\epsfig{file=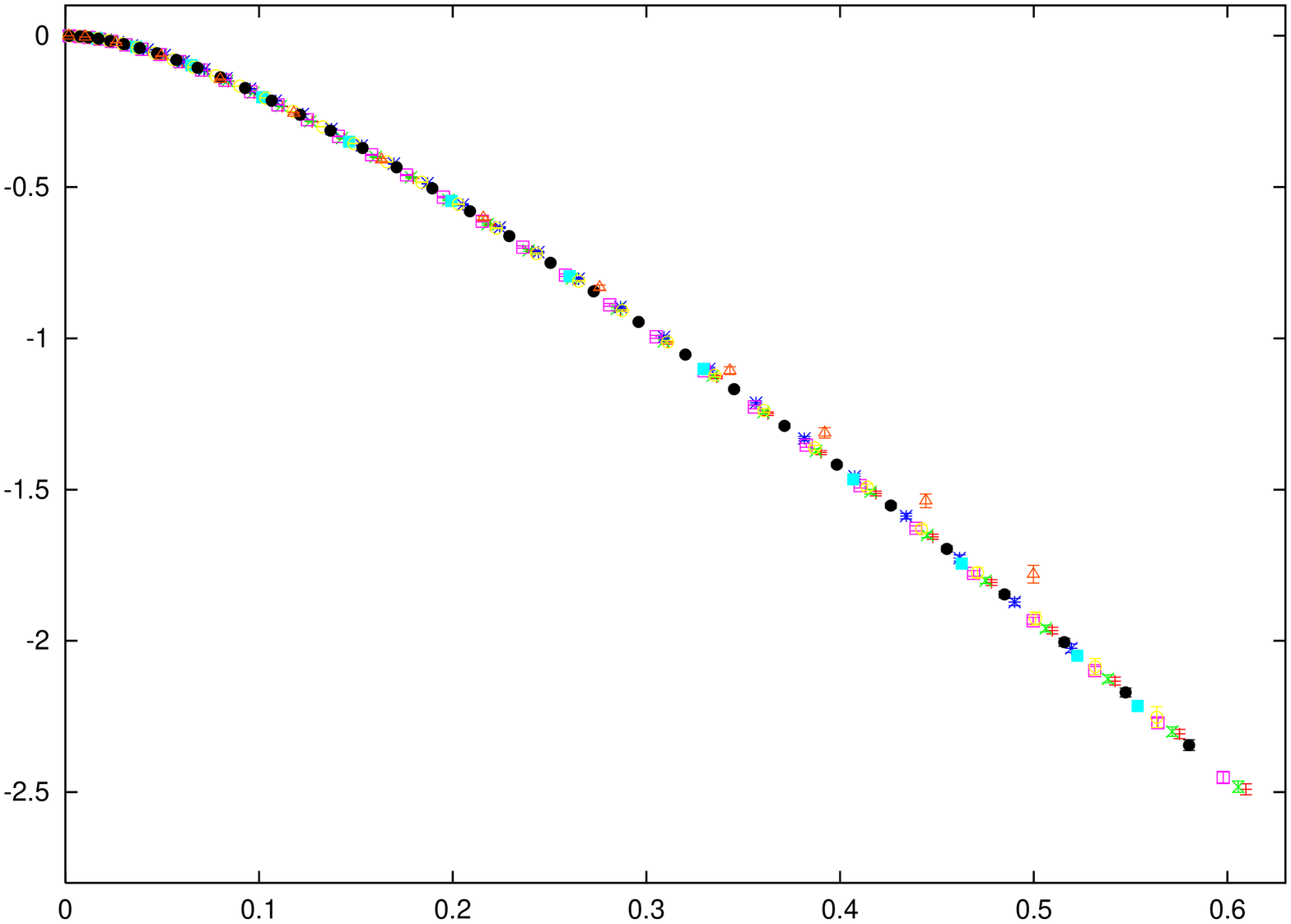, width=.35\linewidth,angle=-90} \epsfig{file=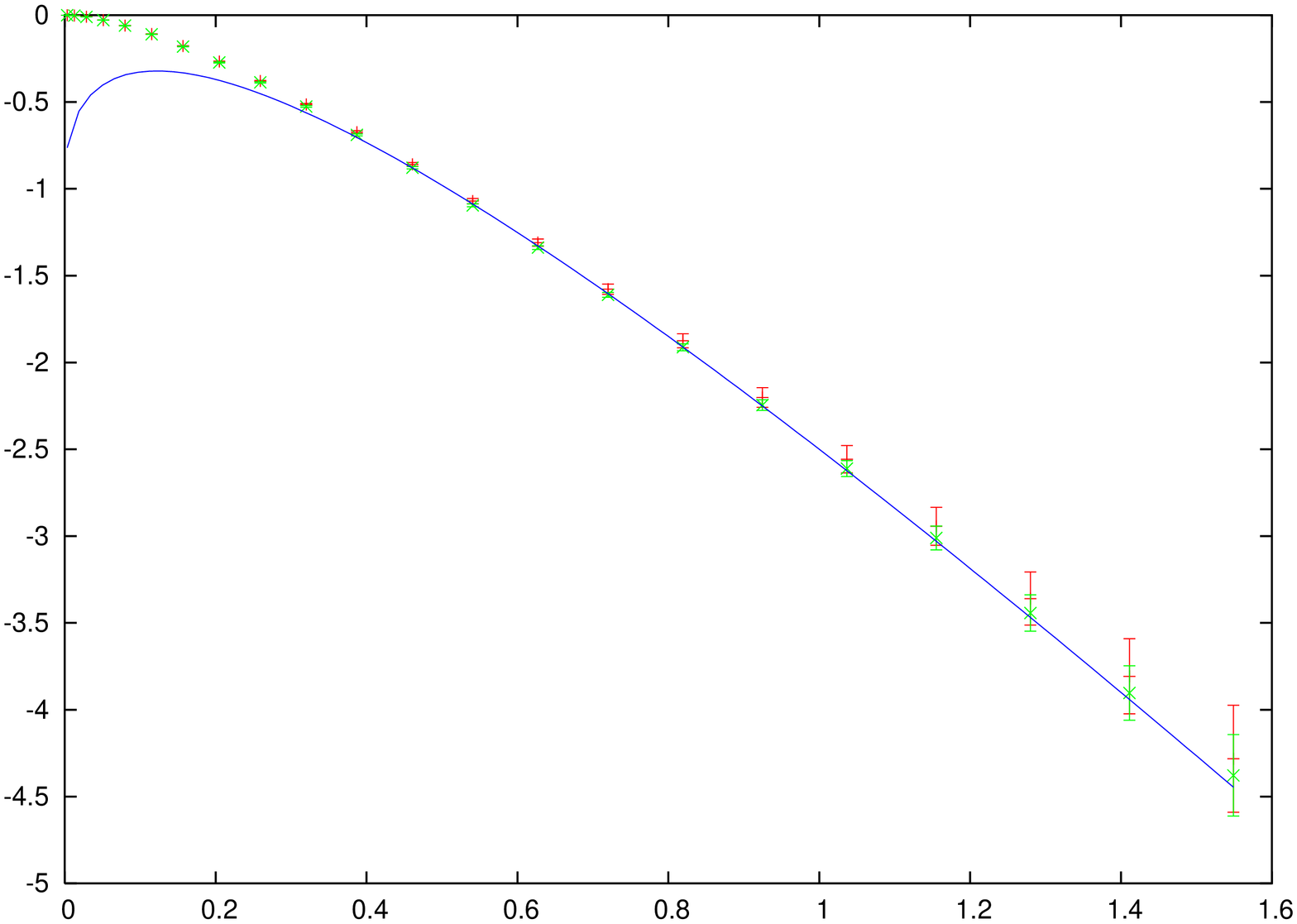, width=.35\linewidth,angle=-90}  
%\captionof{figure}
\caption{Logarithm of a Wilson loop as a function of its area for meron ensembles  $N_M = 50$. Left : Ensembles with  
different values of the  meron size ($0.01 \leq \rho \leq 0.16 $)  and  of the coupling constant ($g^2=8, 32$). The area has been rescaled 
 with  values of $\lambda$ in the range $0.31 \leq \lambda \leq 0.44.$  Right: Ensembles of variable size uniformly distributed in the 
interval  $0.03 \le \rho\le0.17$ (red) and of fixed size $\rho=0.10$  (green).}  
\label{wilrho}
%\end{center}
\end{figure}

The Wilson loops evaluated in  the meron and instanton ensembles with a wide range of  pseudoparticle numbers exhibit a universal 
behavior as  demonstrated by Figs.~\ref{wilompar} and \ref{wiloins}. 
After rescaling the area ${\cal A}\rightarrow \lambda \, {\cal A}$,
the values of the Wilson loop lie on a universal  scaling curve, where deviations from scaling
are within the statistical uncertainties. To account for the rescaling of the area, we generalize the Ansatz  (\ref{fit}) for the logarithm 
of the Wilson loop, 
\begin{equation}
  \label{fitzins}
  \ln  \langle W \rangle = \omega + \tau \,\sqrt{\lambda}\, {\cal P} -\sigma \,\lambda\,{\cal A} \,.
\end{equation}
For merons, the values of the universal parameters are 
\begin{equation}
  \label{fitmer}
\omega = -0.95,\quad \tau=1.34,\quad  \sigma = 12.7 \, .  
\end{equation}
%\end{figure}
%\begin{table}
A central result of this work is that for both merons and instantons, for sufficiently large size, the logarithm of
the Wilson loop decreases linearly with increasing area. This area law clearly demonstrates confinement in these
pseudoparticle  ensembles. The physical scale for each of the pseudoparticle ensembles is obtained by identifying  the value of 
$\lambda \sigma$ with the the value $4.4$ fm$^{-2}$, i.e., the unit of length (u.l.) for each of the ensembles is determined by $\lambda$.
For merons,    
\begin{equation}
 \label{ulfmer}
1\, \mbox{u.l.}= 1.70  \sqrt{\lambda}\, \mbox{fm}\, .
\end{equation}
The corresponding parameters for the instanton ensembles are 
\begin{equation}
  \label{fitins}
\omega = -0.52,\quad \tau=1.18,\quad  \sigma = 19.0, \quad  1\, \mbox{u.l.}= 2.08  \sqrt{\lambda}\, \mbox{fm}\, .  
\end{equation}
\vskip -.3cm
%\end{table}
\vskip -.1cm
Fig.~\ref{wilrho}  displays the dependence of our results on the pseudoparticle size and coupling constant.   Whereas changes in the  
Wilson loops induced by changes in  the pseudoparticle size $\rho$  can be accounted for  to a large extent by a change in the 
physical scale, variations in the coupling  constant by a factor of 4 modify the shape beyond changes in the scale, and a tendency  
 is visible for a slightly smaller string tension in the ensembles with variable pseudoparticle size. We note that in the update process
 for the ensembles with variable size, the sizes of the pseudoparticles are kept fixed.
%\begin{center}
%\hspace{2cm} 
On the basis of the identification of the value of the string tension extracted from the Wilson loops with the empirical value, we can  
compare our results for various observables with those obtained in other approaches, in particular with lattice gauge results.   
The values of the scaling parameter $\lambda$ together with other properties of meron and instanton ensembles are given in 
Tables \ref{medy12} and \ref{inst1}.  
\begin{table}[h]
\begin{center} 
\begin{tabular}{||r|r|r|r|c|c ||c|c|c|c||}  \hline\hline
$N_{M}$&$\langle s\rangle\;\;$&$\lambda$&$\chi^{1/4}$&$C_s(0)$&$ C_{\tilde s}(0)$ &$n_{M}$  &$\rho$& $\langle  s\rangle\;\;$&$\chi^{1/4}$ \\ \hline 
-&-&-&-&-&-& [fm$^{-4}$] & [fm$^{-1}$] & [fm$^{-4}]$  & [MeV]  \\ \hline \hline
1000&2742 &$1.25$&1.19 &$3.67\cdot 10^6$ & $2.49\cdot 10^6 $ & 4.8 & $ 0.30 $ & $ 210 $ & 118 \\ \hline
500&1744  &$1.$ &$1.01  $& $2.12\cdot 10^6$ & $1.45 \cdot 10^6$ & 3.8 & $ 0.27$ & $209 $ & 120\\ \hline
200&897 &0.69 &0.9 &$9.7 \cdot 10^5 $ & $7.0\cdot 10^5 $ & 3.2 & $ 0.23$ & $226 $ & 124\\ \hline
100&507 &0.52  &0.77  &$4.8 \cdot 10^5 $ & $3.6\cdot 10^5 $ & 2.8 & $ 0.20$ & $ 225$ &124\\ \hline
50 &249 &$0.37 $ &0.66  &$2.23\cdot 10^5 $&$1.78\cdot 10^5 $ & 2.7 & $ 0.16 $ & $218$ & 132\\ \hline
\hline
\end{tabular}
%\captionof{table}
\caption{Properties of meron ensembles:  Vacuum expectation values of the action density $s$ defined in (\ref{act0}) and 
of the the topological susceptibility $\chi$ (\ref{susc})  (also in physical units), values of the scaling parameter $\lambda$ (\ref{fitzins}), 
and values of the correlation functions of the action density (\ref{C_s}) and the topological charge density (\ref{C_q}) at zero separation
 with the standard choice of the parameters (\ref{stco}) for ensembles of $N_M$ merons  and with meron density $n_M= N_M/L^4$.}
\label{medy12}
\end{center}
\end{table}
%\begin{figure}
This first discussion of physical observables focuses on the action density and the topological susceptibility. Note that we refer in 
Tables \ref{medy12} and \ref{inst1} to results that will be explained in detail in section 5.  
From lattice SU(2) calculations for 
the action density the value   $\langle s\rangle /\sigma^2 = 4.5$ 
\cite{Campostrini:1983nr} has been deduced. QCD sum 
rule results \cite{SHVZ79,NARI96} range from $\langle s\rangle /\sigma^2  = 4.5 $ to $10$. 
The necessity of subtracting divergent terms to define the continuum limit makes  the 
uncertainty of these values of the action density, or equivalently of the gluon condensate, hard to assess.
\begin{table}[h]  
\begin{center} 
\begin{tabular}{||r|r|r|r|c|c||c|c|c|c||c|}  \hline\hline
$N_{I}$&$\langle s\rangle\;\;$&$\lambda$&$\chi^{1/4}$&$C_s(0)$&$ C_{\tilde s}(0)$ &$n_{I}$  &$\rho$& $\langle  s\rangle\;\;$&$\chi^{1/4}$ \\ \hline 
-&-&-&-&-&-&[fm$^{-4}]$ &[fm$^{-1}$]&[fm$^{-4}]$  &[MeV]  \\ \hline \hline
500&5430   &$1.0 $   &$1.81  $& $ 2.7\cdot 10^7$ & $ 1.6\cdot 10^7$&$1.68 $ &0.33& $291 $ &162\\ \hline
200& 2490 & 0.66    & 1.49 & $9.7 \cdot 10^6$ & $6.6\cdot 10^6 $& 1.54 &$ 0.27$ & $307 $ & 164\\ \hline
100&1350 &0.48  & 1.37  &$5.1 \cdot 10^6 $ & $3.8\cdot 10^6 $&1.45& $ 0.23$ & $ 314$ &180\\ \hline
50 &651  &$0.32$ &1.15  &$ 2.3\cdot 10^6 $& $1.9\cdot 10^6 $&1.64& $ 0.19 $ & $340$& 190\\ \hline
\hline
\end{tabular}
\end{center} 
%\captionof{table}
\caption{Properties of instanton ensembles, as defined in Table \ref{medy12}.}
\label{inst1}
\end{table}
For instance, in the lattice calculations of Ref.~\cite{Campostrini:1983nr},
the divergent contributions are about a 
factor of 20-500 larger
  than the extracted value of $\langle s\rangle$. As discussed above,  the 
pseudoparticle  action density also contains
 contributions which in the limit of vanishing pseudoparticle size $\rho \rightarrow 0$ become singular and should therefore be subtracted. 
 Thus a more relevant quantity to be compared with the lattice  and sum rule results is the ``mean-field value'' of the action density 
[cf. Eqs.~(\ref{bgin},\ref{bgme})]. We find  
\begin{equation}
  \label{sb2s}
\frac{s_B}{\sigma^2}\approx  8\; (10) \quad {\rm for\  merons\  (instantons)}\,.  
\end{equation}
This discussion also shows  that the presence of undamped  fluctuations in the field configurations of the stochastic ensemble 
[cf. Eq.~(\ref{si2s})]  is in severe conflict  with the lattice and sum rule results. 
The topological susceptibility is directly related to the $\eta'$ mass by the
Veneziano-Witten formula \cite{Witten79,Veneziano79} and is well measured in lattice QCD. From the  results in 
Tables \ref{medy12} and \ref{inst1} we conclude 
$$\chi^{1/4}/\sigma^{1/2} \approx 0.31\,,\quad \xi=1\,;\quad
0.42 \le \chi^{1/4}/\sigma^{1/2}\le 0.48\,,\quad \xi=2\,.$$ 
Both the meron and the instanton ensembles yield values of the susceptibility of the correct order of magnitude. The value for the 
instanton ensembles  agrees within its 7\% error bars with the SU(2) lattice result \cite{LUTE01}  $\chi^{1/4} / \sigma^{1/2}  \sim 0.48$. 
\subsection{Small Size Wilson Loops and the U(1) Limit}
Fields generated by superposition of a finite number of pseudoparticles are finite. Fluctuations do not occur with arbitrarily small 
wavelength. They are effectively cut off at length scales of the order of the pseudoparticle size or the average distance between 
neighboring pseudoparticles. As a consequence,  the values of Wilson loops calculated in pseudoparticle ensembles vanish with 
vanishing loop size. This presence of an intrinsic cutoff prohibits  a straightforward comparison of Wilson loops of pseudoparticle 
ensembles with field theoretic treatments in which for  small sizes,   Wilson loops approach the limit of the Maxwell theory and yield 
Coulomb-like static quark-antiquark potentials that diverge with decreasing distance of the sources.  Although  both stochastic and 
dynamically correlated ensembles suffer from this deficiency, the stochastic ensemble with its much larger fluctuations appears to 
exhibit a Coulomb like behavior for areas  ${\cal A} \le 0.01$  (cf. Figs.~\ref{sco500sd} and \ref{smalwil1})  and only for values 
${\cal A} \le 0.002$,  does the Wilson loop show the quadratic dependence on the area.   In the dynamically correlated ensembles 
we cannot identify such an intermediate Coulomb-like regime.  For a meaningful interpretation we therefore have to account 
for the presence of this intrinsic ultraviolet cutoff and we will compare  with  Wilson loops in Maxwell theory in which the necessary 
 ultraviolet regulator is kept finite. \newline 
For small loop size  we expand the exponential in the definition of the Wilson loop in SU(2) Yang-Mills theory
\begin{equation}
  \label{wilosmsi}
  W = \frac{1}{2} \,\mbox{tr}\left\{ P \exp {\rm i}g \oint_{\cal C}A_{\mu}(x){\rm d}x^\mu\right\}\approx 1-\frac{3 g^2}{8} \oint_{\cal C}\tilde{A}_{\mu}(x)
{\rm d}x^\mu \oint_{\cal C}\tilde{A}_{\mu}(y){\rm d}y^\mu \, .
\end{equation}
Here, $\tilde{A}$ denotes one of the color components of the gauge field. For small loop size we compute the vacuum expectation
 value of the  Wilson loop by identifying  $\tilde{A}$ with an abelian gauge field and find
\begin{equation}
\langle W \rangle \approx 1 - \frac{3 g^2}{8} \int J_{c\,\nu} (x) K (x-y) J^{\nu}_{c} (y) {\rm d}^{4} x\, {\rm d} ^{4} y\, ,
\label{wqed}
\end{equation}
where 
\begin{displaymath}
J^{\mu}_{c} (x) = \int \, \delta ^{4} (x - x_{c}(s)) \; \frac{{\rm d} x^{\mu}}{{\rm d}s}\,{\rm d}s
\end{displaymath}
is the current of a charge transported along the curve
${\cal C}$.  The Euclidean gauge field propagator $K$ in Lorentz gauge  is given by the inverse of the 4-dimensional Laplacian. 
The integral (\ref{wqed}) requires regularization. We choose a heat kernel regularization and obtain for  the regularized propagator
 in Maxwell theory  
\begin{equation}
K(x)  = \int \frac{{\rm d}^{4}k}{(2 \pi)^{4}} \;{\rm e}^{-\alpha k} \frac{{\rm e}^{ikx}}{k^{2}} 
=\frac{1}{4 \pi^{2} x^{2}} \; \left [ 1 - \frac{\alpha}{\sqrt{\alpha^{2} + x^{2}}} \; \right ] \, .
\label{regpro}
\end{equation}
We consider  rectangular loops with side lengths $R$ and $T$ and  denote with 
\begin{equation} 
\omega(R,T)= \int^{T}_{0} {\rm d}x_{0} \int^{T}_{0} {\rm d} y_{0} \, K \big ( \sqrt{(x^{0} - y^{0})^{2} + R^{2})} \big )
\label{defom}
\end{equation}
the contribution to $W$ of the  time-components of the  two currents on opposite sides of the rectangle in the expression (\ref{wqed}). 
The Wilson loop is obtained from the 4 contributions of time and space components on the same and on  opposite sides of the rectangle, 
\begin{equation}
\langle W \rangle \approx 1 - \frac{3 g^2}{8}[\,\omega(0,T)+\omega(0,R)-\omega(R,T)-\omega(T,R)\,]\, .
\label{wqed2}
\end{equation}
The integral (\ref{defom}) can be evaluated in closed form with the result
\begin{equation}
 \omega(R,T)= -\frac{1}{2\pi^2}\Big[\ln \frac{\alpha + \sqrt{\alpha^{2} + R^{2} + T^{2}}}
 {\alpha + \sqrt{\alpha^{2} + R^{2}}} - \frac{T}{R} \arctan T \, R \; 
 \frac{\sqrt{R^{2} + T^{2} + \alpha^{2}} - \alpha}{R^{2} \sqrt{R^{2} + T^{2} + \alpha^{2}} + \alpha T^{2}}\Big]\, . 
\label{wqed3}
\end{equation}
\begin{figure} [h]
\begin{center}
\vskip -.5cm
\epsfig{file=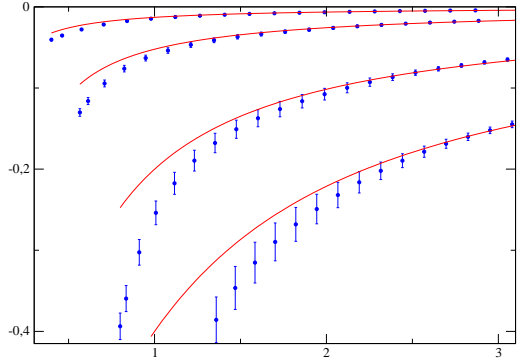, width=.4\linewidth,angle=-90}
\vskip -.2cm
\caption{ Logarithm of rectangular Wilson loops of fixed area (${\cal A}= 0.01, 0.02, 0.04, 0.06 $) as a function of the perimeter for an
 ensemble with 500 instantons of          variable size $\rho \le 0.1$. The curves are obtained from Eqs.~(\ref{wqed2}, \ref{wqed3}) with 
the value (\ref{alg2}) of the regulator.  } 
\label{smwins}
\vskip -.8cm
\end{center}
\end{figure}
Performing the limit $\alpha \rightarrow 0$, we obtain for $R\ll T$
\begin{equation} \langle W \rangle \approx 1 + \frac{3 g^2}{16 \pi^2}\,\Big[\frac{\pi T}{2 R}+1 -\frac{R+T}{\alpha}+2 \ln \frac{R}{2\alpha}\Big]\, ,
\label{wqedas}
\end{equation}
which contains both the Coulomb potential and the self-energy of the static charges.
On the other hand,  in the  limit of small loops and  for fixed $\alpha$,  $ \langle W \rangle - 1$ tends to zero with the square of the area 
\begin{equation} 
\langle W \rangle \approx 1 - \frac{g^2}{2} \Big(\frac{3 R T }{8 \pi \alpha^2}\Big)^2\, , 
\label{wqed4}
\end{equation}
as  is the case for Wilson loops in pseudoparticle ensembles.  The small Wilson loops in the stochastic ensemble in Fig.~\ref{smalwil1} 
 display these two different limits.   The regime  (\ref{wqed4}) of vanishing loop size is restricted to values of the area of the order 
of $10^{-3}$ and is followed by the Coulomb-like behavior associated with the limit (\ref{wqedas}) of vanishing regulator size, which 
extends up to values of the area of the order $10^{-2}$. With the reduction in the fluctuations, the dynamically correlated ensembles
 do not explicitly show the Coulomb-like behavior. Nevertheless on the basis of this analysis we can understand the small loop limit 
of the dynamically correlated ensembles and make the connection to the perturbative U(1) limit.   
  
For the comparison of  the pseudoparticle Wilson loops  with Wilson loops  of the Maxwell theory (\ref{wqed2},\ref{wqed3}) at finite 
 $\alpha$, we have determined the value of the regulator by a fit of expression (\ref{wqed2}) to the Wilson loops in the   pseudoparticle
 ensembles.
The results are shown in Fig.~\ref{smwins}. The value of the regulator is 
\begin{equation}
  \alpha= 0.15 \, . 
\label{alg2}
\end{equation}
The heat kernel regulated expression for the Wilson loop (\ref{wqed2}) reproduces the trend of the numerical results. Quantitative 
agreement is obtained for large values of the perimeter, i.e., for $R\ll T$. 
%###################################################################
\subsection{Large Size Wilson Loops and the String Limit}  
Results from lattice gauge theories (cf. e.g. \cite{LUWE02}) support the conjecture that large Wilson loops can be described by a an
effective string theory. In the string picture, one assumes the Wilson loop is determined by the string partition function  \cite{amop84}
\begin{equation}
W(R,T) = \int {\rm d}[x_{\perp}] {\rm e}^{ -\sigma_{cl} R T -S_{qf}(x_{\perp})},\quad
S_{qf} = \int_0^R\int_0^T {\rm d}\xi_1\,  {\rm d}\xi_2\, \partial_{\alpha}x_{\perp}^{\mu}\, \partial_{\alpha}x_{\perp}^{\mu}\, .  
\label{str1} 
\end{equation}
Apart from the classical string  energy $\sigma R$, this expression accounts for the small quantum fluctuations around the classical
 string. The integral over the fluctuating transverse string coordinates $x^{\mu}(\xi_1,\xi_2)$ is given by the determinant of the 
two-dimensional Laplacian, 
\begin{equation}
\label{str21} 
{\rm e}^{-FT}=\int {\rm d}[x_{\perp}] {\rm e}^{-S_{qf}(x_{\perp})}= \frac{1}{\det(-\Delta)} \, . 
\end{equation}
For  computation of  $\det(-\Delta)$, one imposes Dirichlet boundary conditions along the boundary of the Wilson loop. 
In terms of the resulting spectrum, the free energy  $F$ for a rectangular loop of side lengths $R,T$ reads
\begin{equation}
\label{str2} 
FT = \sum^{\infty}_{m,n = 1} \ln \left ( \frac{\pi^{2}m^{2}}{T^{2}} 
+ \frac{\pi^{2} n^{2}}{R^{2}} \right ) \, . 
\end{equation}
In $\zeta$ function  regularization \cite{DIFI83}, the following finite expression
\begin{equation}
\label{zeta} 
\det(-\Delta) = \frac{1}{\sqrt{2 R}}\eta \Big({\rm i} \frac{T}{R}\Big)\,
\end{equation} 
in terms of  the Dedekind $\eta$ function 
$$ \eta(z) = {\rm e}^{{\rm i}\pi z /12} \prod_{n=1}^{\infty}(1-{\rm e}^{2{\rm i}\pi n z}) $$ 
is obtained. The prefactor in the $\eta$ function yields the well known L\"uscher term  \cite{LUWE02}, $\pi T/12R$, which dominates 
the logarithm of the Wilson loop  in the limit of large $T/R$. 

For the purpose of analyzing the Wilson loops in pseudoparticle ensembles, we need a regularization scheme that makes explicit the 
role of the cutoff. 
In heat kernel regularization, the expression for the free energy reads  
\begin{equation}
T F (\lambda) = \sum^{\infty}_{m,n = 1} \ln \left ( \frac{\pi^{2}m^{2}}{T^{2}} + 
\frac{\pi^{2} n^{2}}{R^{2}} \right ) \; {\rm e}^{- \lambda \sqrt{\frac{\pi^{2}m^{2}}{T^{2}} + 
\frac{\pi^{2} n^{2}}{R^{2}}}} \, .
\label{sthk}
\end{equation}
The leading term in $\lambda^{-1} $ can be evaluated in closed form 
\begin{eqnarray*}
T F (\lambda)   & \approx &   \frac{TR}{\pi^{2}} \int {\rm d} p \, {\rm d} q \, \ln (p^{2} + q^{2}) \; 
{\rm e}^{- \lambda \sqrt{p^{2} + q^{2}}}   \\
& \approx & \frac{4 T R}{\pi} \left ( - \frac{\rm d}{{\rm d} \lambda} \right ) \int^{\infty}_{0} {\rm d} p \; 
{\rm e}^{- \lambda p} \ln p = \frac{4 T R}{\pi}  \frac{1}{\lambda^{2}} \; 
\big ( {\bf C} - 1 - \ln \lambda \big ) \, .
\end{eqnarray*}
In the limit $\lambda \rightarrow 0$, a quadratically divergent correction to the string tension is obtained. As follows from dimensional
 arguments, the subleading linearly divergent (up to logarithms) terms  are multiplying $R$ and $T$.  \newline
For comparison with the  large  Wilson loops evaluated in the pseudoparticle ensembles, we keep  $\lambda$ finite and introduce 
counterterms associated with the divergent contributions. We fit  our numerical results for the Wilson loop with the  Ansatz 
\begin{equation}
\ln W = - T F (\lambda) + a+ b(T+R)+ \sigma_{0} R T\, ,
\label{sthk2l}
\end{equation}
with the  counterterms $a,b$  and $\sigma_0$ depending  on $\lambda$.
\begin{figure}[h] 
\vskip -.5cm
\begin{center}
\epsfig{file=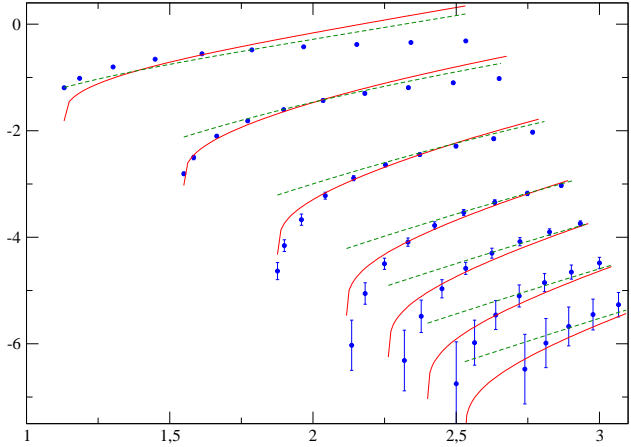, width=.5\linewidth,angle=-90}
\vskip -.4cm
\caption{Logarithm of rectangular Wilson loops of fixed area ${\cal A}$ as a function of the perimeter for an ensemble with 500 instantons of     
variable size $\rho \le 0.1$. The 7 values of the area vary between 0.08 and 0.40.  The  solid and dashed curves are  obtained from
 Eqs.~(\ref{wwl1}) and (\ref{sthk2l}) with the values of the parameters given in Eqs.~(\ref{instwp2}) and (\ref{instwp3}) respectively.} 
\label{instas}
\vskip -.3cm
\end{center}
\end{figure}
\begin{figure} [h]
\epsfig{file=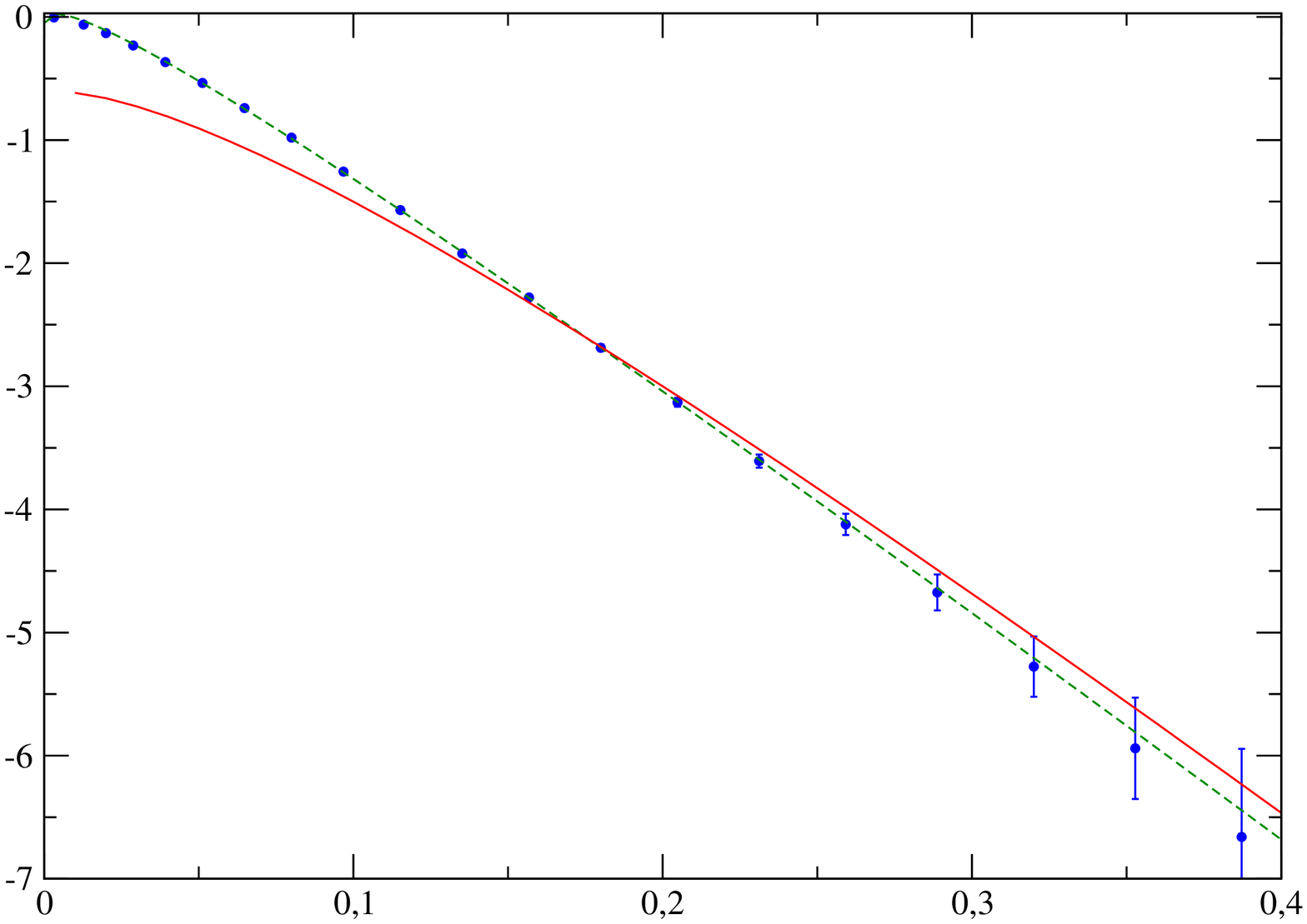, width=.4\linewidth,angle=-90}
\epsfig{file=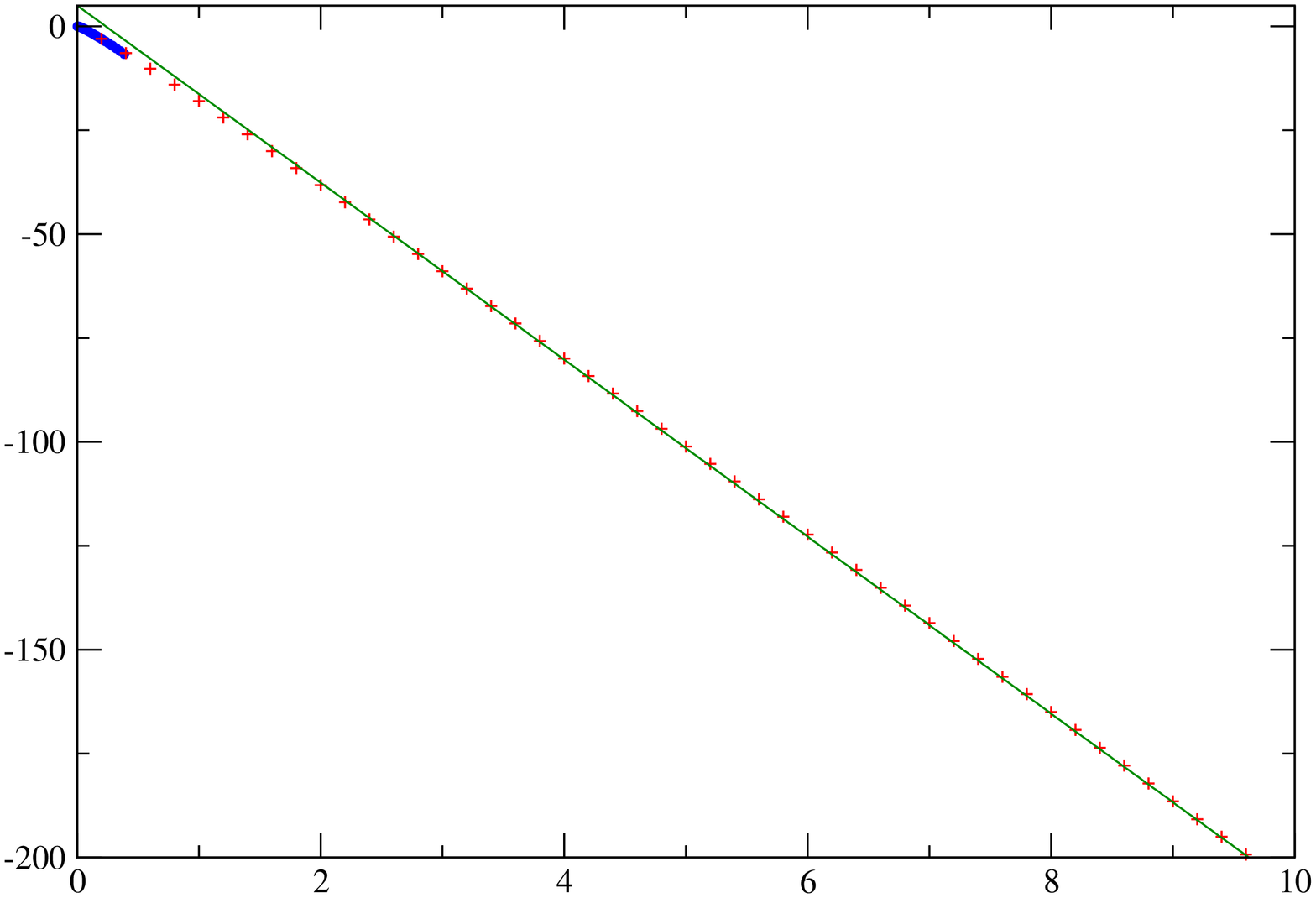, width=.4\linewidth,angle=-90}
\vskip -.3cm
\caption{ Left: Logarithm of rectangular Wilson loops with aspect ratio 1:2 as a function of  the area for an ensemble with 500 instantons of 
variable size $\rho \le 0.1$. The solid curve has been obtained from a fit of $\ln W$ in Eq.~(\ref{wwl1}) to the data in Fig.~\ref{instas} 
with the resulting values given in  (\ref{instwp2}) while  the dashed curve shows our standard  fit (\ref{stp1}) to the data points with
 aspect ratio 1:2 in the present figure with the extracted values (\ref{stp2i}).   
Right: Asymptotic extraction of the string constant from Wilson loops as a function of the area. Notice the much larger scale of the area.
% Numerical results for an ensemble with 500 instantons in the upper left corner (${\cal A}\le 0.4$). 
Crosses represent the function
  (\ref{wwl1})  with the values of the parameters  (\ref{instwp2}). The  straight line is obtained from fitting the asymptotics of 
(\ref{wwl1}) (crosses). The extracted value of the string tension is given in  Eq.~(\ref{stringas}).} 
\label{lawlm1}
\vskip -.3cm
\end{figure}
We have applied this procedure to Wilson loops of ensembles with 500 instantons and merons  respectively with variable size 
($\rho \le 0.1$). The fit to the pseudoparticle results yields  the following values of the parameters,  
\begin{equation}
 \lambda =0.25,\; a=-0.7 ,\; b=1.8,\; \sigma_{0}= -14\;.  
\label{instwp3}
\end{equation}
As  Fig.~\ref{instas} shows,  the string model defined by Eqs.~(\ref{sthk2l},\ref{instwp3}) leads essentially to a linear dependence 
of the logarithm of the Wilson loops as a function of the perimeter with a very mild dependence of the slope on the value of the
 area. For  sufficiently large aspect ratios of the loops, the string model reproduces   the numerical results for the pseudoparticle
 ensembles. However the model  misses the  change in slope close to  the threshold, i.e., in the limiting case of a square. For fixed
 area ${\cal A}$, the slope of the Wilson loop as a function of the perimeter  ${\cal P}$ becomes in general infinite if  the Wilson loop 
depends on $|T-R|$. 
We find
$$\frac{\partial}{\partial {\cal P}} |T-R|^{\mu} = \frac{\mu}{4}
 |T-R|^{\mu-2}{\cal P}\, ,$$
i.e., a divergent slope at threshold is obtained for $\mu<2$. The presence of such a term is indicated by the rapid change of the 
numerical results with  ${\cal P}$ at the threshold  ${\cal P}=4\sqrt{{\cal A}}$. To account for this structure, we modify minimally 
our Ansatz (\ref{sthk2l}),  
\begin{equation}
\ln W = - T F (\lambda) + a+ b\sqrt{T^2-R^2}+ \sigma_{0} R T,
\label{wwl1}
\end{equation}
such that neither the asymptotic behavior at large area nor, for fixed area,  at large perimeter is changed. In turn, the changes in  fit 
parameters are very small [cf. Eq.~(\ref{instwp2})]
\begin{equation}
 \lambda = 0.25,\; a=-0.7 ,\; b=1.8,\; \sigma_{0}= -13\;.  
\label{instwp2}
\end{equation}
As  Fig.~\ref{instas} demonstrates, the Ansatz (\ref{wwl1}) catches the essential properties of the pseudoparticle Wilson loops of 
sufficiently large size. Deviations at large values of the perimeter must be expected. The string picture  breaks down if the smaller
 length $R$ of the rectangle is of the order of the diameter of the flux tube,
$$R\approx 2r_0,$$
 and therefore the perimeter has to satisfy 
\begin{equation}
{\cal P} \le {\cal P}_{\rm max}=\frac{{\cal A}+4r_0^2}{r_0} \, .
\label{plim}
\end{equation}
With $r_0\approx 0.15 \,\mbox{fm} $ (corresponding to 0.07 in our units)  we obtain for ${\cal A}=0.15$ (second curve from the top in 
Fig.~\ref{instas}) $ {\cal P}_{\rm max} = 2.4$, a value which, as can be seen from Fig.~\ref{instas},  is of the correct order of magnitude.
    
The  origin of the observed threshold behavior is not obvious. It could be a peculiarity of the effective regularization in pseudoparticle
 ensembles.   Within the string model description, to lowest order, the string partition function [Eq.~(\ref{sthk})] does not depend on the
 variable $|T-R|$. One can show that the derivative of the $\zeta$ function (\ref{zeta}) with respect to the perimeter is finite at threshold.
 Infinite slopes of the Wilson loops however appear in higher order corrections  which have been  calculated  in
 $\zeta$ function regularization \cite{DIFI83}. It  would be interesting to extend our analysis and include such corrections within
 heat kernel regularization.
 
Having determined the counterterms (\ref{instwp2}), we extract the value of the string constant $\sigma$ from the asymptotic 
behavior of $\ln W$ [Eq.~(\ref{wwl1})]   
\begin{equation}
\sigma RT = \lim_{ R T \to \infty} - T \; F (\lambda) + \sigma_{0} RT\,.
\label{sthk3}
\end{equation}
Fig.~\ref{lawlm1} illustrates the procedure. On the left hand side our standard procedure for determining the string constant by 
fitting the Wilson loop at constant aspect ratio (1:2) is shown. The fit with the standard Ansatz 
\begin{equation}
\ln W = \omega + \tau {\cal P} -\sigma {\cal A}
\label{stp1}
\end{equation}
and extracted values  
\begin{equation}
 \omega=-0.1,\; \tau=0.58,\; \sigma= 20.5\, 
\label{stp2i}
\end{equation}
yields an excellent fit to our numerical results. The extracted value of the coefficient $\tau$ is  about a factor two smaller than 
the slopes of the dashed curves in Fig.~\ref{instas}, and does not adequately describe the numerical results.
 
The Ansatz based on the string model (\ref{wwl1}), which also accounts for the dependence on the perimeter for fixed area,  
describes the pseudoparticle Wilson loops  in a limited region only. Clearly the string model is not suited to describe the perturbative
 region of small loops. On the right hand side we show the extraction of the string constant from the asymptotics. Within the 
uncertainty of the procedures, the value  
\begin{equation}
 \sigma= 21\, 
\label{stringas}
\end{equation}
agrees with the value in (\ref{stp2i}).
%\vskip -.5cm
%\hspace{-1cm}
\begin{figure}
\epsfig{file=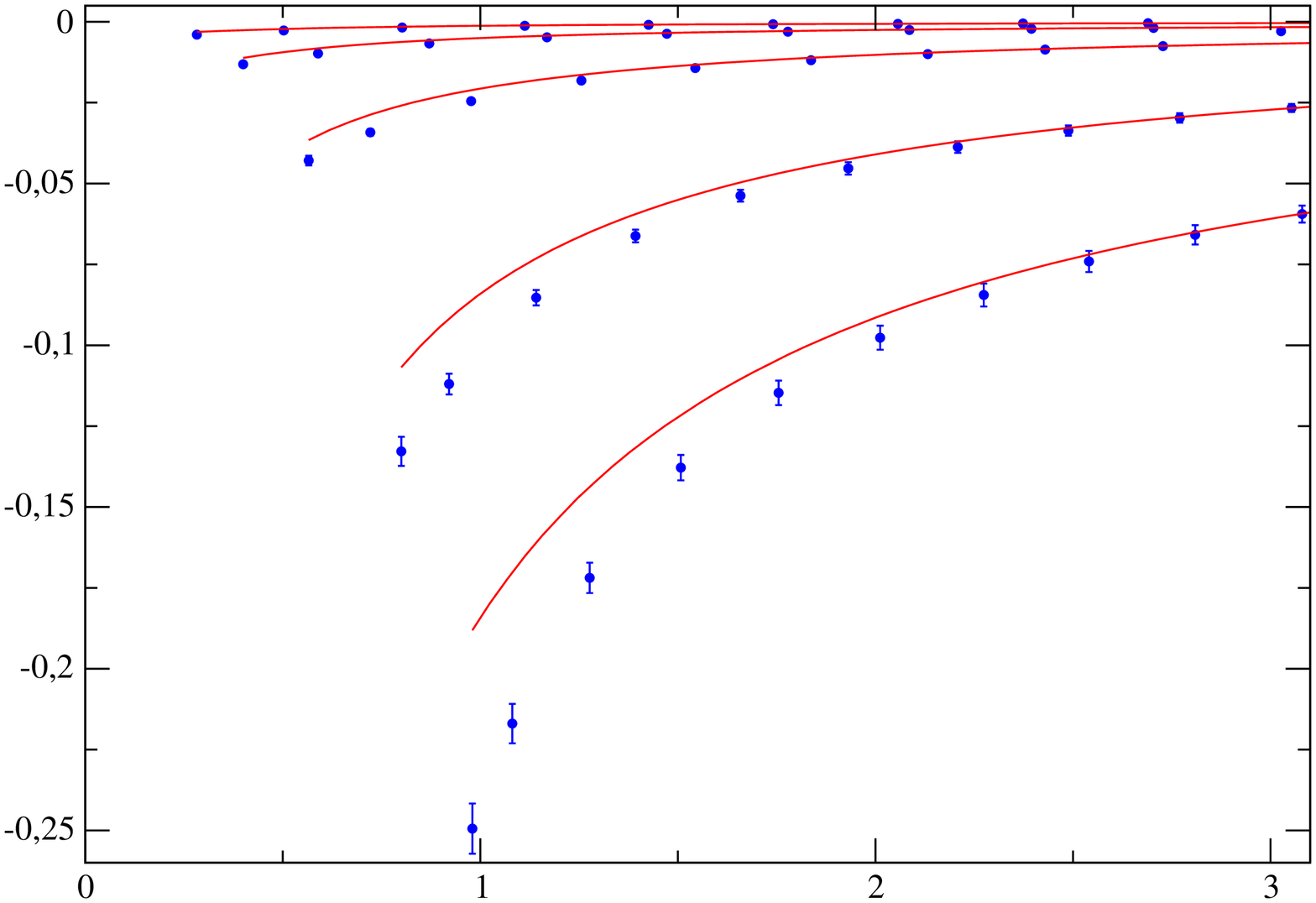, width=.4\linewidth,angle=-90}\hspace{-.5cm}\epsfig{file=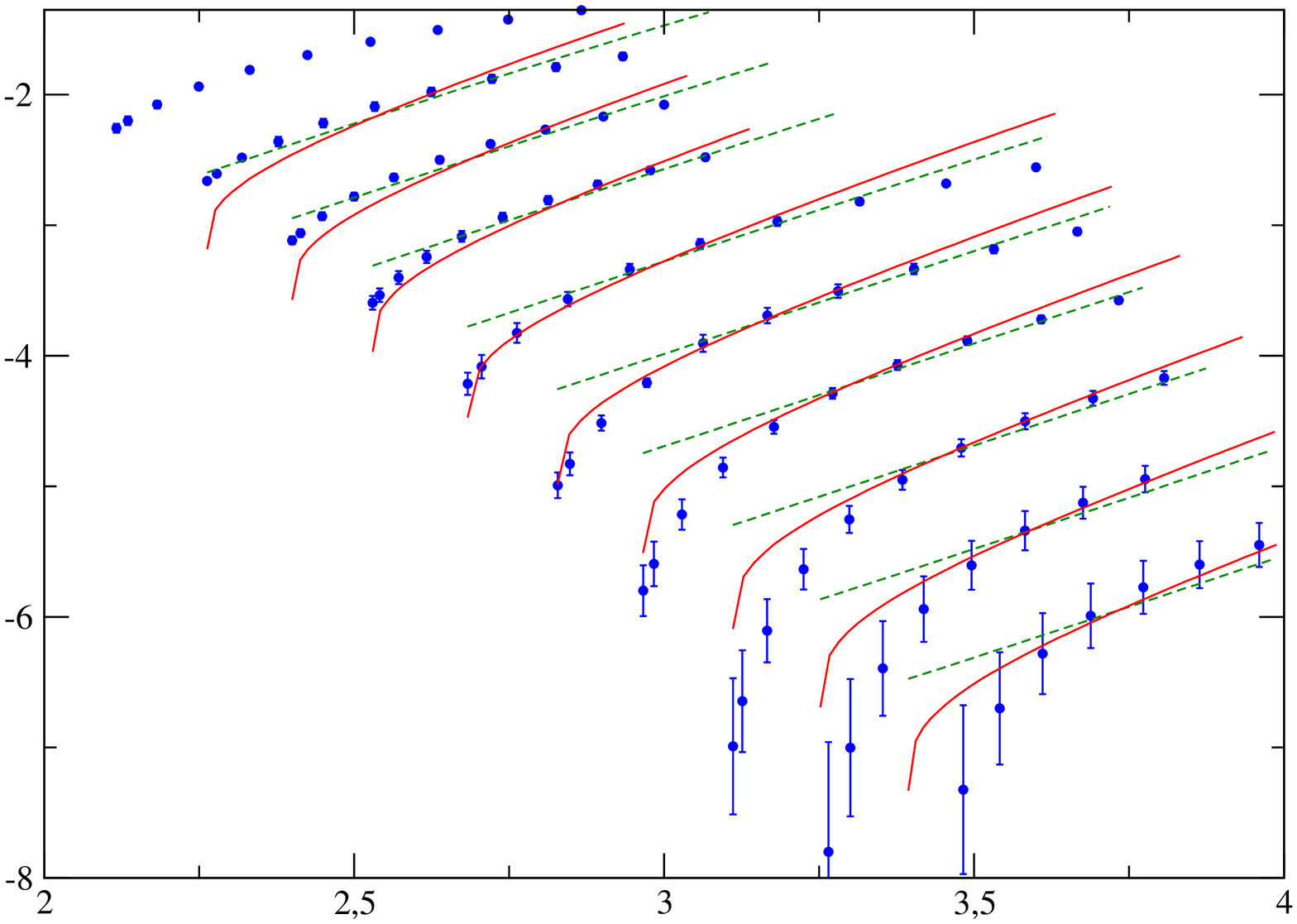, width=.4\linewidth,angle=-90}
%\vskip -.2cm
%\captionof{figure}
\caption{Logarithm of rectangular Wilson loops of fixed area as a function of the perimeter for an ensemble with 500
 merons of variable size $\rho \le 0.1$. Left: ${\cal A}= 0.005, $ $0.01,\, $ $ 0.02, 0.04, 0.06 $.  The curves are obtained from
 Eqs.~(\ref{wqed2}, \ref{wqed3}) with the value (\ref{alg3}) of the regulator.  Right:  The 10 values of the area vary between 0.08 and 0.72.  
The  dashed and solid  curves are  obtained from Eqs.~(\ref{sthk2l}) and (\ref{wwl1})  with the values of the parameters given in
 Eqs.~(\ref{sthk5}) and (\ref{wwl2}), respectively.} 
\label{smwl}
\end{figure}

Very similar results are obtained for the Wilson loops in meron ensembles. In Fig.~\ref{smwl}, the results of the analysis 
are presented for small and large Wilson loops respectively. The value of the heat kernel regulator for the small Wilson loops 
[cf. Eqs.~(\ref {wqed2},\ref {wqed3})]  is
\begin{equation}
  \alpha= 0.20 \, . 
\label{alg3}
\end{equation}
The parameters for the string model Ansatz (\ref{sthk2l}) have been determined to  be
\begin{equation}
 \lambda = 0.21,\; a=-1.2,\; b=0.9,\; \sigma_{0}= 0.8\,
\label{sthk5}
\end{equation}
and for the modified Ansatz (\ref{wwl1})  
\begin{equation}
 \lambda = 0.21,\; a=-1.1,\; b=0.9,\; \sigma_{0}=1.3\, .
\label{wwl2}
\end{equation}
The string tension extracted from the asymptotics of Eq.~(\ref{wwl1}) is
\begin{equation}
\sigma=12.5\, ,
\label{wwl3}
\end{equation}
in agreement with the value deduced from the parameterization (\ref{stp1})
\begin{equation}
 \omega=-0.5,\; \tau=0.94,\; \sigma= 12.8\, .
\label{stp2}
\end{equation}
The consistency of the results of these two different methods is an important indicator of the robustness
of our determination of the string tension.
\subsection{Wilson Loop Distributions and Higher Representation Wilson Loops } 
Wilson loop distributions and  Wilson loops in higher representations further characterize the confining Yang-Mills dynamics. In 
particular, by lattice calculations, 
an intermediate regime of Casimir scaling has been established for both SU(2) and SU(3) Yang-Mills theories and deviations from 
Casimir scaling by string breaking have been identified, cf. \cite{bali00,defo00,katr02,krade03}. More recently, 
the Wilson loop distributions have been introduced \cite{blnt05} as quantities that contain the full information on the Wilson
loops in higher representations.  The study of Wilson loop distributions revealed an unexpectedly simple property of the 
Yang-Mills dynamics. These distributions  can be described as a result of a diffusion process.  Here we will present  the Wilson
 loop distribution for pseudoparticle ensembles and discuss the related expectation values of Wilson loops in higher representations.
 We will compare our results with the diffusion model  for the Wilson loop distribution.  In this model the Wilson loop distribution 
$ p(\cos \vartheta,t) $
is a solution of the diffusion equation on the group manifold $S^3$,
\begin{equation}
  \label{dieq}
\left(\frac{\partial}{\partial t}-\Delta_{S^3}\right)\,\frac{p(\cos \vartheta,t)}{\sin \vartheta} = \frac{1}{\sin^2 \vartheta}\,\delta(\vartheta)\delta(t)\, ,  
\end{equation}
and is given by
\begin{equation}
  \label{spre}  
p(\cos \vartheta,t)= \frac{2}{\pi}\,\theta(t)\,\sum_{n=1}^{\infty}n\,\sin n\vartheta \,{\rm e}^{-(n^2-1)t} \,.
\end{equation} 
\begin{figure}[h]
\epsfig{file=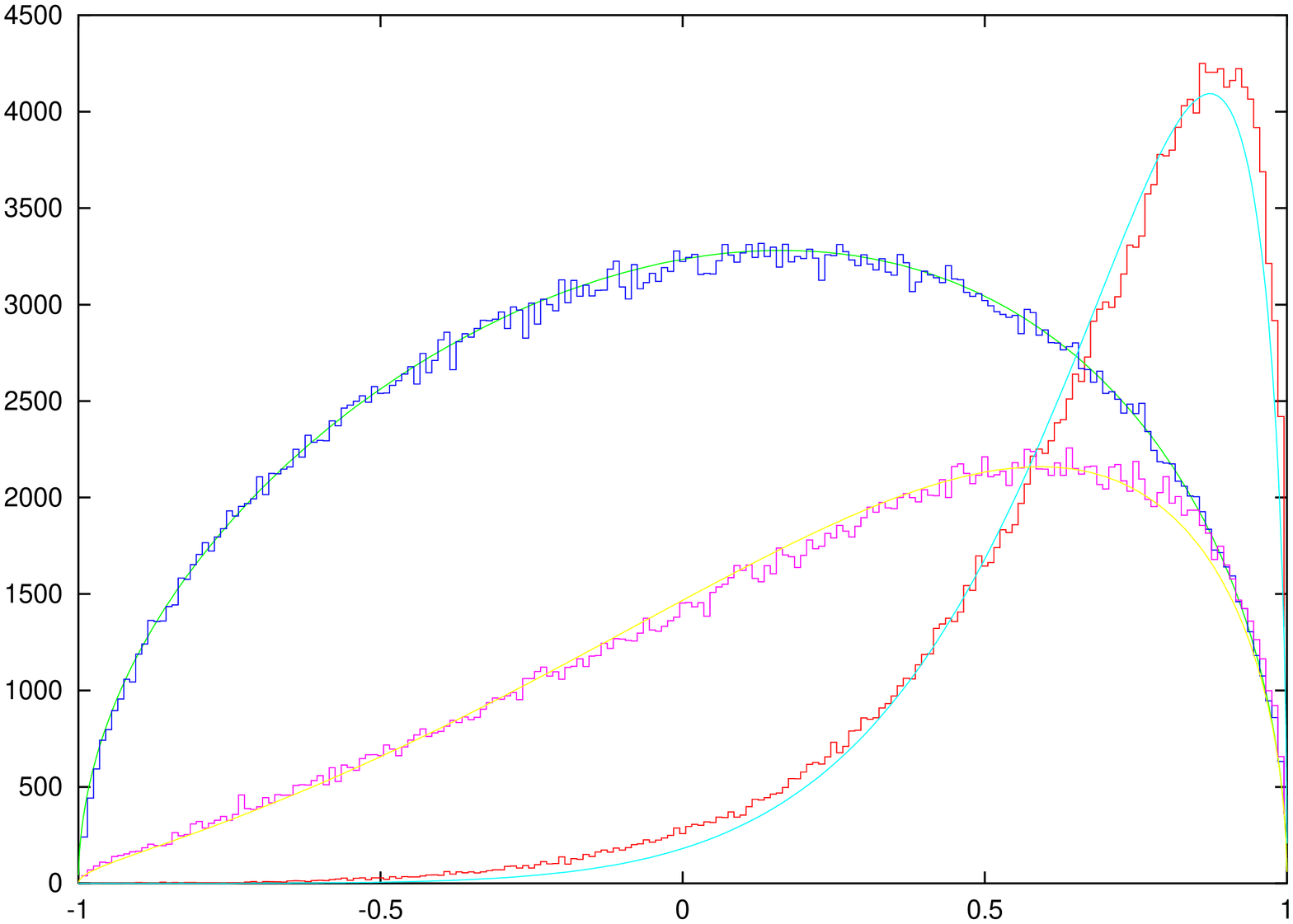, width=.33\linewidth,angle=-90}\epsfig{file=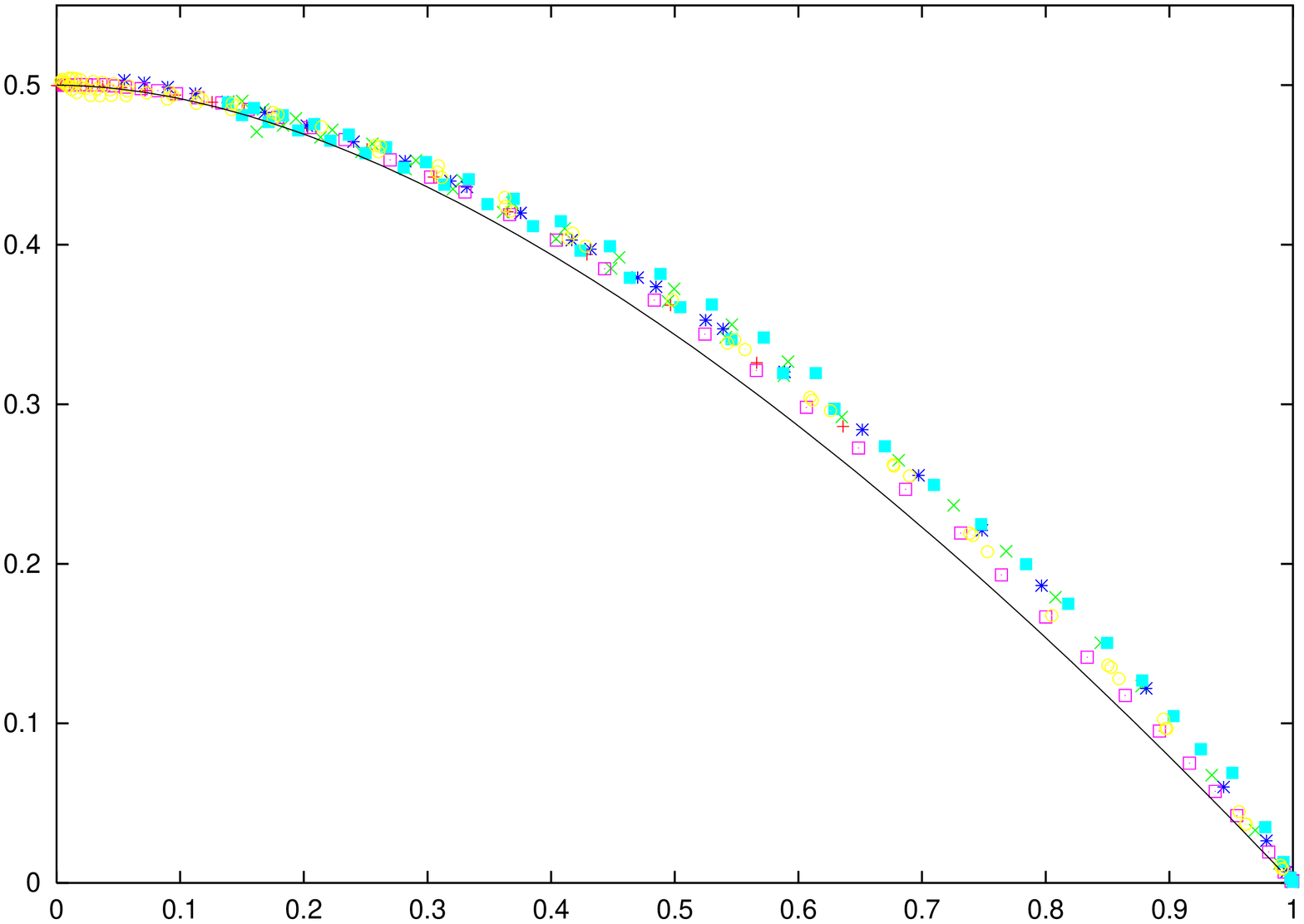, width=.33\linewidth,angle=-90}
%\captionof{figure}
\caption{Left: Wilson loop distribution for an ensemble of configurations containing 500 merons. The sizes of the loops 
are $0.48\times 0.24, 0.72\times 0.36, 1.\times 0.5 $ with average Wilson loop values $0.66,0.26,0.04$. 
These results  are  fitted with the distribution $p(\cos \vartheta,t)$ (\ref{spre}) with  parameters  $t=0.13, 0.44, 1.05$ respectively.  
%with normalizations  $1400,2500,9250$. 
Right: Variance of the Wilson loop  
  $\Delta w$  (\ref{diffvar}) as a function of the average $\langle w \rangle$ for ensembles of merons with $N_{M} = 50, 100, 200, 500, 1000, 3200$ 
together with the results of  the diffusion model  Eq.~(\ref{diffvar}).}
\label{diffwdis}
\end{figure} 
The diffusion model makes no assumption about the connection between the ``time'' $t$  and the Wilson loop size. For given size 
of the loop, we determine the value of the time $t$ by fitting the expectation value of the Wilson loop. As demonstrated in the left 
part of Fig.~\ref{diffwdis}, the Wilson loop distributions are well described by the diffusion model. In particular, the change in shape 
from the strongly peaked distribution for small loops to the Haar measure for large loops is correctly reproduced. These results are
 very close to results obtained in lattice gauge calculations  \cite{blnt05} and, together with our other results, strongly support the 
treatment of the Wilson loop dynamics in the pseudoparticle approach. 

\begin{figure}
\hspace{-1cm}\epsfig{file=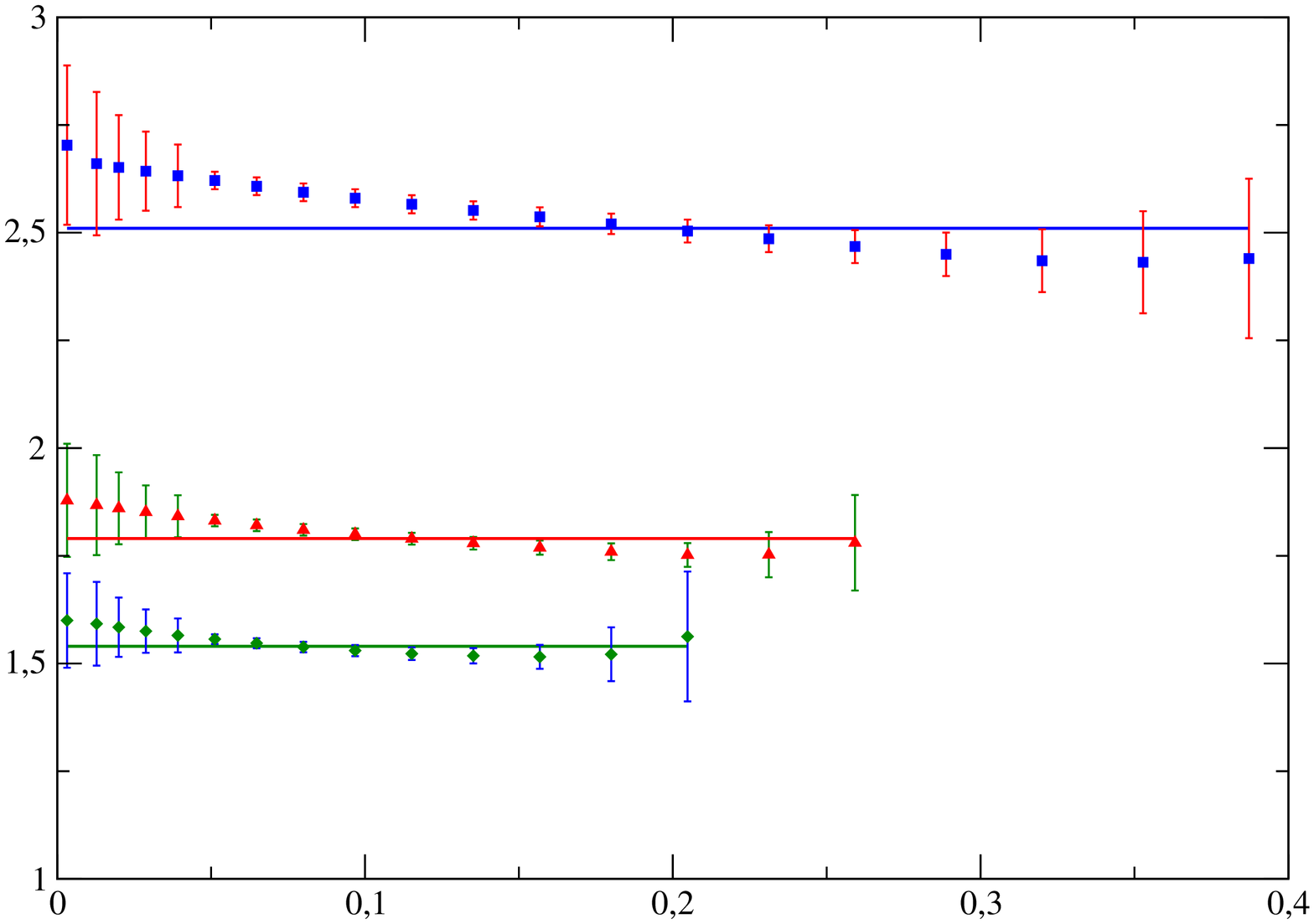, width=.45\linewidth,angle=-90}\hspace{-.8cm}\epsfig{file=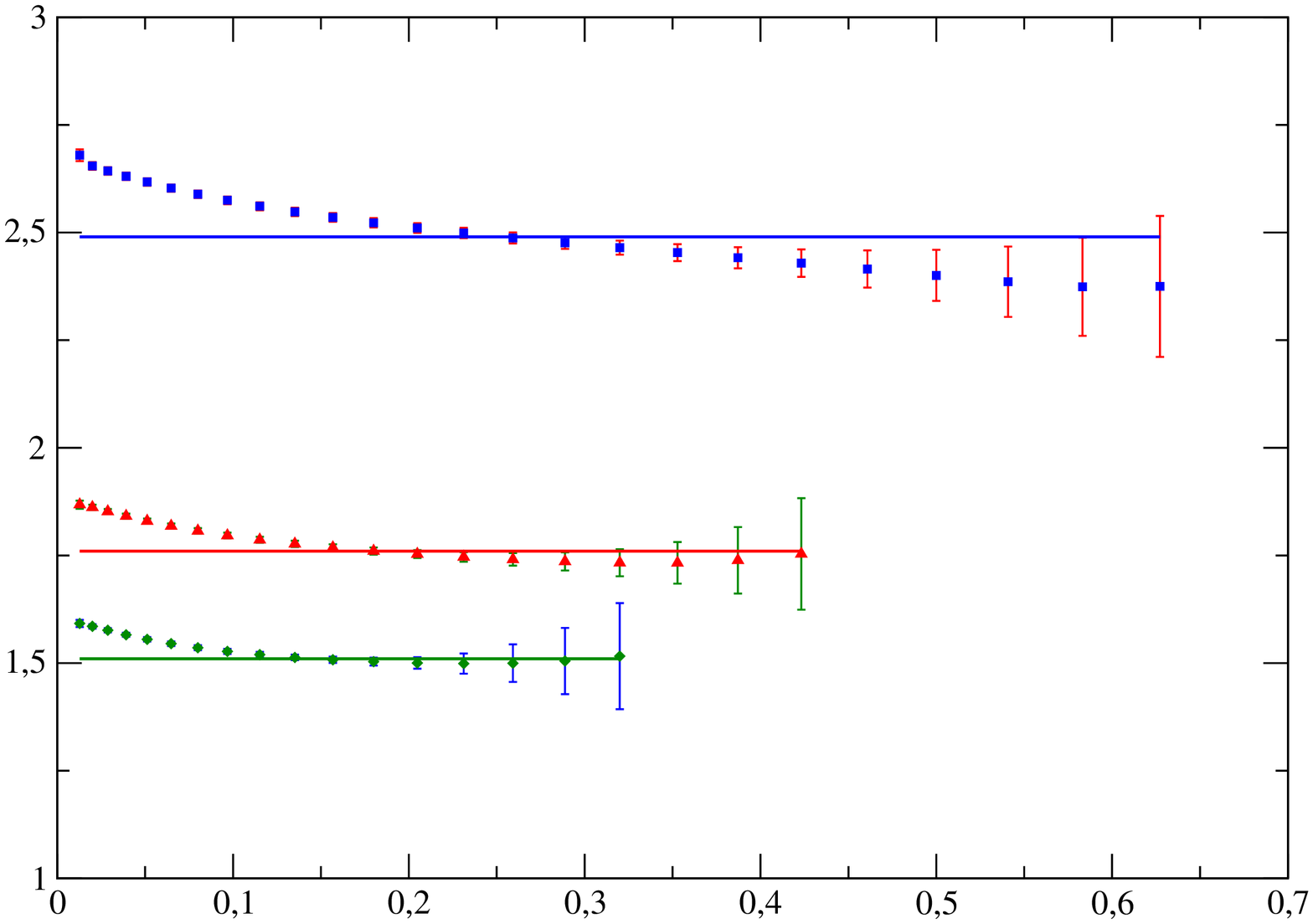, width=.45\linewidth,angle=-90}
\caption{Ratios of logarithms of Wilson loops of merons in consecutive representations $R_j$ [Eq.~(\ref{cassc})] for $j=1/2, 1, 3/2$,  $N_M=500$ 
(left) and $N_M=200$ (right).}
\label{rtwlj}
\end{figure}

Given the distribution of Wilson loops,  the variance of  Wilson loops is easily computed, 
\begin{equation}
  \label{diffvar}
  \Delta w =\sqrt{\langle w^2 \rangle -\langle w \rangle^2}=\frac{1}{2}\sqrt{1-4\langle w \rangle^2+3 \langle w \rangle^{8/3}}\,, 
\end{equation}
as well as  the expectation value 
\begin{equation}
  \label{wex}
\langle W_{j} \rangle = {\rm e}^{- 4j(j+1) t }\,
\end{equation}
of Wilson loops in  the ($2j+1$)-dimensional representation, 
\begin{eqnarray}
W_{j} (\vartheta) = \frac{1}{2 j + 1} {\rm tr} \; \exp\{ 2i \vartheta \left(
\begin{array}{ccc}
\!\!-j\! & & \\[-0.5em]  & \!\ddots\! & \\[-0.5em] & & \!j\! 
\end{array}
\right ) \} =  \frac{\sin (2 j+1)\vartheta}{(2j+1)\,\sin \vartheta}  \,. 
\label{wj}
\end{eqnarray} 
The right hand side of  Fig.~\ref{diffwdis} demonstrates the universality of the variance of the Wilson loop for meron ensembles
 with meron numbers $N_M$ varying by up to a factor 60. Agreement with the diffusion model within 10\% or less  is found. 
\begin{figure}
\begin{center}
\vskip -.5cm
\epsfig{file=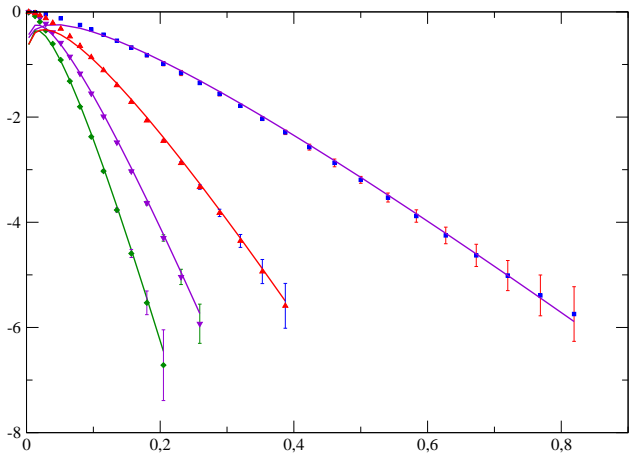, width=.5\linewidth,angle=-90}
%\captionof{figure}
\caption{Logarithm of a Wilson loop as a function of its area for meron ensemble with  $N_M =500$ in $j=1/2, 1, 3/2,$ and $2$
representations.
Also shown are curves corresponding to the parameterizations  (\ref{fitzins})  with the values of the parameters  given in Table \ref{hrwl}.}
\label{wlj}
\end{center}
\end{figure}
The numerical results for Wilson loop expectation values  in different representations are shown in Fig.~\ref{wlj}  together with 
fits using the parameterization  (\ref{fitzins}).  With increasing dimensionality of the representation, the negative slope of the 
logarithm of the Wilson loop expectation values increases by about a factor 5 (cf. Table \ref{hrwl}). 
%\end{figure}
\begin{table}[h]
\begin{center} 
\begin{tabular}{ |c|c|c|c||} \hline\hline
$j$ &$\omega$&$\tau$& $\sigma$  \\ \hline \hline
1/2&$-0.72$ &$1.10$ &$11.5$ \\ \hline
1&$-0.99$ &$1.89$ &$24.5$ \\ \hline
3/2&$-0.85$ &$2.24$ &$37.5$ \\ \hline
2&$-0.96$ &$2.95$ &$54.5$ \\ \hline
\hline
\end{tabular}
%\captionof{table}
\caption{Parameters for the fit (\ref{fitzins}) of Wilson loops in the representation $j$  for a meron ensemble with $N_M=500$
 ($\lambda=1$).}
\label{hrwl}
\end{center}
\end{table}
\vskip-.5cm
According to the diffusion model Eq.~(\ref{wex}), Wilson loops in different representations exhibit Casimir scaling, i.e.,  the values of the
 string tension, or more generally the interaction energy of static charges in higher representations, are proportional to $j(j+1)$. We have 
\begin{equation}
\label{cassc}
R_j=\frac{\ln \langle W_{j+1}\rangle}{\ln \langle W_{j}\rangle}=\frac{(j+1/2)(j+3/2)}{j(j+1)} \, . 
\end{equation}
 Casimir scaling of the Wilson loops in two meron ensembles is examined in Fig.~\ref{rtwlj}. On the average, we find that Casimir 
scaling is satisfied within 8\%. We observe a decrease in $R_j$ by about this amount. It appears that Casimir scaling is exact for small 
Wilson loops. Also these results are compatible with the findings in lattice gauge calculations where the validity of Casimir scaling has
 been demonstrated for SU(2)  \cite{amop84} and  SU(3)
  \cite{bali00} Yang-Mills theory.  Indications for deviations from Casimir scaling for sufficiently large loops have been obtained  
 in \cite{defo00} and \cite{katr02} and string breaking has been observed in \cite{krade03}. Due to poor statistics in the computation of
 large loops, we have not been able to
establish or rule out string breaking in the pseudoparticle ensembles. 
\vskip-.7cm
\section{Correlation Functions}
%
%%%%%%%%%%%%%%%%%%%%%%%%%%%%%%%%%%%%%%%%%
%
Correlation functions are useful tools to study susceptibilities, response functions, and excitations of many particle systems and quantum
 field theories. In the case of Euclidean time, they also enable the measurement of  the energies of low-lying excited states.  In this
 section, we will use appropriately chosen operators to study the topological susceptibility and low glueball masses of our effective theory,
 and compare them with corresponding lattice results.  We begin with a discussion of methodology, describe the operators used to probe 
the system, discuss several related theoretical issues,  present results for the  $0^+$, $1^+$, $1^-$, $2^+$, and $2^-$ glueballs, and discuss a 
correlation function that reflects confinement.  
\subsection{Correlation Functions of the Action-  and Topological Charge Density}
As an introduction, it is useful to begin by considering two correlation functions that we can analyze to a large extent analytically, the
  correlation functions of the action density [Eq.~(\ref{act0})],
\begin{equation}
   \label{C_s}
C_s(x) = \langle [s(x)-\langle s\rangle ] [s(0) - \langle s \rangle ]
\rangle , 
\end{equation}
and of the topological charge density [Eq.~(\ref{tocd})],
\begin{equation}
   \label{C_q}
C_{\tilde s}(x) = \langle {\tilde s}(x) {\tilde s}(0) \rangle .
\end{equation}
The action density $s(x)$ and topological charge density $\tilde{s}(x)$ are the simplest examples of gauge invariant quantities. They are 
scalar or pseudoscalar quantities and bilinear in the field-strength.  
The explicit expressions of these quantities  for  a single meron or instanton  are given in Eqs.~(\ref{smde1}) and (\ref{smde2}).
The normalization of the correlation functions  yields the variance of the action and topological density respectively. 
Many results from this section have been tabulated in Tables  \ref{medy12} and  \ref{inst1} for previous reference and discussion, and 
one can see from these tables that
fluctuations of the action are  of the same order as the average value, 
$$ \sqrt{C_S(0)} \sim  \langle s \rangle .$$
The  large fluctuations are due to the large differences in the values of the action density inside the pseudoparticles  and in the background.
In the center,
$$ s_0(0)=\tilde{s}_0(0) = \frac{12 \xi^2}{\rho^4} $$
which,  in the ensemble of fields with 500 merons for example, is an order of magnitude larger than the average action density. The  
square of the  action and of the topological density is concentrated in the core of the pseudoparticles.  Therefore the  strength of the 
fluctuations and more generally the correlation function can be estimated by the incoherent sum over the  single pseudoparticles. In 
such an incoherent single pseudoparticle approximation, we define the correlation function of, for instance,  the topological density by
\begin{equation}
  \label{spco}
C_{\tilde{s}}^{\rm incoh}(x)= \frac{1}{V}\sum_{i=1}^{N_P}\int {\rm d}^4 z_i\, \tilde{s}_0 (y-z_i) \,\tilde{s}_0 (y+x-z_i)=
 \frac{N_P}{V}\int {\rm d}^4 y \ \tilde{s}_0 (y)\, 
\tilde{s}_0 (y+x)\, ,
\end{equation}
and similarly for any other observable. 
In this approximation, the integration over the color orientation  gives a multiplicative factor which is canceled by the normalization of
 the partition function.
Using the explicit expressions (\ref{smde1}) and (\ref{smde2}), we obtain the following estimates for the fluctuations in meron 
ensembles ($N=N_M$)
\begin{equation}
  \label{acnorm}
C_s^{\rm incoh}(0) =  \frac{1551 \pi^2}{280}\frac{N_M}{\rho^4\,V}\, ,\quad 
  C_{\tilde{s}}^{\rm incoh}(0)  =  \frac{177\pi^2 }{35}\frac{N_M}{\rho^4\,V}\, ,\quad \xi=1\, ,
\end{equation}
and for instantons
\begin{equation}
  \label{actonorm}
C_s^{\rm incoh}(0)=C_{\tilde{s}}^{\rm incoh}(0)=\frac{384\pi^2 }{7}\frac{N_I}{\rho^4\,V}\, ,\quad \xi=2\, .
\end{equation}
Comparison with the numerical results in Tables  \ref{medy12} and \ref{inst1}  shows that the general trends in the values
 of the correlation functions for vanishing separation are correctly reproduced by this incoherent superposition of pseudoparticles.
 The similar values of $C_s(0)$ and $C_{\tilde{s}}(0)$, the difference by an order of magnitude for meron and instanton ensembles,
 the dependence on the pseudoparticle size and number density agree typically within 20-30 \% with the numerical results.
 Although the pseudoparticle ensembles  are far from being random, the fluctuations are dominated by the variations of the densities 
on the small scale of the meron size where  the longer range correlations between pseudoparticles have little effect.
 
In the following, we will compare  our numerical results for the correlation function with the single pseudoparticle predictions 
[cf. Eq.~(\ref{spco})] which we have evaluated numerically for various observables. Besides the values for zero separation 
[Eqs.~(\ref{acnorm}, \ref{actonorm})], the asymptotic behavior can also be determined analytically. We find for meron ensembles
\begin{equation}
  \label{asyapm}
  t \rightarrow \infty: \quad C_s^{\rm incoh}(t) \rightarrow \frac{9 \pi^2}{2} \frac{N_m}{V}\frac{\ln t/\rho}{t^4}\,,\quad C_{\tilde{s}}^{\rm incoh}(t) 
\rightarrow 24 \pi^2 \frac{N_M}{V}\frac{\rho^2}{t^6}\,,\quad \xi=1\, , 
\end{equation}
and for instanton ensembles
\begin{equation}
  \label{asyapi}
  t \rightarrow \infty: \quad C_s^{\rm incoh}(t)= C_{\tilde{s}}^{\rm incoh}(t) \rightarrow 24 \pi^2 \frac{N_I}{V}\frac{\rho^4}{t^8}\,,\quad \xi=2\, . 
\end{equation}
The integral over the topological charge density correlation function yields the topological susceptibility
\begin{equation}
\label{susc}
\chi = \Big(\frac{1}{8\pi^2}\Big)^2 \int {\rm d}^4x \, C_{\tilde s}(x)\, ,
\end{equation}
an important quantity of the strong interaction. In  Tables  \ref{medy12} and  \ref{inst1}   the values of $\chi^{1/4}$ for the various
 ensembles are given. Once more, these values essentially arise from an incoherent sum of single pseudoparticle contributions. 
For the calculation of $\chi$  we need  the curvature $\beta$, 
$$ \beta = -\frac{(C_{\tilde{s}}(0))^{\prime\prime}}{2C_{\tilde{s}}(0)}= -  \frac{1}{2C_{\tilde{s}}(0)}\int {\rm d}^4 x\, \tilde{s}(x)\,
\partial^2_t \,\tilde{s}({\bf x},t),$$
and by 
approximating the single pseudoparticle correlation function by a Gaussian of the same curvature we obtain 
$$\chi^{1/4}= \sqrt{\frac{1}{8\pi\beta}}\,C_{\tilde{s}}(0)^{1/4}\,$$
resulting in
\begin{equation}
  \label{bechi0}
\beta^{\rm incoh}  = \frac{184}{177}\frac{1}{\rho^2}\, ,\quad \xi=1\,;\quad \beta^{\rm incoh} = \frac{4}{3\, \rho^2}\, ,\quad \xi=2\,,
\end{equation}
 and
the following expressions for meron and instanton ensembles
\begin{equation}
  \label{bechi}
\Big(\chi^{\rm incoh}\Big)^{1/4}= 0.52\,\Big( N_M/V\Big)^{1/4}\,,\quad \xi=1\,;\quad \Big(\chi^{\rm incoh}\Big)^{1/4}
= 0.83 \,\Big(N_I /V\Big)^{1/4}\,,\quad \xi=2 \,. 
\end{equation}
The $N^{1/4}$ dependence is easily understood on rather general grounds. If we assume the 
topological charge in volume $V$ to be given by the difference in the number of pseudoparticles and anti-pseudoparticles  
$N - \bar{N}$, the fluctuation in the 
charge is expected to  be $\sim \sqrt{N_M} $  and hence $\chi \sim N$. The estimate (\ref{bechi})  deviates from  the numerical results
 by  about  10 \%. For small pseudoparticle sizes, the approximation of the distribution by a Gaussian becomes invalid and deviations
 of the order of 20-30 \% occur.

 The case of the topological charge density
correlation function is particularly interesting.
Since ${\tilde s}\sim E\cdot B$, it acquires an ${\rm i}$ in Euclidean space and the correlation function
is negative for nonzero $x$, $C_{\tilde s}(x) \sim -\sum_n  |\langle
0 | {{\tilde s}}  | n \rangle  |^2 {\rm e}^{-E_n x} $.  The positive topological
susceptibility arises from a positive contact term at at the origin \cite{Vicari99}.
\begin{figure}
\hspace{3cm}  \epsfig{file=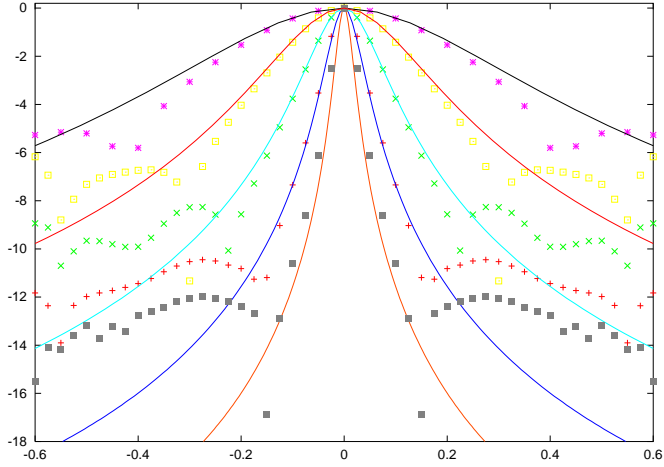, width=.4\linewidth,angle=-90}  
\caption{The correlation function $|C_{\tilde{s}}(x)|$ for $N_{M}= 50$, $\rho =0.01, 0.02, 0.04, 0.08, 0.16$
in comparison with the single-meron correlation functions. 
Note that in every case the correlation function has a node beyond which it is negative.}
\label{cpr116}
\end{figure} 
As Fig.~\ref{cpr116} shows, the pseudoparticle approximation to the path-integral exhibits these properties. The contact term is 
approximated by the single pseudoparticle  distribution [Eq.~(\ref{spco})], and after the change of sign, the correlation function
 exhibits a new length scale.  
 For $\rho=.02$, the results follow the single meron correlation function until $\tilde{s}(x)$ has decreased to a value of about 
$5\cdot 10^{-5}$. Only at this level does a new length scale become important.  We see here in detail how with decreasing cutoff $\rho$,  
the strength remains in the short range peak to produce a topological susceptibility essentially independent of the value of the cutoff.
 
In section 4, we have already compared the measured topological susceptibilities for instanton and meron ensembles with
 lattice results. The instanton result, $0.42 \le \chi^{1/4}/\sigma^{1/2}\le 0.48$, agrees within roughly 10\% with the SU(2) lattice
 result \cite{LUTE01}, whereas the meron result $\chi^{1/4}/\sigma^{1/2} \approx 0.31$ is significantly lower.  From the developments
 in this section, we now see the origin of this difference. The estimate (\ref{bechi}) of the ratio of 1.6 of the topological susceptibilities 
($\chi^{1/4}$) of instantons and merons arises from the fact that the topological charge of instantons is twice that of merons, and 
this ratio is only slightly reduced when expressed in physical units.
\begin{figure}
\hspace{-1.1cm}\epsfig{file=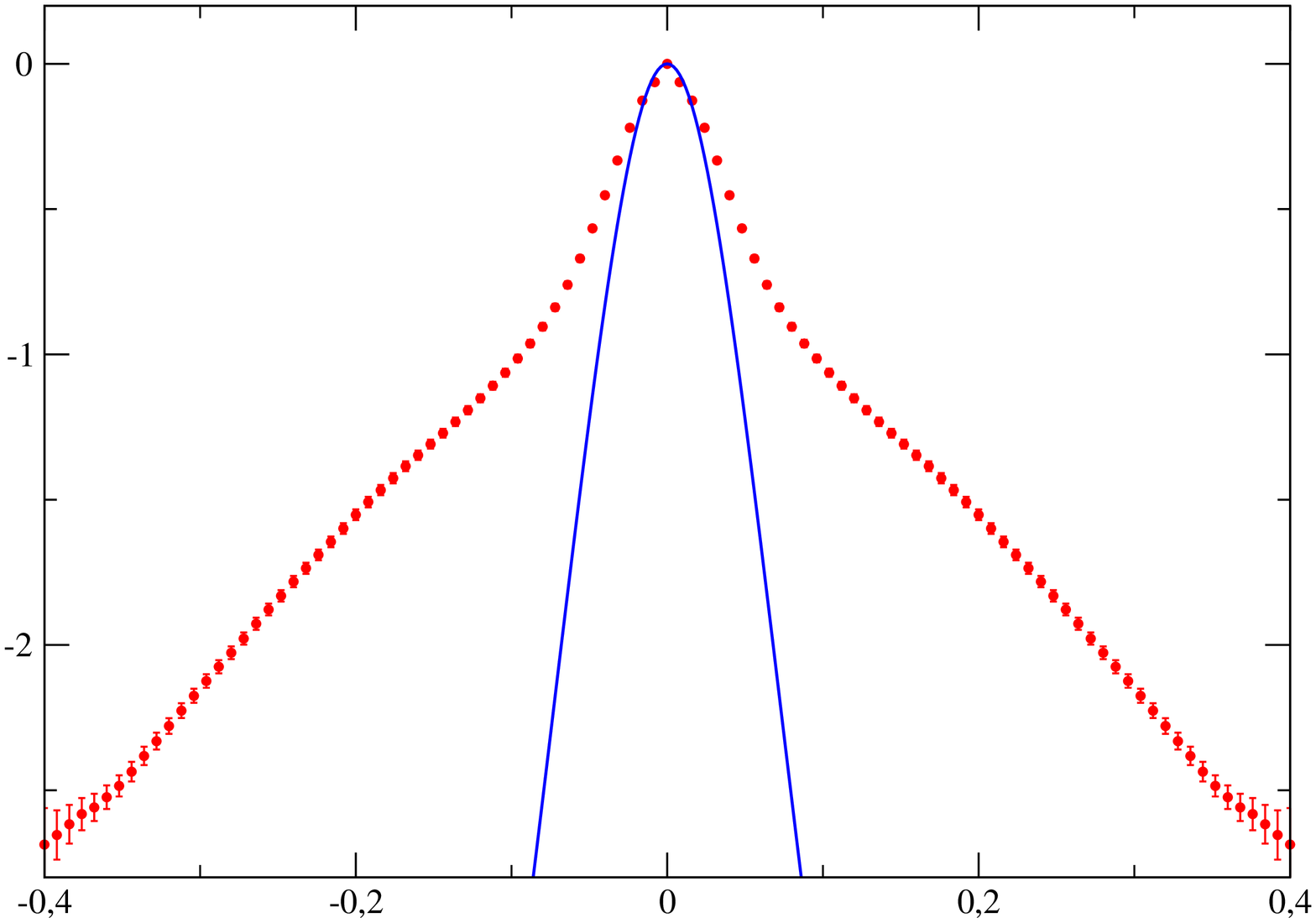, width=.45\linewidth,angle=-90}\hspace{-.7cm}
\epsfig{file=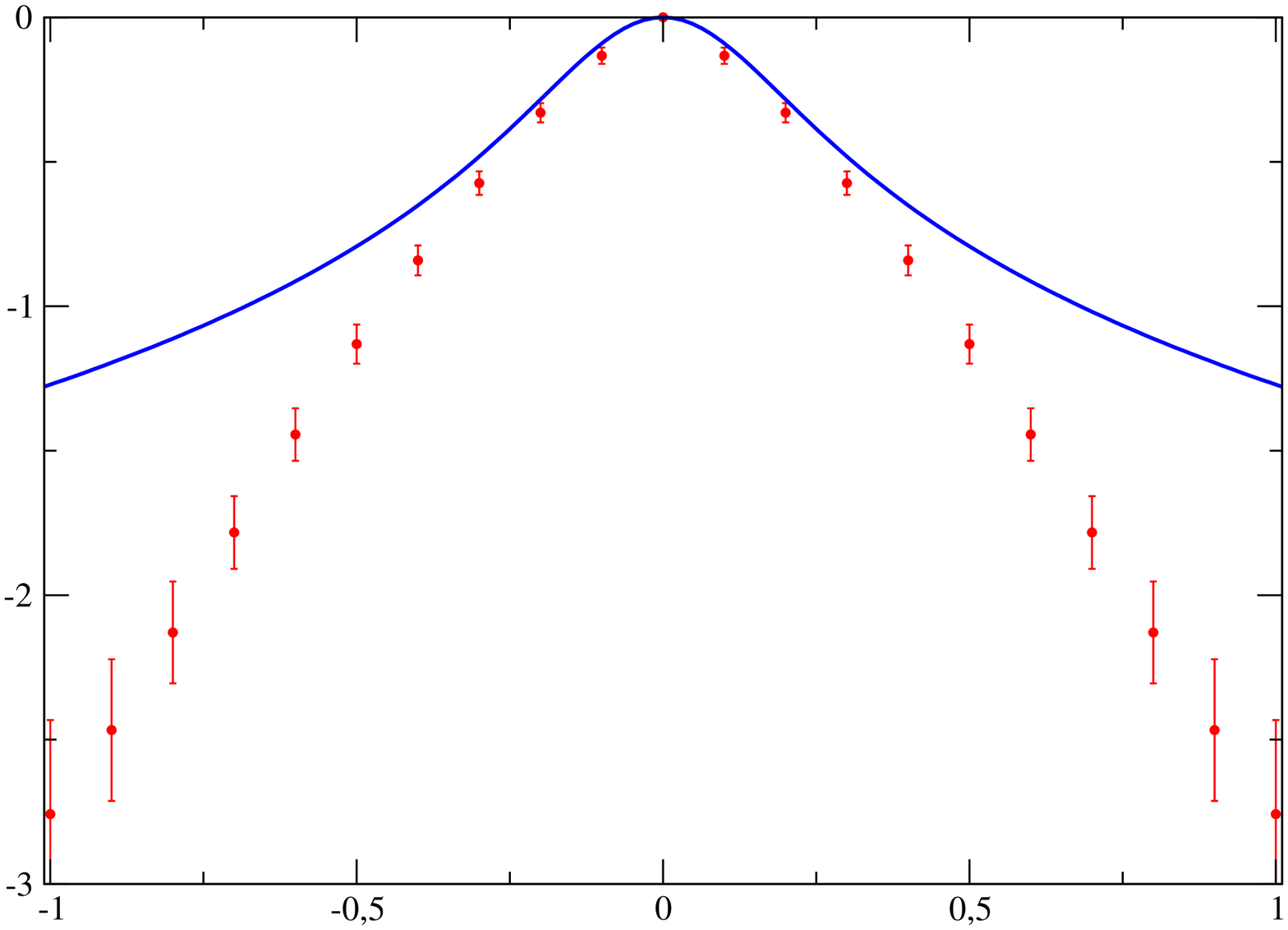, width=.45\linewidth,angle=-90}
\vspace{-.5cm}
\caption{Emergence of the hadronic scale in the correlation function $C_s(x)$ in ensembles  with 50 instantons  of  size $\rho=0.05$ (left)
  and 800 merons  ($L=4$) of  size $\rho=0.2$  (right), in comparison with the respective single pseudoparticle correlations functions.}
\label{acdens}
\end{figure}
 Fig.~\ref{acdens} shows results for the action density correlation function that illustrate the emergence of a hadronic scale, 
that is, a scale generated by  dynamics that is significantly different from the scale set by the size of the pseudoparticles. If an 
effective theory has the correct degrees of freedom, we should expect it to be able to generate scales larger or smaller than 
that of the underlying pseudoparticles.  The left figure demonstrates the superposition of instantons of small size ($\rho=0.05\ {\rm fm}$) 
to generate a much larger correlation length, a familiar phenomenon arising in many-body systems.
The right figure shows an example of the opposite case, where superimposing merons of size 0.2 fm actually  generates a hadronic
 scale that is smaller than the  meron size. To emphasize the very slow decay [cf. Eq.~(\ref{asyapm})] of the  single meron correlation 
function, its profile is also shown in the figure.  The emergence of the hadronic scale is a rather subtle effect generated by destructive
 interference among the pseudoparticles. To demonstrate the dominance of this hadronic scale  over a large distance  despite the 
weak decay of the building blocks,  we had to use a  meron ensemble in a volume 16 times larger than usual and containing 16 times
 more  pseudoparticles.

The successful generation of the negative correlation function for the topological charge density and  of hadronic scales both 
larger and smaller than the pseudoparticle scales for the action density are important indicators of the potential of the 
pseudoparticle effective theories to describe low energy dynamics.  Based on this success, we now use correlation functions
 to explore the low lying glueball spectrum.
%%%%%%%%%%%%%%%%%%%%%%%%%%%%%%%%%%%%%%%%%
\subsection{Correlation Functions and Glueball Masses}
%%%%%%%%%%%%%%%%%%%%%%%%%%%%%%%%%%%%%%%%%
To the extent to which the effective degrees of freedom succeed in
representing the full path integral,  the integration yields a transfer matrix
and one may write a Euclidean correlation function as a sum over eigenstates,
\begin{equation}
\langle {\cal O}(x) {\cal O}(0) \rangle  \sim \sum_n \langle \Omega | {\cal O}
| n \rangle {\rm e}^{-E_n x} \langle  n | {\cal O} | \Omega \rangle  \ .\label{expsum}
\end{equation}
As in lattice gauge theories, glueball masses may be extracted from the exponential decay at sufficiently large distances. 
In the case of the action density, we thus expect for large  separation $x$
that $C_s(x)  \sim |\langle n_0 | s  | 0 \rangle  |^2 {\rm e}^{-m_0 x} $, where $m_0$ is the mass of the
lowest $0^{++}$ glueball. Similarly, the asymptotic decay of the pseudoscalar correlation function is expected to be determined 
by the lowest  $0^{-+}$ glueball mass.
%%%%%%%%%%%%%%%%%%%%%%%%%%%%%%%%%%%%%%%%%
\subsubsection{Methodology}
To facilitate the extraction of single exponential terms in  Eq.~(\ref{expsum}), it is advantageous to project onto states with zero 
momentum so that the continuum of states of non-zero momentum associated with  each intrinsic excited state is eliminated, 
and the gap between the desired state and the next higher energy state is as large as possible.  In addition, although in a
 rigorously translationally  invariant system it would be sufficient to project either the source operator at the time origin or the 
sink operator at $t$, since the current calculation is only approximately translationally invariant, it is beneficial to project both the 
source and the sink.  Hence, we will use the doubly momentum projected correlation functions
\begin{equation}
\hat{C}_{\cal O}(t) = \int_{V} {\rm d}^3 x \int_{V} {\rm d}^3 y \langle 0|{\cal O}(t,{\bf x}){\cal O}(0,{\bf y})|0\rangle\, .
\label{chat}
\end{equation}
The time direction in 4-dimensional space is defined here by the coordinate over which one does not integrate. The integration  
over the 3-dimensional cube $V$ is limited to  $\ell \approx 0.5 L...0.7 L$ in  order to avoid surface effects. 

 In order to accurately determine glueball masses, it is necessary to identify a substantial region of pure exponential decay  between 
short distance artifacts associated with the intrinsic size $\rho$ of our effective degrees of freedom  and large distance artifacts 
associated with the finite volume edge effects and the fact that  beyond some point the  fluctuations arising from finite statistics 
are  comparable with the magnitude of the decaying exponential. In general, the region between the short distance and large 
distance artifacts is quite limited, and the identification of the exponential regime is ultimately subjective.  Unfortunately, this precludes 
quantifying statistical and systematic errors, and in the following discussion we will simply explain our rationale for selecting the 
fiducial regions we use to extract the slopes defining the exponential decay of each state, and give representative values, or ranges 
of values for each slope. Fig.~\ref{acdens} is a good example of a case in which there is clear separation between the short range 
artifacts associated with the intrinsic size $\rho$ and a substantial region of nearly exponential decay.  Similarly, in Fig.~\ref{cpr116}, 
as previously noted, one observes the expected negativity in  $C_{\tilde s}(x)$ beyond $ 2 \rho$, indicating that one is beyond the range 
of short distance artifacts.  However, in this case, it is also clear that the fluctuations become so large at larger distances that there is 
no correspondingly clean region of exponential decay, and we are not able to extract a mass from the pseudoscalar correlation 
functions in the present work.
 
Given the difficulties in determining masses in many cases, we have also required 
that we obtain consistent results ($<\pm 10 \%)$ for ensembles with different meron sizes or ensembles with different number of 
pseudoparticles.  When possible, we have also used different operators to create states of the same quantum numbers, and 
sought consistency between these results.  Finally, for spin 1 and 2 excitations we required  deviations from rotational invariance 
to be small  in the fiducial region. 
\subsubsection{ Operators}
%%%%%%%%%%%%%%%%%%%%%%%%%%%%%%%%%%%%%%%%%
To calculate glueball masses, we use correlation functions based on two kinds of operators:  local operators constructed from 
products of two or three field-strength operators, and non-local Wilson loop operators.  They are closely related to gluonic 
operators used in recent lattice calculations of the glueball spectrum and matrix elements (see  \cite{morn06} for the relationship
 to the lattice symmetry group and further refinements).  We first introduce the local operators and discuss their behavior under
 interchange of Euclidean space and time axes, and then describe the Wilson loop operators.  Properties of the operators and 
calculations based on them are summarized in Table~\ref{cor-ops}.

\noindent{ \bf Local operators with two or three field-strengths}

We have considered the complete set of gauge invariant operators consisting of products of 2 or 3 field-strength operators, 
which we  represent in terms of magnetic and electric fields in order to make the transformation properties under 3-rotations 
explicit.  We use the following set of observables consisting of two field operators 
\begin{equation}
  \label{obs2}
s= \frac{1}{4}({\bf E}^2 + {\bf B}^2)\,, \; T_+={\bf E} \otimes {\bf E} +  {\bf B} \otimes{\bf B} \,, \;
 T_-={\bf E} \otimes {\bf E} -  {\bf B} \otimes{\bf B} \,, \; T_a ={\bf E} \otimes {\bf B} +  {\bf B} \otimes{\bf E} \, ,
\end{equation}
with the traceless tensors      
\begin{equation} ({\bf X}\otimes {\bf Y})_{ij}= X_i^a Y_j^a-\frac{1}{3}{\bf X}^a{\bf Y}^a\delta_{ij}\, ,
\end{equation}
and the following observables consisting of products of 3 field-strengths
\begin{eqnarray}
  \label{obs3}
{\bf V}_p &=&  \epsilon^{abc} ({\bf E}^{a}\cdot {\bf B}^{b}) \; {\bf B}^{c}\,,\;{\bf V}_a =  \epsilon^{abc} ({\bf E}^{a}\cdot {\bf B}^{b}) \; 
{\bf E}^{c},\,\nonumber \\ T_3 &=& \epsilon^{abc}\large[({\bf B}^a \times {\bf B}^b)\otimes {\bf E}^c + {\bf E}^c\otimes ({\bf B}^a \times 
{\bf B}^b)\large]\, .
\end{eqnarray}
To improve the statistics we have summed the correlation functions for vectors and tensors over the spatial components, e.g.
$$\langle ({\bf X}\otimes {\bf Y})(x) ({\bf X}\otimes {\bf Y})(y) \rangle = \sum_{j\ge i=1}^{3} \langle  ({\bf X}\otimes {\bf Y})_{ij}(x)({\bf X}\otimes 
{\bf Y})_{ij}(y)\rangle $$
and have used the equality of the components as a test for rotational invariance of the ensembles.

Our subsequent calculations  all refer to instanton ensembles, for which the fiducial interval for extracting  masses is larger. In cases
 where masses could be extracted from both kinds of  ensembles, the results agreed within the uncertainty of the procedure. We 
observe that the instanton fields [Eqs.~(\ref{fssm}, \ref{dfssm})] satisfy  
$$B^a_i = E^a_i= -4\frac{1}{(x^2+1)^2}\delta_{ai}\, .$$     
As a consequence, gauge invariant combinations consisting of products of two or three field-strength operators have to be either scalar 
or pseudoscalar. For merons this is only true after momentum projection. If the momentum projection is not complete, the presence of
single meron contributions will complicate the analysis of the correlation functions.

In addition to  the action density and the three tensors [cf. Eq.~(\ref{obs2})], we can form a second scalar, ${\bf B}^2-{\bf E}^2$, the 
Hamiltonian density transformed to Euclidean space, and the Poynting vector, ${\bf E}\times {\bf B}$.  Although if integrated over 
all space, the vacuum is an eigenstate of the resulting operators, our ensembles of pseudoparticles do not yield exact eigenstates 
of the Hamiltonian and projection to zero momentum is not exact. Hence, we obtain non-vanishing correlation functions.  However, 
we find that the fluctuations in the Hamiltonian, i.e., the value of the correlation function for vanishing $|t|$, are  one to two orders of
 magnitude smaller than those of the action,  depending on the ensemble.  In the case of the integrated Poynting vector, the momentum 
operator, we find very strong dependence of the correlation functions on the parameters of the ensembles,  which highlights the 
unphysical origin of the correlations. For these reasons we do not consider these correlations further.  Finally we remark that the 
``singular'' behavior (cf. Fig.~\ref {cpr116}) of the pseudoscalar (${\bf E}\cdot{\bf B}$) correlation function $C_{\tilde {s}}$  prevents a 
meaningful extraction of a mass  after momentum projection.

{\bf Behavior under interchange of Euclidean space and time axes}

In order to understand some of the results obtained subsequently, it is useful to note
that by relativistic covariance, electric and magnetic  fields can be transformed into each other. For vanishing separation, the 
distinction between electric and magnetic field depends on the choice of the coordinate system. We illustrate the equivalence 
by considering the following contribution to the correlation function of $T_a$ 
$$ \langle t_{ij}(x_4)\rangle =\langle B_i E_j({\bf 0},x_4 )\,B_i E_j(0)\rangle =\langle \tilde{F}_{4i}F_{4j}({\bf 0},x_4 )\,\tilde{F}_{4i}F_{4j}(0)\rangle\, .$$
Here color indices are  suppressed and no summation over $i$ and $j$ is carried out. Under  a transformation which rotates the $x_4$ 
and $x_i$ coordinates into each other
\begin{equation}  
R:\; x_i \rightarrow x_i^{\prime}= -x_4\,,\quad x_4 \rightarrow x_4^{\prime}=  x_i\, 
\label{rot1}
\end{equation}
we find, assuming 4-dimensional rotational invariance in the ensembles 
\begin{equation} R:\; \langle t_{ij}(x_4)\rangle=\langle \tilde{F}_{4i}F_{ij}(x^{\prime} )\,\tilde{F}_{4i}F_{ij}(0)\rangle\,,\quad x^{\prime}_k =
\delta_{ki}x_4\, , \quad i\neq j\, .\label{rot3}
\end{equation}
For vanishing separation ($x_4=0$), this result, if expressed by electric and magnetic fields 
\begin{equation} 
\langle B_i E_j(0)\,B_i E_j(0)\rangle = \sum _{k=1}^3 \epsilon^{ijk}\langle B_i B_k(0)\,B_i B_k(0)\rangle \,, \quad i\neq j\, .
\label{rot2}
\end{equation}
together with the definition (\ref{obs2}) of the relevant tensors  is seen to yield the identity 
\begin{equation}  
C_{T_+}(0)= C_{T_a}(0)\label{rot0}
\end{equation}
for the off-diagonal elements.  The diagonal elements of the traceless tensor can be connected to the off diagonal ones by rotations 
not involving the 4-coordinates. We  note that the rotation (\ref{rot1}) has no counterpart in Minkowski space.  
This method is not directly applicable for non-vanishing separations since  electric and magnetic fields are defined with respect to  
the direction of separation chosen as time axis 
$$E_{\nu} = \hat{z}_{\mu}F_{\mu\nu}\,,\quad B_{\nu} = \hat{z}_{\mu}\tilde{F}_{\mu\nu}\, .$$
Here $z$ denotes the 4-vector defining the separation and $\hat{z}$ the corresponding unit vector.
Only with this definition  electric and magnetic field do not  mix under 3-dimensional rotations in the space transverse to the direction 
of separation, i.e., under rotations satisfying
$$ R_{\mu\nu}z_{\nu}= z_{\mu}\, .$$
In this way the resulting operators  project  on states with definite angular momentum and parity. This redefinition invalidates  the 
above derivation for finite separations. More complicated relations between the two correlation functions can be derived which 
however defy a direct numerical application.  As we discuss below,  it is striking that our numerical results for the two  momentum 
projected correlation functions  
$C_{T_+}(x)$ and $C_{T_a}(x)$ exhibit essentially identical normalizations and a strong similarity in shape. Significant deviations occur
 only at the level of about 0.01 of the value at zero separation. It would be interesting to investigate these correlation functions in lattice 
calculations. In the derivation of the identity (\ref{rot0}), we made use of the transformation properties of the field-strength under 
rotations by 90 degrees, and one may therefore expect that such relations remain valid on the lattice with the discrete symmetry of a 
hypercube. 

It will also be important subsequently to note the close relationship of  $T_-$ to the Poynting vector. Using the same arguments as 
above, it follows that the off-diagonal matrix elements $T_-$ are related to the Poynting vector  by the rotation (\ref{rot1})
$$ {\bf P} = {\bf E}^a \times {\bf B}^a\, ,$$
with the consequence that
\begin{equation}
  \label{t-p}
C_{T_-}(0)= \langle T_{-ij}(0)T_{-ij}(0) \rangle = \sum_{k,l=1}^{3}\epsilon^{ijk}\epsilon^{ijl}\langle  P_k(0) P_l(0)\rangle \, \quad i \neq  j\,, \, i,j 
\ \mbox{fixed}      .        \end{equation}

\noindent{\bf Wilson loop operators}

In addition to the local operators discussed above, it is also interesting to consider correlation functions of Wilson loops.  This will 
enable us to investigate the role of the locality of the operator in the determination of the glueball masses and to  make contact  
with the operators used in lattice gauge calculations. 

We denote the direction in which the Wilson loops are separated as time direction and will perform 
 weighted spatial averages of the Wilson loops in order to project on states of definite spin. We  consider circular loops of radius $r $
and denote the unit vector normal to the plane of the circle by ${\bf n}$ and the coordinate of the center of the circle by $ {\bf x}_{0} $,
\begin{equation} 
W_{r} ({\bf n}, {\bf x}_{0}, t) = \frac{1}{2} \; \mbox{Tr P} \; {\rm e}^{{\rm i} g \oint A_{\mu} (x) {\rm d}x^{\mu}}  \,.                 
 \label{wlc1}
\end{equation} 
If applied to the vacuum, this operator generates  states of different momenta and different spins. 
Projection of $W({\bf n},{\bf x}_0, t) $ on definite spin $\ell, m$ is obtained by
\begin{equation}
W^{\ell m}_{r} ({\bf x}_{0}, t) = \int {\rm d} \Omega_{{\bf n}}  W_{r} ({\bf n}, {\bf x}_{0}, t)\, 
Y^{*}_{\ell m} ({\bf n})                                 \label{wlc2}
\end{equation}
and on zero momentum by spatial averaging 
\begin{equation}
W^{\ell m}_{r} (t) = \frac{1}{V} \int {\rm d}^{3} x_{0} W^{\ell m}_{r} ({\bf x}_{0},t) \,.  \label{wlc3}
\end{equation}
Translational invariance in Euclidean 4-space and 3-dimensional rotational invariance impose
constraints on the correlation functions. We have 
\begin{eqnarray}
& & \langle (W^{\ell m}_{r} (t) - \langle W^{\ell m}_{r} (t))(W^{* \ell^{\prime}m^{\prime}}_{r}
(t^{\prime}) - \langle W^{* \ell^{\prime}m^{\prime}}(t^{\prime}) ) \rangle \nonumber\\
& & = \delta_{\ell \ell^{\prime}} \delta _{m m^{\prime}}
\left[ \langle W^{\ell}_{r} (t - t^{\prime}) W^{\ell}_{r} (0)\rangle - \langle W^{\ell}_{r} \rangle ^{2} \delta _{\ell, 0} \right] 
= C_{\ell} (t - t^{\prime}) \; \delta_{\ell \ell^{\prime} } \delta _{m m^{\prime}} \,.   \label{wlc4}
\end{eqnarray}
We have used the fact that by rotational and translational invariance the correlator 
\begin{equation}
\langle \int {\rm d}^{3} x_{0}\, W_{r} ({\bf n}, {\bf x}_{0}, t) \int {\rm d}^{3}  y_{0}\, W^{*}_{r} 
({\bf n}^{\prime}, {\bf y}_{0}, t^{\prime}) \rangle = W_{r} ({\bf n}\cdot {\bf n}^{\prime}, t - t^{\prime}) \label{wlc5}
\end{equation}
can  depend only on the angle between the two orientations of the Wilson loops.
Similarly
\begin{equation}
\langle (W^{\ell m}_{r} ({\bf x}_{0}, t) - \langle W^{\ell m}_{r} ({\bf x}_{0}, t) \rangle ) 
(W^{* \ell^{\prime}m^{\prime}}_{r} ({\bf x}_{0}, t^{\prime}) - \langle W^{* \ell^{\prime} m^{\prime}} 
({\bf x}_{0}, t)) \rangle 
= D_{\ell} (t - t^{\prime}) \delta_{\ell \ell^{\prime}} \delta _{m m^{\prime}} \,.   \label{wlc6}
\end{equation}
We finally observe that [cf. Eq.~(\ref{wlc1})]
\begin{equation}
W_{r} ({\bf n}, {\bf x}_{0}, t) = W_{r} (- {\bf n}, {\bf x}_{0}, t),      \label{wlc7} 
\end{equation}
and therefore the correlation functions (\ref{wlc6}, \ref{wlc7})  vanish for odd $\ell $. 
\vskip 0.2cm
In order to improve the statistics, we actually do not make use of the $m $-independence of the 
correlation functions but rather sum over all values of $m $ in Eqs.~(\ref{wlc4}) and
(\ref{wlc5}). 
In order to relate the  Wilson loop correlation functions to the field-strength correlation functions, we consider the limit of small
size Wilson loops, i.e., $r \ll \rho$. In this limit and for fixed orientation ${\bf n}$
we have
\begin{equation}
  \label{wlbb}
  r \rightarrow 0 :\quad  \;W_{r} ({\bf n}, {\bf x}_{0}, t) = g^2 \pi^2 r ^4  ({\bf n}\cdot{\bf B}^a({\bf x}_{0}, t))^2\, . 
\end{equation}
Integration over the orientation [Eq.~(\ref{wlc2})] yields 
\begin{equation}
  \label{wlbb0}
  r \rightarrow 0 :\quad  W^{0,0}_{r} ({\bf x}_{0}, t) = \frac{\sqrt{4 \pi}}{3}g^2 \pi^2 r ^4  ({\bf B}^a({\bf x}_{0}, t))^2\, ,\quad  
\end{equation}
whereas  the $\ell=2$ projection gives, for instance, for the $m=0$ component  
\begin{equation}
  \label{wlbb2}
  r \rightarrow 0 : \quad W^{2,0}_{r} ({\bf x}_{0}, t) = \frac{\pi}{10}g^2 \pi^2 r ^4 [(B_3^a({\bf x}_{0}, t))^2-\frac{1}{3}({\bf B}^a({\bf x}_{0}, t)^2)] \, .  
\end{equation}
As expected, at small sizes, the angular momentum projected operators reduce for $\ell=0$ to $({\bf B}^a)^2$  and for $\ell=2$ to  $ {\bf B}
\otimes {\bf B}$,
\begin{equation}
  \label{wlbb2p}
({\bf B}\otimes {\bf B})_{ij}=B_i^a B_j^a- \frac{1}{3} {\bf B}^a {\bf B}^a \delta_{ij}= \frac{1}{2}(T_+ - T_-)_{ij}   \, .  
\end{equation}
Electric field operators do not appear since the Wilson loops considered are located in spatial planes $({\bf e}_4 \cdot {\bf A^a}=0)$.

For subsequent reference, all the operators used to calculate correlation functions, salient properties, and the results of 
measurements are summarized in Table~\ref{cor-ops}.
\begin{table}[h]
\begin{center} 
\begin{tabular}{|c|c|c|c|c|c|}  \hline     
state &name& operator & $x_i  \leftrightarrow x_4 $ &  figures (slope) \\  \hline \hline
  $0^+$ &S & ${1/4}({\bf E}^2 + {\bf B}^2)$    &    & \ref{accor}  ($-4.0$) \\ \hline
 $0^+$ & $W^{00}_r$ & $\int {\rm d} \Omega_{{\bf n}}  W_{r} ({\bf n}, {\bf x}_{0}, t)\, 
Y^{*}_{0 0} ({\bf n}) $    &    &  \ref{wlbbf}  (-) \\ \hline 
 $1^-$ & ${\bf V}_p$ &  $  \epsilon^{abc} ({\bf E}^{a}\cdot {\bf B}^{b})\; {\bf B}^{c} $   &    &  \ref{Vpa}  ($-9\; ....-10$) \\ \hline 
 $1^+$ & ${\bf V}_a$ & $\epsilon^{abc} ({\bf E}^{a}\cdot {\bf B}^{b})\; {\bf E}^{c}$    &    &   \ref{Vpa}   ($-10.5\;....-11$)\\ \hline 
 $2^+$ & $W^{2m}_r$ & $\int {\rm d} \Omega_{{\bf n}}  W_{r} ({\bf n}, {\bf x}_{0}, t)\, 
Y^{*}_{2 m} ({\bf n}) $    &    &  \ref{wlbbf}  ($-7.5\;....-7.8$) \\ \hline 
% $2^+$ & $ T_-$ &$  {\bf E} \otimes {\bf E} -  {\bf B} \otimes{\bf B}$     &  $ {\bf p} = {\bf E}^a \times {\bf B}^a $  &   23(4.2*), 25 (4.5*), 27(4.5*) \\ \hline 
 $2^+$ & $ T_-$ &$  {\bf E} \otimes {\bf E} -  {\bf B} \otimes{\bf B}$     &  $ {\bf P} = {\bf E}^a \times {\bf B}^a $  &  \ref{scal}  (-), \ref{cor567p}   (-), \ref{T-}  (-) \\ \hline 
 $2^+$ &   $ T_+$  & ${\bf E} \otimes {\bf E} +  {\bf B} \otimes{\bf B} $  &  $ T_a$ & \ref{T+}  ($-8.0$), \ref{cor567p}  (-) \\ \hline 
 $2^+$ & $ T_a$ & $  {\bf E} \otimes {\bf B} +  {\bf B} \otimes{\bf E}$   &    $ T_+$  & \ref{scal}  (-), \ref{Ta}   ($- 8.3$), \ref{cor567p}   (-)\\ \hline 
 $2^-$ & $ T_3$ & $ \epsilon^{abc}\large[({\bf B}^a \times {\bf B}^b)\otimes {\bf E}^c + {\bf E}^c\otimes ({\bf B}^a \times {\bf B}^b)\large]  $ &    & 
\ref{T3}   ($-9.3$) \\ \hline 
 \end{tabular}
\caption{Summary of operators used to calculate glueball masses and the results of measurements.  The slope, specifying the glueball
 mass, for each figure is shown in parentheses.}
\label{cor-ops}
\end{center}
\end{table}
%%%%%%%%%%%%%%%%%%%%%%%%%%%%%%%%%%%%%%%%%
\subsubsection{Correlation Functions of Field-Strength Operators}
%%%%%%%%%%%%%%%%%%%%%%%%%%%%%%%%%%%%%%%%%
With the preceding definitions and discussion, we now present  the results of our calculations of correlation functions in instanton 
ensembles. 

\begin{figure}[h]
%\epsfig{file=ee+bbfit2.ps, width=.6\linewidth,angle=-90}
%\begin{center} 
\vskip.1cm\hspace{3cm}\epsfig{file=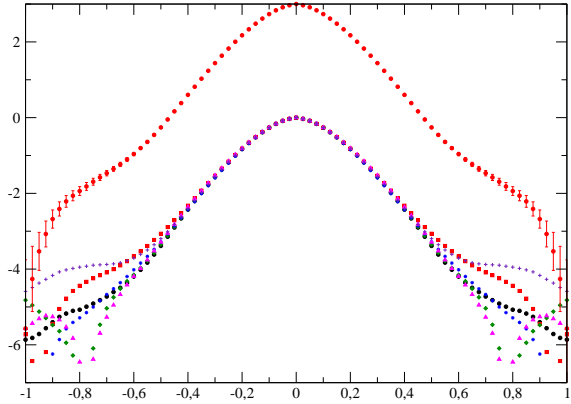, width=.45\linewidth,angle=-90}
\vskip -.4cm
\caption{Summed correlation function of $T_+$ [Eq.~(\ref{obs2})] and its decomposition into its 6 components (denoted by different 
symbols) $T_{+\,ij}, \;i \le j$  in an ensemble with 50 instantons of variable size.}
\label{T+}
\end{figure}

\noindent{\boldmath $T_+$} \\
We begin with the  correlation function of the $2^+$ operator $T_+$ [cf. Eq.~(\ref{obs2})] in Fig.~\ref{T+}.
 For small separations, the shape of the correlation function (curvature) still reflects the size of a single instanton, even though the  
single instanton correlation functions vanishes. Now the instanton size appears at small separations via the interference of  some 
``mean-field'' with the gauge field of a single instanton. In the intermediate regime $.2\le |t| \le .6 $ approximately, a common slope can 
be associated with the correlation functions in these various ensembles. In this particular case of the $T_+$ correlation function, we 
extract a slope of 8.   For larger separations we observe strong fluctuations in the correlation functions. As is shown in the lower 
part of the  figure, these fluctuations are connected to violations of the rotational symmetry.    
After reducing the diagonal components of the tensor by 1/3, all 6 components of the tensor are identical if the ensemble exhibits 
exact rotational symmetry. Comparison of the two figures strongly suggest that the differences in the correlation functions for the 
different ensembles for $|t| >  .5 - .6 $ are a consequence of the violations of the rotational symmetry in the individual ensembles.

 As a further consistency check  we also have compared correlation functions of ensembles which differ in the number of pseudoparticles.
\begin{figure}[h]
%\hspace{-1.1cm}\epsfig{file=insac04b.ps, width=.45\linewidth,angle=-90}\hspace{-.7cm}
\hspace{3cm}\epsfig{file=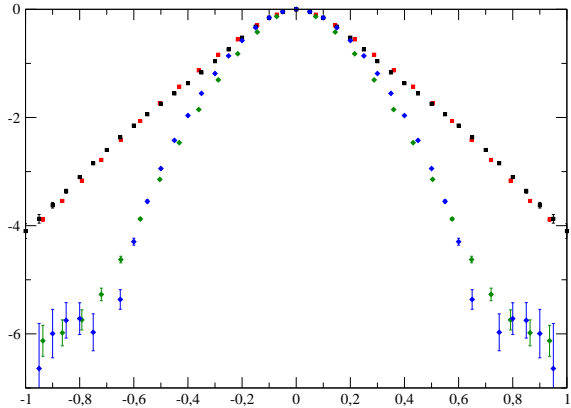, width=.45\linewidth,angle=-90}
%\vspace{-.5cm}
\caption{Demonstration of scaling of the $T_-$ (red and black squares, upper curves) and $T_a$ (green and blue diamonds, lower curves)
correlation function for ensembles with 50 and 200 instantons of size $\rho=0.16$.}
\label{scal}
\end{figure}
In  Fig.~\ref{scal} we have rescaled the values of $t$ for the correlation function of the ensemble with 50 instantons  by a factor 1.44  as
suggested by the analysis of the Wilson loops (cf. Table \ref{inst1}). As is seen, the extracted slopes or masses scale as expected with 
$\sqrt{\sigma}$ within 10\% or better. 

%%%%%%%%%%%%%%%%%%%%%%%%%%%%%%%%%%%%%%%%%
\noindent{\bf S}\\
The intermediate states contributing to the correlation functions of the action density are $0^+$ states. 
\begin{figure}[h]
\hspace{3cm}\epsfig{file=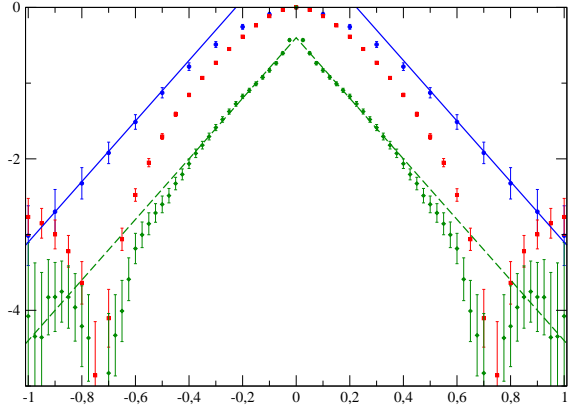, width=.45\linewidth,angle=-90}\vspace{-.3cm}
\caption{Action density correlation function $C_s(x)$ of ensembles with 50 instantons of size $\rho=0.16$ (red squares), $\rho=0.04$ 
(green diamonds), and 800 instantons ($L=4$) of size $\rho=0.16$ (blue circles). The slope of the straight lines is $-4.0$.}
\label{accor}
\end{figure}  
Difficulties arise in the numerical evaluation  of this correlation function due to the large contribution from the intermediate 
ground state. The  subtraction  of the vacuum expectation value [cf. Eq.~(\ref{C_s})] makes the correlation function  $C_s$ 
particularly sensitive to the violations in translational and rotational symmetry.  A stable and unambiguous result could be 
obtained only by constructing ensembles in a larger volume in which the violation of translational invariance is significantly 
weaker. We have used the ensemble of pseudoparticles discussed above in the context of the thermodynamic limit and calculated 
the action density correlation function in a 16 times larger  volume with 800  pseudoparticles. The result of this calculation is 
shown in Fig.~\ref{accor}.
\begin{figure}[h]
%\hspace{-1.1cm}\epsfig{file=insac04b.ps, width=.45\linewidth,angle=-90}\hspace{-.7cm}
\hspace{3cm}\epsfig{file=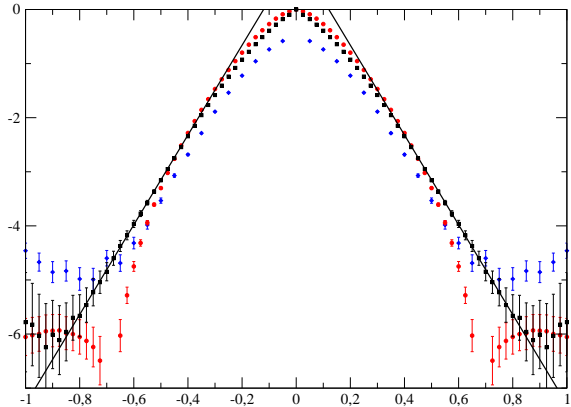, width=.45\linewidth,angle=-90}\vspace{-.3cm}
\caption{Correlation function of $T_a$ for different ensembles. 800 instantons ($L=4$) of size $\rho=0.04$ (blue diamonds), 50 instantons of size 
$\rho=0.04$ (black squares) and of variable size (red circles). The value of the slope is $-8.3$.}
\label{Ta}
\end{figure}   
\vskip -.1cm   
Unlike for the other correlation functions, very different results are obtained for the action density correlation in the two
``equivalent'' ensembles. We note that the correlation function in the small volume displays a zero at $|t|=0.75$ independent 
of the instanton size.  In the larger volume, these fluctuations violating translational invariance are reduced resulting in a 
reasonable behavior of the correlation function. As illustrated in Fig.~\ref{accor} for sufficiently small instanton sizes, where 
the single instanton contribution is restricted to small separations, a fiducial interval exists where a meaningful value of the 
mass can be extracted also in the  ensemble 
 with the smaller volume, and we indeed obtain the same slope.

 \noindent{\boldmath $T_+, T_-, T_a$}\\
The correlation functions of the 3 tensor operators %\vskip.1cm
\begin{figure}[h]
\hspace{2.5cm}  \epsfig{file=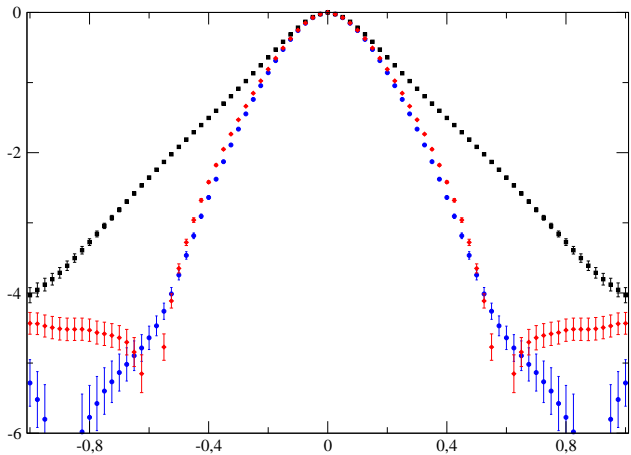, width=.5\linewidth,angle=-90}  
\caption{Tensor correlation functions $T_+$ (blue circles), $T_-$ (black squares), and $T_a$ (red diamonds) of an ensemble with 
50 instantons of size $\rho=0.1$.}
\label{cor567p}
\end{figure} 
 bilinear in the field-strength [cf. Eq.~(\ref{obs2})] with the $2^{+}$ ($T_{\pm}$) and $2^{-}$ ($T_a$) excitations appearing as intermediate
states are 
shown in  Fig.~\ref{cor567p}. Surprisingly, not the correlation functions of  operators with the same  but rather those with opposite 
parity are very similar in shape and actually possess the same normalization. Our numerical results yield $C_{T_+}(0)= C_{T_a}(0)$ 
within the statistical accuracy and furthermore, this  equality  persists at finite separation within a few percent for the corresponding 
values of the momentum projected (\ref{chat}) correlation functions. As discussed in the previous section, the equality at zero separation 
follows from Eq.~(\ref{rot0}), arising from the equivalence of the two operators under interchange of $x_t$ and $x_i$. 

 We now turn to the extraction of glueball masses from these correlation functions and discuss first the case  of the negative parity 
$2^{-}$ state excited by the  tensor $T_a$.  As Fig.~\ref{Ta} shows, a sufficiently large fiducial volume exists for extracting  masses
 for the two ensembles  and the two values agree  within 5\% with an average of 8.3.  We will come back to this state when discussing 
products of three field-strength operators.
 
Similarly, the analysis of the correlation function of $T_+$ already discussed above (cf. Fig.~\ref{T+}) yields for the same ensembles
  a common slope of about 8. The discussion of the Wilson loop correlation function will provide additional information on the 
corresponding $2^+$ state.
 
Also in the $T_{-}$ correlation function, $2^+$ states appear as intermediate states. As the left side  %\vskip-.8cm
\begin{figure}[h]
\hspace{-.8cm}\epsfig{file=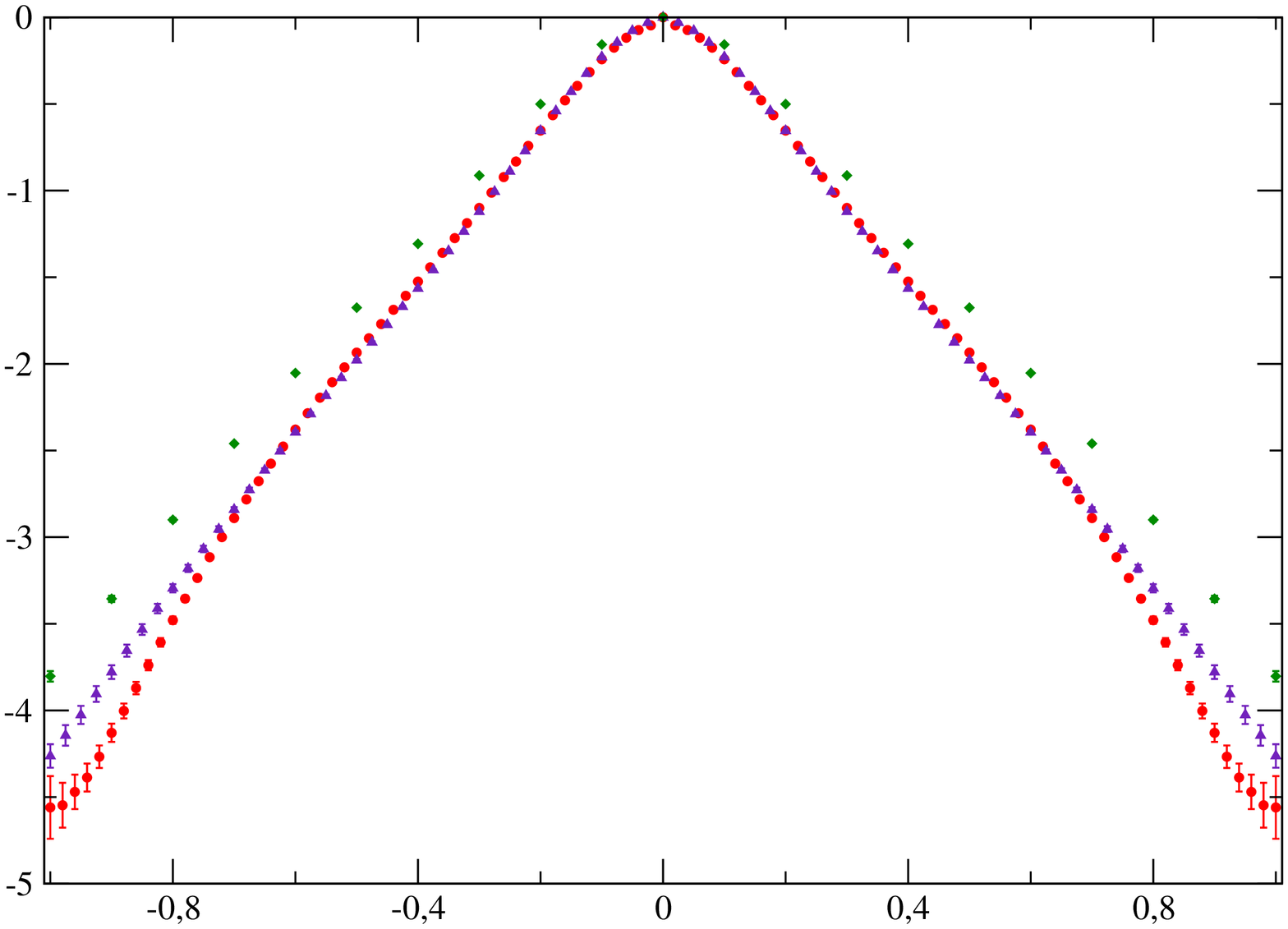, width=.45\linewidth,angle=-90}
\hspace{-1cm}\epsfig{file=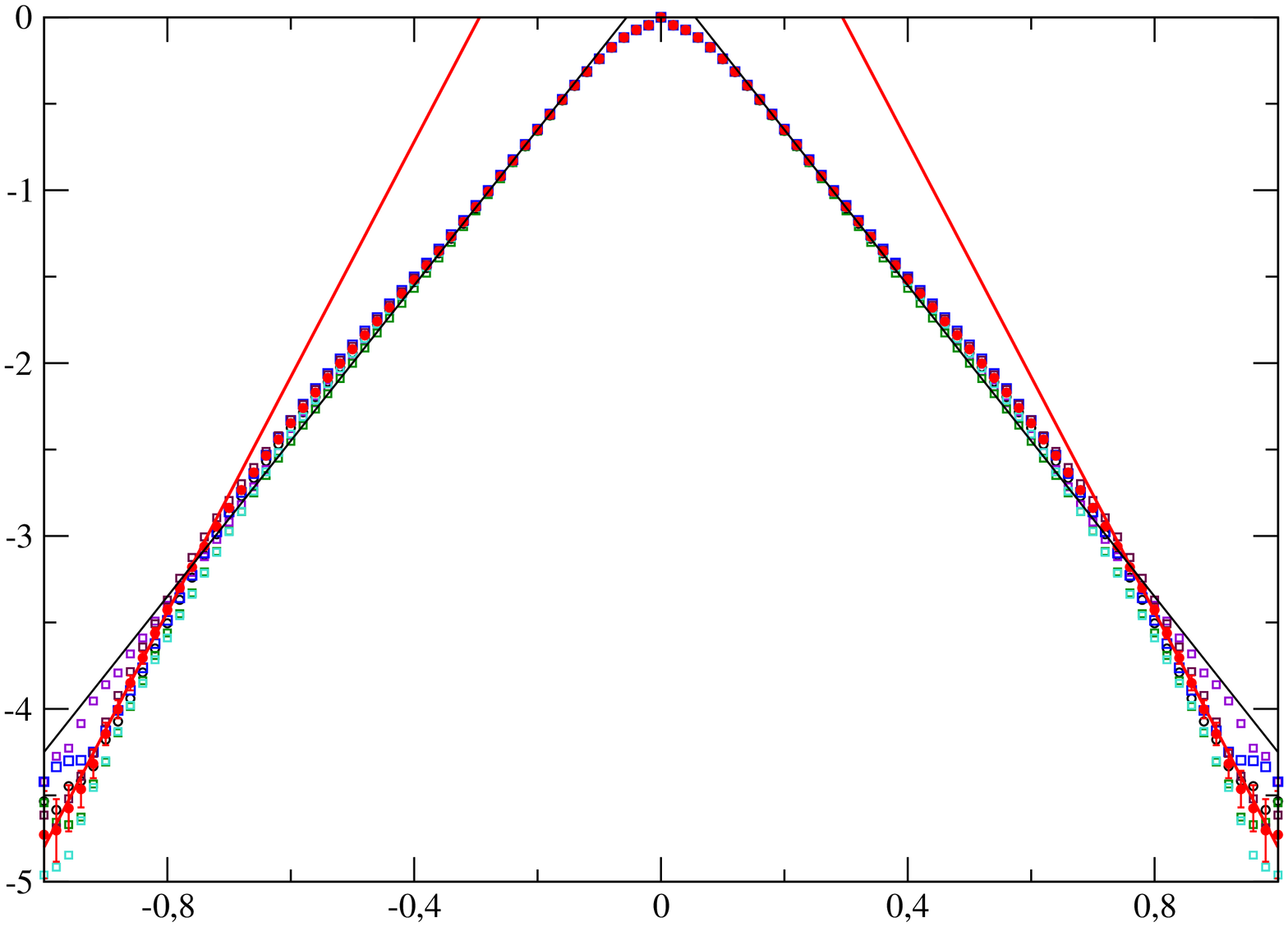, width=.45\linewidth,angle=-90}\vspace{-.3cm}
\caption{Left: Correlation function of $T_-$ for different ensembles ($N_I, \rho$) = (50, 0.16) - green, (50, 0.08) - red, (50, variable $\rho$) - indigo.
Right:  Correlation function of $T_-$
(red) and components for the ensemble (50, 0.08). The values of the slopes are 4.5 and 6.8.}
\label{T-}
\end{figure}
%\vskip.1cm
of Fig.~\ref{T-} shows,  this correlation function is rather exceptional,  in that the fiducial intervals for the different ensembles extend to 
distances of up to 1. 
A fit to these correlation functions for $|t|\leq 0.6$, the fiducial interval also used for the other correlation functions,  yields an average 
value of 4.1 for the slope. At larger values of $|t|$, the slopes of the correlation functions for the different ensembles start to deviate.  
As is seen from the right side of this figure, the correlation function for the (50, 0.08) ensemble displays an  increase in  slope by a factor 1.5. 
Furthermore, as the decomposition in different  components indicates, this change of slope is not accompanied by a significant violation 
of rotational invariance.  
A change to a larger slope  also occurs though more moderately  around $t=0.6$ in the (50, 0.16)  (from 3.7 to 4.9) and in the (50, variable
 $\rho$)  (from 4.1 to 5.2) ensembles. For determination of a glueball mass, we  have to use  the larger values of the slope as more realistic. 
The change to a larger slope with increasing separation suggests   that for this particular operator, the regime in which the time evolution
 is governed by a transfer matrix has not been reached in the region $|t| \leq 0.6$ where the smaller slopes are extracted.  As for the 
$T_+$ - $T_a$ relation, our numerical results for the  momentum projected correlation functions of  $ {\bf P}$ and $T_-$ agree for  $|t| 
\leq .4$ within $6\%$, indicating that, like the correlation function of the integrated Poynting vector, the correlation function of the 
integrated $T_-$ operator has no interpretation in terms of glueballs in this regime.
 Despite their rather complicated structure, the correlation functions of $T_-$ exhibit the expected scaling behavior as shown in 
Fig.~\ref{scal}. However the strong dependence of the larger slope on the instanton size (4.9 - 6.8) prevents extraction of a glueball 
mass from this correlation function. Given the comparatively small fiducial interval $(0.7\leq |t| \leq 1)$, we cannot rule out that several 
states contribute with weights depending on the instanton size. In this case we can only deduce an upper  limit for the mass
\begin{equation}
m_{2^+}\leq 4.9\, .
\label{ulm2}
\end{equation}

\noindent{\boldmath $T_3$}\\
We turn now to a discussion of correlation functions associated with observables composed of products of 3 field-strength operators.
 We will study excitations of $1^{\pm}$ states and reexamine the excitation of $2^{\pm}$ states. Since the field-strengths are evaluated 
at the same point, very rapid variations in the vicinity of the center of the pseudoparticles result. 
\vskip -.1cm
\begin{figure}[h]
\hspace{3cm}\epsfig{file=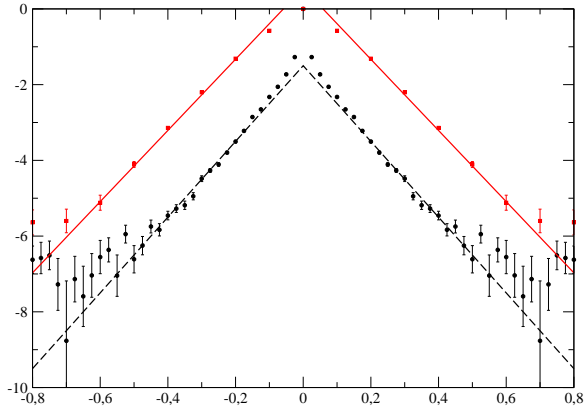, width=.45\linewidth,angle=-90}
\caption{$T_3$ correlation function for the ensemble of 50 instantons of variable size (black circles) and 800 instantons ($L=4$) of size 
$\rho=0.16$ (red squares), the value of the slopes is $-9.3$.}
\label{T3}
\end{figure}
Due to these rapid variations  significant results could not be obtained for the $0^{\pm}$ excitations which  contain single instanton 
contributions. In the case of the $1^{\pm}$ and $2^{\pm}$ excitations, where single pseudoparticles do not contribute, the variations 
due to the interference with the ``mean field'' are still too large for generating reliable results for the smallest instanton size ($\rho=0.04$) 
considered. Consequently the fiducial intervals are in general significantly smaller than in the case of products of two field-strength 
components.
The correlation function of the $2^-$  operator $T_3$ shown in Fig.~\ref{T3} displays  at small $|t|$   rapid variations  and a strong 
dependence on the individual ensembles. In particular, the ensemble with variable $\rho$ exhibits for small $|t|$ an almost singular 
behavior. Beyond this region, a sufficiently large fiducial interval exists where a common slope can be extracted for the two ensembles. 
The value of 9.3 differs by 10\% from the value 8.3 in Fig.~\ref{Ta}. The average of these two values will be used subsequently.
\begin{figure}[h]
\hspace{-1cm}\epsfig{file=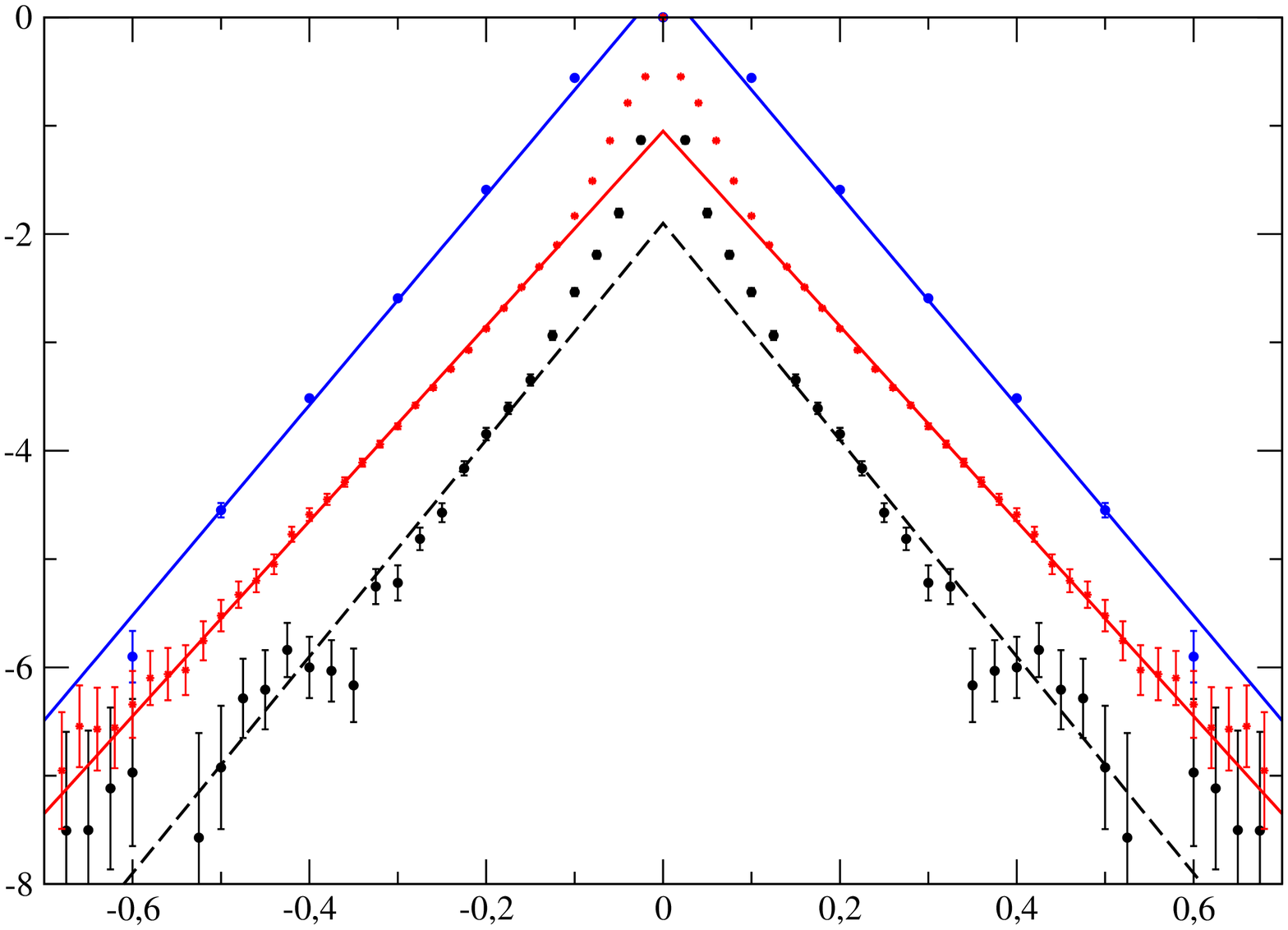, width=.45\linewidth,angle=-90}\hspace{-.9cm}\epsfig{file=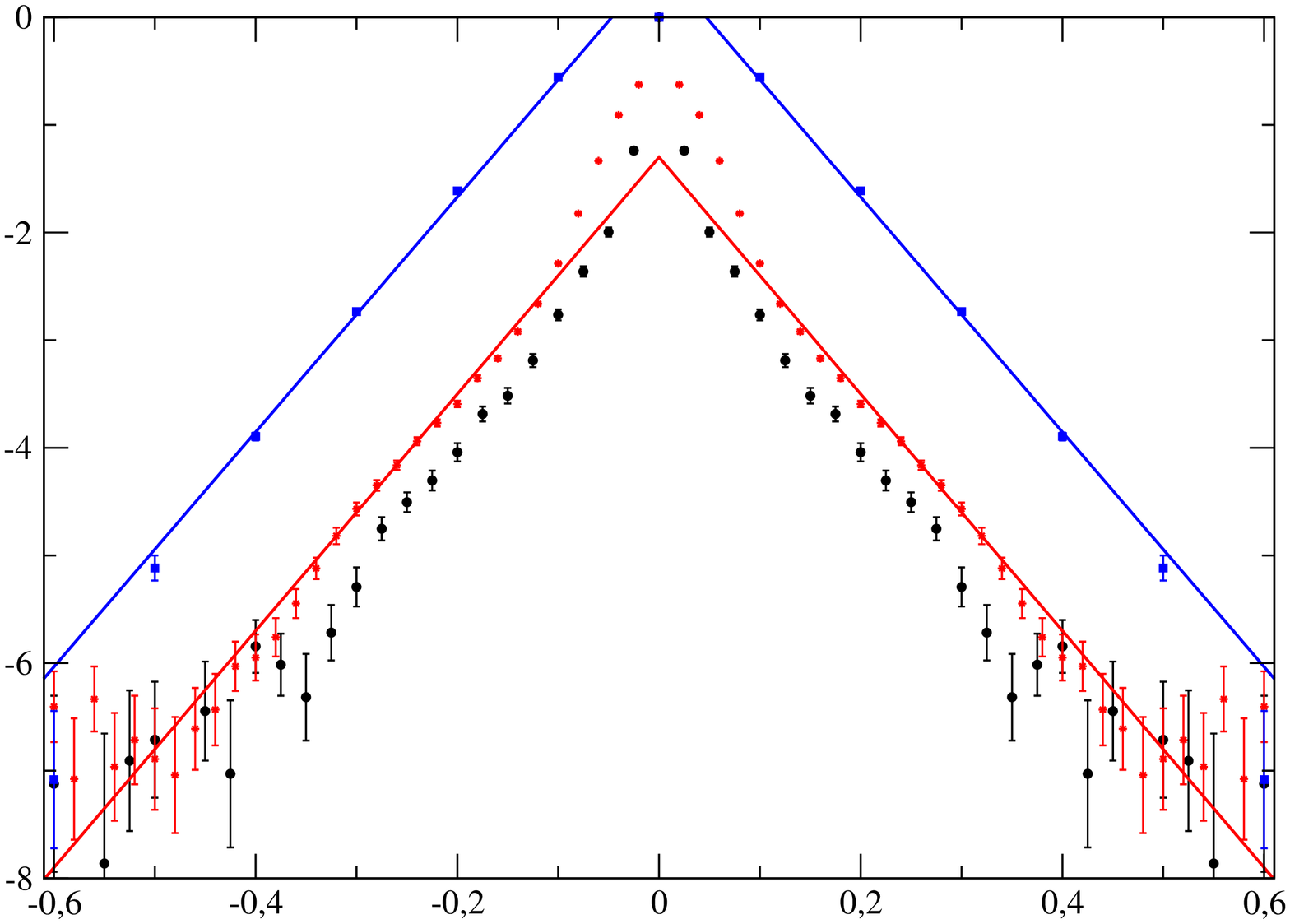, width=.45\linewidth,angle=-90} 
\caption{Axial  $V_a$ (left) and polar vector $V_p$ (right) correlation functions for ensembles of 50 instantons of variables size 
(black circles ), of size $\rho=0.16$ (blues squares) and $\rho=0.08$  (red stars). The values of the slopes are $-9.0, -9.7, -10.0$  ($V_a$) 
and $-10.5, -11.0$  ($V_p$).}
\label{Vpa}
\end{figure}  
A meaningful extraction of a slope parameter of  the correlation functions for  the   $2^+$ operator $({\bf E}\times{\bf E})\otimes{\bf B}$
was not possible and is not discussed here.

\noindent{\boldmath $V_a, V_p$}\\
We finally turn to a discussion of the spin 1 correlation function in Fig.~\ref{Vpa}. For both, the $V_a$ and $V_p$ [cf. Eq.~(\ref{obs3})] 
correlation functions, the three ensembles exhibit, beyond the small $|t|$ region, sufficiently large fiducial intervals for extracting
the slopes reliably. The average values of these slopes are used later. 
\begin{figure}
\hspace{-1cm}\epsfig{file=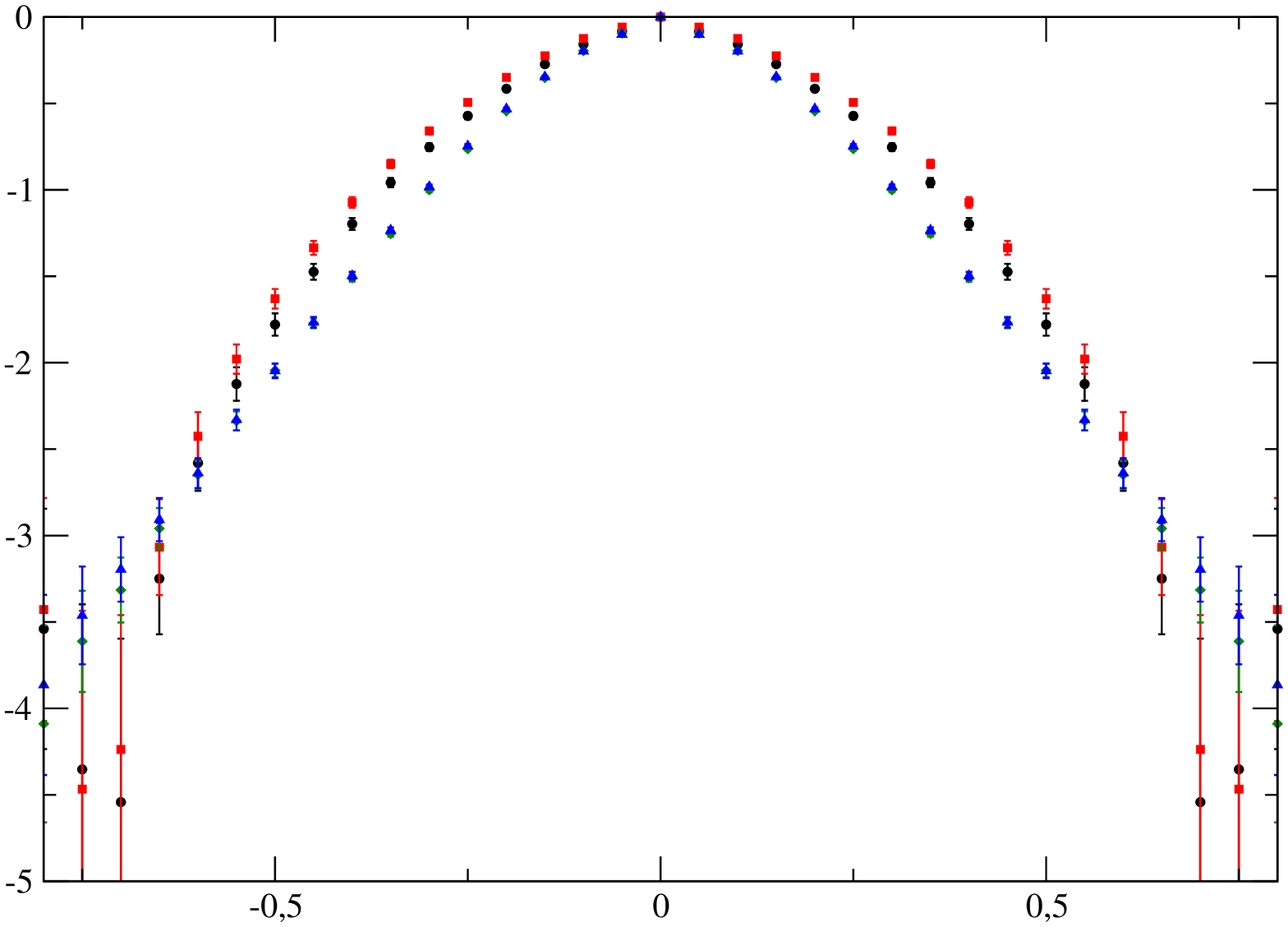, width=.45\linewidth,angle=-90}\hspace{-.5cm}\epsfig{file=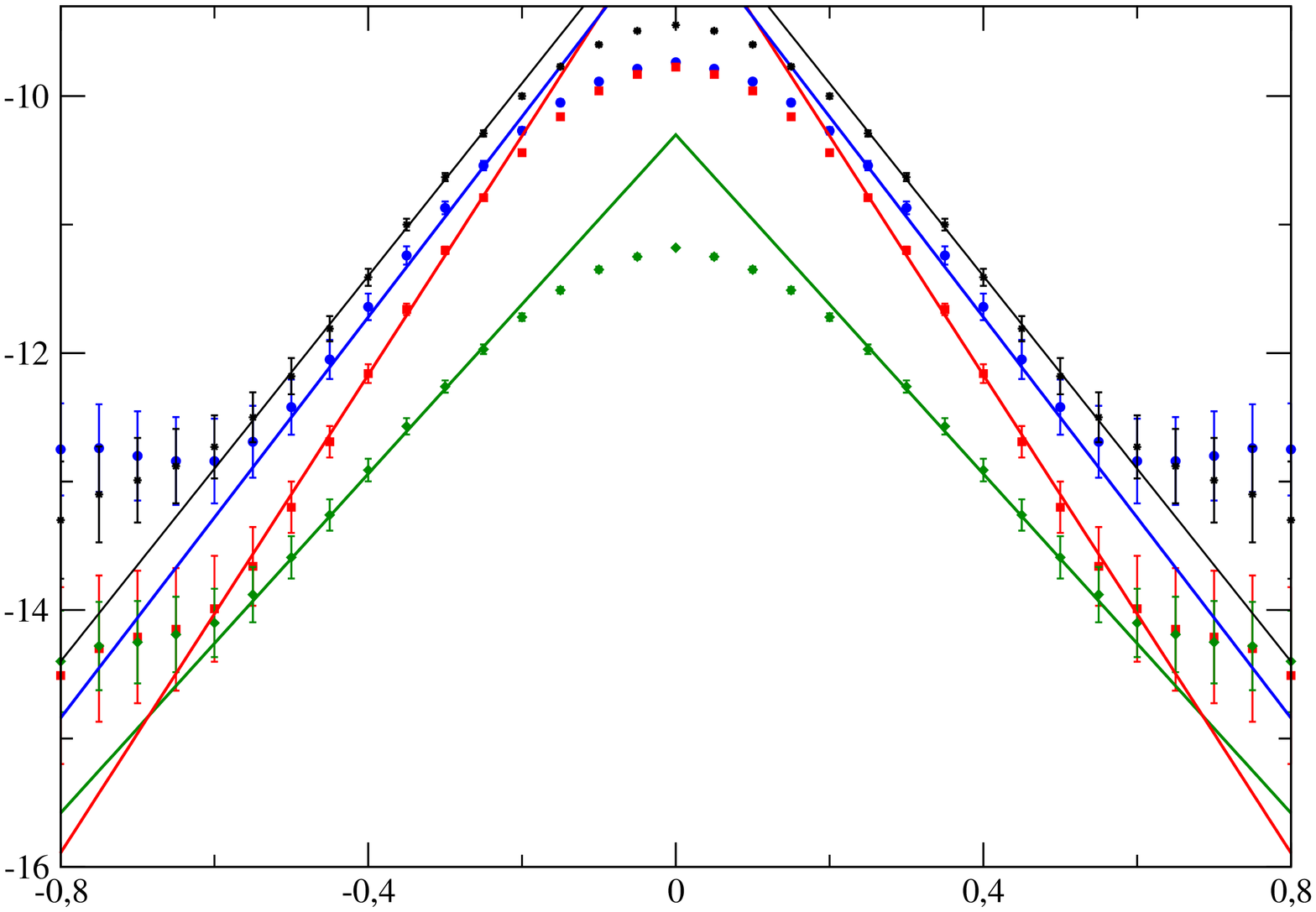, width=.45
\linewidth,angle=-90}    
\caption{Left: Wilson loop correlation functions $ D_{\ell}(t)$ (\ref{wlc4}) for  $\ell =0$ (red squares), $\ell=2$ (blue triangles),  
$N=50, \rho=0.16$ and $\langle r \rangle = 0.1$ compared with field-strength correlators of ${\bf B}^2$ (black circles) and 
${\bf B}\otimes {\bf B}$ (green diamonds). The two $\ell=2$ correlation functions are virtually indistinguishable. 
Right: Wilson loop correlation functions $ D_{2}(t)$ (\ref{wlc4}) for $N=50, \rho=0.08$ and $\langle r \rangle = 0.2, 0.3,0.4,0.5$ 
(green diamonds, blue circles, blacks stars, red squares). The extracted slopes are $-6.6, -7.8, -7.5, -9.3$. }
\label{wlbbf}
\end{figure} 

\noindent{\boldmath $W^{00}_r, W^{2m}_r$}\\
For various ensembles, we have investigated Wilson loop correlation functions for loops of different radii. We have considered
 both Wilson loops of fixed radius and homogeneously distributed radii. For an appropriate choice of the parameters, the 
 observables have been found  to be insensitive to this choice. Due to the higher computational complexity as compared to
 the correlation functions of the local operators, less configurations and for a given configuration fewer points for the momentum
 projection had to be used. The use of Wilson loop correlation functions  did not improve the determination of the $0^+$ glueball mass,
 since also in this case difficulties due to  the subtraction of the vacuum expectation value with the concomitant sensitivity to deficits
 in rotational and translational invariance occur. As shown by the right side of Fig.~\ref{wlbbf}, slopes (or masses) could be extracted from the
 $2^+$ correlation functions with an average value of about 8 and an uncertainty of about 15\%.  Form this figure we also can read
 off  the relative normalization of the correlation functions for the different sizes of the Wilson loops. As compared to the correlation 
functions for  $\langle r \rangle=0.2$ the strength of the coupling to the $2^+$ excitations is enhanced by  a factor of 5.7 for the 
correlation function with  $\langle r \rangle=0.4$. One therefore might expect a larger systematic uncertainty in the slope extracted
 for $\langle r \rangle=0.2$. As the figure shows, the  size of the Wilson loop that produces the maximal overlap with the exited 
$2^+$ state is 0.4 and the extracted slope essentially does not change when varying the size from 0.3 to 0.4. Adopting the rational 
that the maximal overlap produces the most meaningful slope, we extract from this correlation function the value  7.7. We observe
 that this result is within $\sim 10 \%$ compatible with the slope extracted from the $T_+$ correlation function. We also note that the
 fact that the overlap is maximal for a Wilson loop of radius $r=0.4$ implies that the size of the state is approximately 0.47 fm, consistent
 with lattice calculations \cite{deke92,loyi06}.
%%%%%%%%%%%%%%%%%%%%%%%%%%%%%%%%%%%%%%%%%
\subsection{The Glueball Spectrum}
\begin{figure}[h]
\begin{center}
\epsfig{file=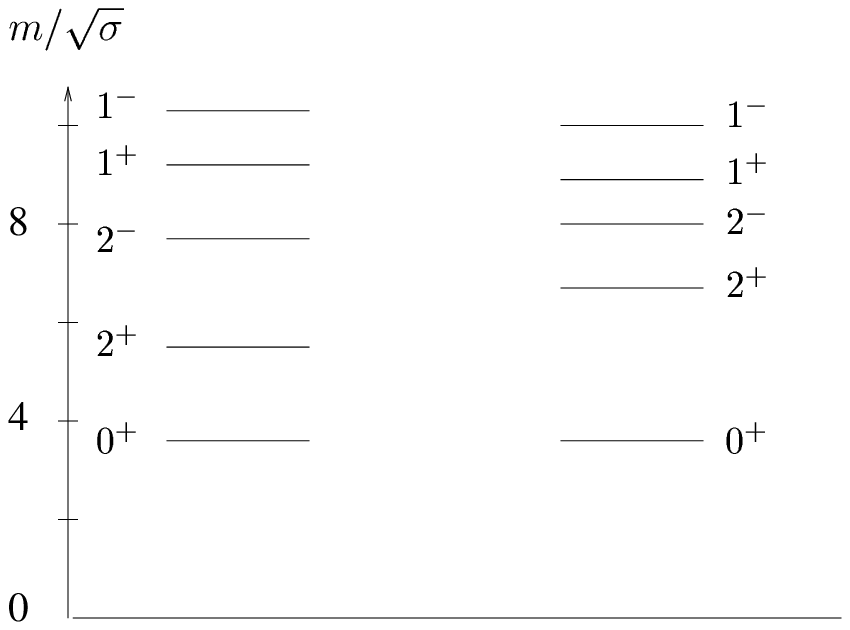,width=8cm,angle=0}
\caption{Spectra of 
$ 0^{+}, 2^{\pm}, 1^{\pm} $ glueball
states of  SU(2) Yang-Mills Theory. 
Left: Lattice gauge theory \cite{tepe98}. $\;$ Right: Masses from 
 pseudoparticle ensembles after rescaling (see text).}
\label{spek}
\end{center}
\end{figure}

%%%%%%%%%%%%%%%%%%%%%%%%%%%%%%%%%%%%%%%%%
The results for the glueball masses  extracted from our investigations of correlation functions of local operators built from
 2 and 3 field-strength operators and from the non-local Wilson loop operators are summarized in Table \ref{glma}. After a
 rescaling, our results are  compared in Fig.~\ref{spek} to the results of  the lattice gauge calculation of Ref.~\cite{tepe98}.
\begin{table}[h]
\begin{center} 
\begin{tabular}{||c|c|c||}  \hline
 \hline
  state &  slope& slope$/\sqrt{\sigma}$ \\ \hline \hline
  $0^+$ & 4.0 & 1.6\\ \hline
%  $2^+ $  &4.6 & 1.8 \\ \hline
 $2^+ $ & 7.7  & 3.1\\ \hline 
  $2^- $  &8.8 &3.5   \\ \hline
 $ 1^+ $& 9.6 &3.9\\ \hline
 $ 1^- $  &10.8 &4.4\\ \hline \hline
\end{tabular}
\caption{Glueball masses. Note that errors are discussed in the text and are not indicated here.}
\label{glma}
\end{center}
\end{table}
%\begin{figure}[h]
%\epsfig{file=spek1.eps, width=.5\linewidth,angle=0}
%\hspace{-1cm}\epsfig{file=bee1+fit3.ps, width=.45\linewidth,angle=-90}\hspace{-.9cm}
%\epsfig{file=eeb1-fit3.ps, width=.45\linewidth,angle=-90} 
%\caption{Spectra of 
%$ 0^{+}, 2^{\pm}, 1^{\pm} $ 
%states of  SU(2) Yang-Mills Theory. 
%Left: Lattice gauge theory \cite{tepe98}. $\;$ Right: Masses from 
% pseudoparticle ensembles after rescaling (see main text).}
%\vskip -3.6cm \hspace{.5cm}{\small $0^+$}\hspace{5.95cm}{\small $0^+$}
%\vskip -1cm \hspace{6.9cm}{\small $2^+$}
%\vskip -1.5cm \hspace{.5cm}{\small $2^+$}
%\vskip -1.1cm \hspace{6.9cm}{\small $2^+$}
%\vskip -1.15cm \hspace{6.9cm}{\small $2^-$}
%\vskip -.33cm \hspace{.5cm}{\small $2^-$}
%\vskip -1.2cm \hspace{6.9cm}{\small $1^+$}
%\vskip -.6cm \hspace{.5cm}{\small $1^+$}
%\vskip -1.cm \hspace{6.9cm}{\small $1^-$}
%\vskip -.6cm \hspace{.5cm}{\small $1^-$}
%\vskip 4.6cm \hspace{-.5cm} 0
%\vskip -2.45cm \hspace{-.5cm} 4
%\vskip -2.49cm \hspace{-.5cm} 8
%\vskip -2.49cm \hspace{-.5cm} $m/\sqrt{\sigma}$
%\label{spek}
%\vskip 7cm 
%\end{figure}
In comparison with the lattice results, the energy scale is too small by roughly a factor of  2. For a more detailed 
comparison we have rescaled our results with a factor of 2.2, which makes the value of the average excitation 
energy agree for the two spectra. Having adjusted the overall scale, we note that the  masses of the $0^+$, $2^-$ 
and $1^{\pm}$ glueballs of the two spectra agree within 5\% while the mass of the $2^+$ state is 25\% larger than 
the lattice value. We find this agreement surprising also in view of the large adjustment of the overall scale. These 
final results do not account for the possible existence of a second $2^+$ with an upper limit (after rescaling) for the 
mass of [cf. Eq.~(\ref{ulm2})]
$$m_{2^+}\le 2 \sqrt{\sigma}\, .$$
As emphasized above, the  smallness of the fiducial intervals did not permit us to firmly establish or rule out the existence of an 
additional low mass $2^+$ state.
  
The correlation functions for meron ensembles yield a very similar spectrum although the uncertainties in the 
extraction of the slopes, for the reasons discussed above, are  significantly higher. In particular, finer details like 
the splitting of the $1^{\pm}$ states could not be determined. As for the instanton ensembles, the existence of a $2^+$ state, 
essentially degenerate with the $0^+$ excitation, cannot be ruled out. The overall scale factor necessary to reproduce the 
average excitation energy of the lattice results is 1.8.    
%%%%%%%%%%%%%%%%%%%%%%%%%%%%%%%%%%%%%%%%%
\subsection{Wilson Loop Correlation Functions and Confinement}
%%%%%%%%%%%%%%%%%%%%%%%%%%%%%%%%%%%%%%%%%
In this concluding paragraph of our studies of correlation functions, we will return to the issue of confinement in the 
pseudoparticle ensembles. The physics of confinement is intimately related to the formation of  flux-tubes or gauge 
strings  and one may wonder whether the area law found in the fit to the Wilson loops  indeed implies that such a 
mechanism is operative also in the pseudoparticle ensembles. To study the dynamics of Wilson loops  we follow the 
investigations in \cite{FL06} and calculate correlation function of loops with fixed orientation. We assume  circular loops 
of equal radius which are parallel to each other and orthogonal to the direction of separation. We do not perform any 
averaging, i.e., we calculate the following correlators [cf. Eq.~(\ref{wlc1})]
\begin{equation}
  \label{grog}
C_r(t)= \langle W_{r} ({\bf n}, {\bf x}_{0}, t)W_{r} ({\bf n}, {\bf x}_{0},0 )\rangle\, .
\end{equation}
\begin{figure}[h]
\begin{center}
\hspace{-.6cm}\epsfig{file=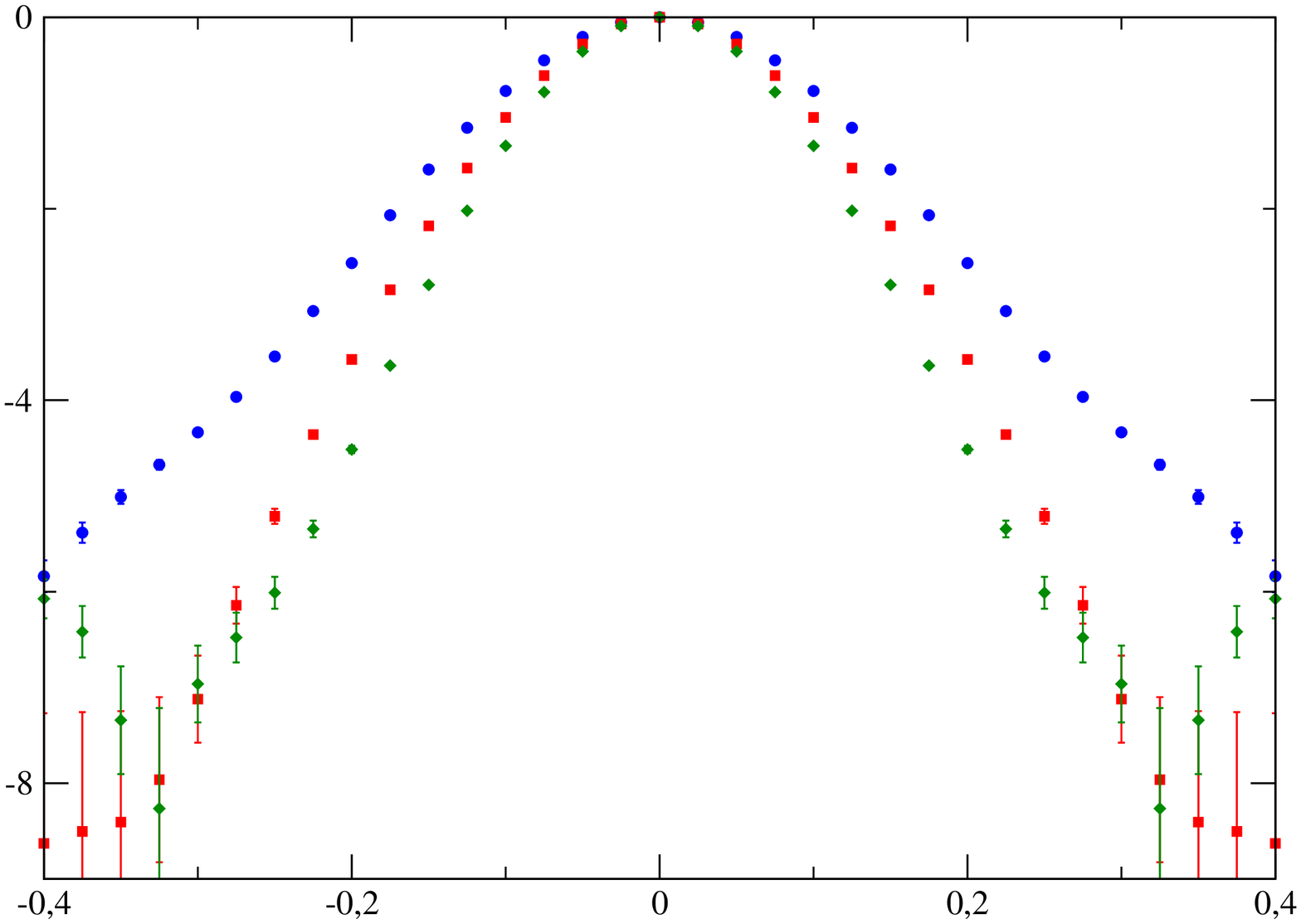, width=.4\linewidth,angle=-90}\hspace{-.1cm}\epsfig{file=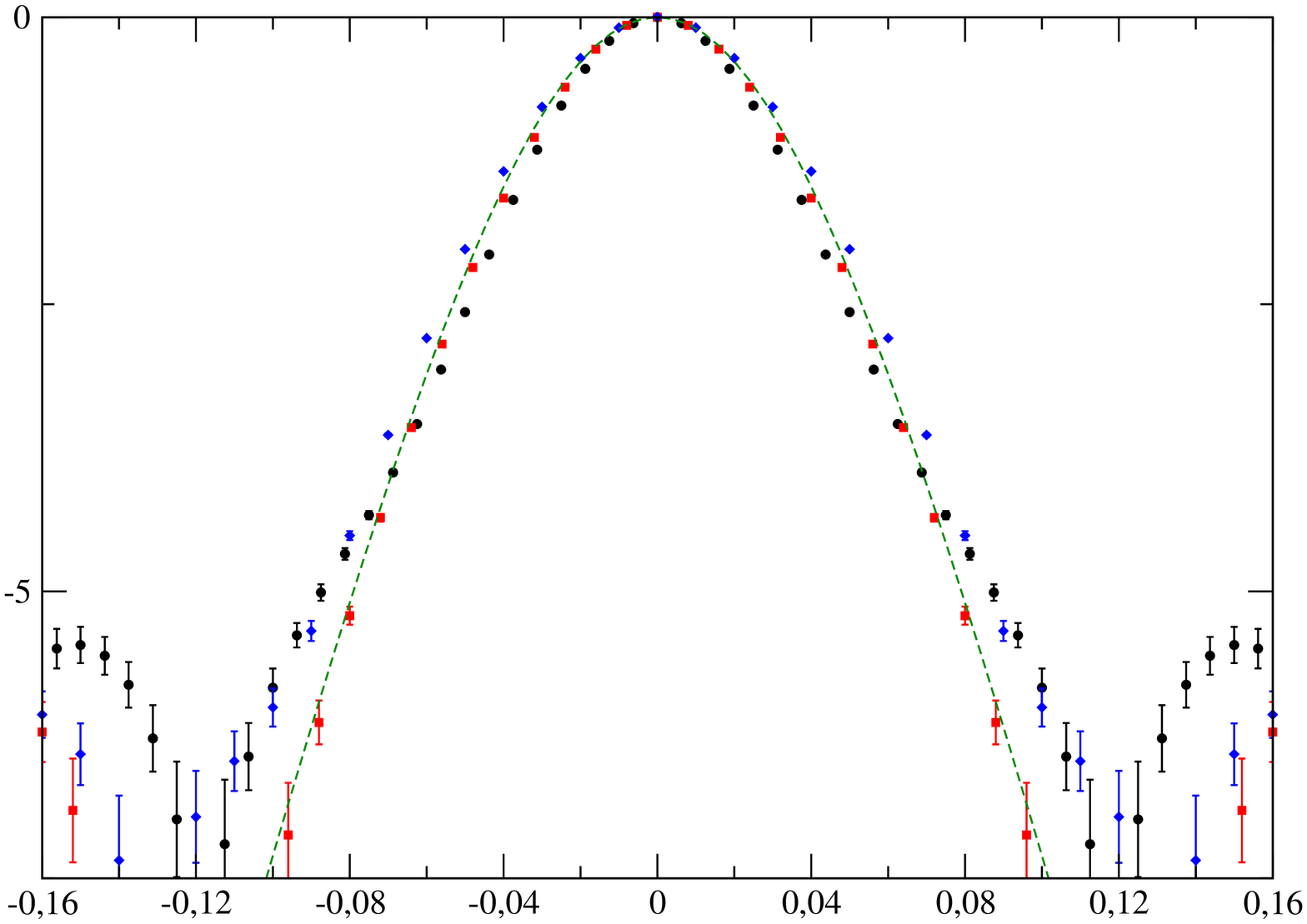, width=.4\linewidth,angle=-90}
\caption{Left: Logarithm of correlation functions of circular Wilson loops with fixed parallel spatial orientation for an ensemble
 of 500 instantons of size $\rho = 0.08$ as a function of separation in time $t$ for 3 values of the Wilson loop radius $r =0.25$ (black circles),
 $r=0.32$ (red squares), $r=0.40$ (blue diamonds) . \,  Right: The same as a function of the variable $rt$.}
\label{grogf}
\end{center}
\end{figure}
The results of the calculations are shown in Fig.~\ref{grogf}. If one assumes that the  application of  the Wilson loop operator to the
 vacuum generates a gauge string with (approximate) energy 
$$E= 2\pi\sigma r\, ,$$
the relevant variable for describing the correlation function is $rt$. As Fig.~\ref{grogf} shows, the three correlation functions are 
described approximatively by a universal curve in terms of this variable. It thus appears that the dynamics in the pseudoparticle 
ensembles is compatible with the formation of gauge strings or flux tubes. The existence of  such a universal curve offers the 
possibility for an alternative determination of the string tension. Due to the limited range of values a direct extraction of the string 
tension is not possible. We have parameterized the distributions by
\begin{equation}
  \label{groog}
\ln W =- \frac{a(rt)^2}{\sqrt{b+a(rt)^2}}\, .
\end{equation}
From the fit to the correlation functions of the Wilson loops shown in the right part of Fig.~\ref{grogf} we  obtain
\begin{equation}
\sigma = \frac{\sqrt{a}}{2\pi}= 17.4 \pm  6. 
\label{sigco}
\end{equation}
The  small  range of relevant values of the variable $rt$ does not permit a more precise determination of $\sigma$. The value (\ref{sigco}) of the string tension agrees within 20\% with that derived from the asymptotics of the Wilson loops (cf. Eqs. (\ref{stp2i}) and (\ref{stringas})).  

At large separations the universality breaks down as expected. For $  rt  \ge  r^2$ it becomes  energetically favorable for the system 
of two loops to couple by annihilation of the gauge string to the vacuum or glueball states. This corresponds to the Gross-Ooguri
phase transition in the large $N_c$ limit \cite{grog98}.
\section{Conclusions}
The  effective theories with pseudoparticle degrees of freedom studied in this work  describe  important properties of QCD.
The regular gauge instantons or the merons with their long-range gauge fields apparently constitute a proper and economical 
set of degrees of freedom in the description of the Yang-Mills dynamics in the infrared. By superposition of these building blocks, 
field configurations are generated with  confinement as an inherent  property. An area law for sufficiently large Wilson loops is 
already obtained in (stochastic) ensembles with the same weight assigned to  all field configurations. However such ensembles 
violate  basic principles. The action of field configurations in such ensembles is diverging in the infrared, concomitantly the ensembles 
do not possess a proper thermodynamic limit and violate translational invariance. By assigning the standard weight given by the 
action to the field configurations, all these problems are cured at once and confinement is preserved. The resulting configurations 
not only confine static color charges but also  the building blocks themselves. Only at the expense  of a logarithmically diverging 
action can a building block  be removed to infinity. The pseudoparticles are strongly correlated over large distances in order to
prevent a logarithmic rise in the action (for this reason a global update procedure had to be employed in our computations). On 
the other hand with its many spikes and valleys in color singlet quantities  like the action density or the topological susceptibility, 
single field configurations  appear strongly disordered. 

Thus, in our investigations, a picture of the Yang-Mills vacuum emerges which is that of a nematic substance -- a medium that is 
strongly correlated in color or ``director'' space ($S^3$ and $\mathbb{R}P^{2}$ respectively) and which at the same time is essentially 
disordered in configuration space (cf. \cite{frle04}). This picture summarizes our findings.  On the one hand, updates of the randomly 
chosen coordinates of the pseudoparticles in addition to the color update have no significant effect.  On the other hand, a color  disordered 
ensemble such as the stochastic ensemble we have studied does not produce  a qualitatively correct model of the Yang-Mills vacuum. 
Clearly, the color correlations in meron and instanton ensembles are at the heart of the phenomenon of confinement.  When a singular 
gauge instanton contributes to 2 Polyakov lines, it produces a linear potential, while singular instantons affecting one Polyakov line 
only are irrelevant. The last part of this argument \cite{Callan:1977gz} gets  modified in our context.  Due to the strong color correlations,
 single instantons still contribute to the potential.  In addition to  confinement of static charges, the color correlated fields reproduce 
remarkably well other important features of the Yang-Mills dynamics. They exhibit proper scaling under variations of the number of 
pseudoparticles  or the size of the system and they reproduce qualitatively characteristic properties such as the values of the gluon 
condensate and the topological susceptibility. Unlike the gluon condensate, the value of the topological susceptibility can be related
 to  the single pseudoparticle properties.  We also have been able to investigate finer details and  for instance to show analytically 
 that characteristic properties  of the Wilson loops in the limits of small and large size, i.e., the U(1) and the string limit, are qualitatively
reproduced. Here the role of the pseudoparticle size as an effective ultraviolet regulator is essential. We also qualitatively 
confirmed the string picture for Wilson loop correlation functions and thus obtained a significant consistency check of the description
 of confinement in term of pseudoparticles.     

The outcome of our investigations of  correlation functions is  more ambiguous. We have demonstrated the emergence of a hadronic 
scale under the usual circumstances  where the size of the building blocks is smaller than the emerging size of the ``many-body'' hadron.
 In the case of merons with their long range field-strength the opposite mechanism is at work and establishes   a  hadronic size  smaller
 than the constituent meron size. To study hadronic excitations,  we have computed momentum projected correlators of local gauge 
invariant operators with definite angular momentum built from field-strength bilinears and trilinears and have also calculated correlation 
functions of momentum and angular momentum projected Wilson loops.  With these computations, we have been able to determine the
 spectrum of glueballs with quantum numbers $0^{+}, 1^{\pm}$ and   $2^{\pm}$. While the structure of the spectrum agrees qualitatively
 with the corresponding lattice results, the masses determined are too small by 
an overall  factor of about 2 for both instanton and meron ensembles. We have not been able to uniquely identify the origin of this 
difficulty. Ambiguities in the determination of the string constant are unlikely to explain it. A smaller value deduced from the fit to
 the Wilson loop would, after identification with the empirical value, result in a higher energy scale. There is indeed some freedom left 
by redefinition of the relevant interval of the area of the loops. A change of the order of 20\% cannot be ruled out entailing a $10\%$ 
change in the energy scale. Corrections of a similar order of magnitude seem to result when using a ``smeared'' Wilson loop \cite{szwa}. 
Neither effect however approaches the necessary factor of 4 in the string tension.
The most likely explanation is that with the pseudoparticle ensembles primarily determined by infrared properties, the 
ultraviolet fluctuations are not adequately described. 
For a description of an object like the scalar 
glueball with a size of approximately $0.2 $ fm  \cite{deke92} the spatial resolution of our field configurations -- possibly limited by a too 
``small'' number of pseudoparticles -- may just be insufficient. In our discussion of small size Wilson loops we indeed have noticed the 
disappearance of the ``Coulomb regime'' with the disappearance  of the disorder in color space. To cure this problem one would have 
to either significantly enlarge the number of pseudoparticles or to introduce additional degrees of freedom which improve the description 
of the vacuum at small scales. Candidates for such additional degrees of freedom are singular gauge instantons which are the center
 reflected partners of the regular gauge instantons. Inclusion of these degrees of freedom would, by construction, avoid the otherwise
 problematic issue of multiple  counting of gauge copies.  

In addition to  such improvements in the description of the Yang-Mills theory, for phenomenological applications quarks have to be
incorporated. Here   the advantages of the continuum formulation of the  effective pseudoparticle theories could be essential.    
First steps in this endeavor have been taken \cite{sz06}. Here one aims to take over  the successful treatment of the dynamics of
quarks in the instanton liquid model \cite{Schafer:1996wv}.  The ultimate goal would be to establish in this way an effective theory 
that exhibits the two nonperturbative basic phenomena of QCD  -- confinement and chiral symmetry breaking.     

\section*{Acknowledgements}
It is a pleasure to acknowledge useful conversations with Thomas Schaefer, Mikhail Shifman,
Edward Shuryak, and Frank Wilczek.  J.N. is grateful for support by an Alexander
von Humboldt Foundation Research Award and for  hospitality at the Institute
for Theoretical Physics III at the University of Erlangen where this research
was initiated. This work was supported in part by
funds provided by the U.S. Department of Energy (D.O.E.) under grant DE-FG02-94ER40818.

\end{document}